\documentclass[12pt]{article}
\usepackage{amsmath, amssymb}
\usepackage{graphicx,epsf,subfigure} 
\usepackage{enumerate}
\usepackage{natbib}
\usepackage{url} 
\usepackage{setspace}

\newcommand{\blind}{1}

\usepackage[top=1in, bottom=1in, left=1in, right=1in]{geometry}

\renewcommand{\part}[1]{\vspace{.1in}\noindent\textbf{Step #1:}}

\newcommand{\bbeta}{\boldsymbol{\beta}}

\newcommand{\btheta}{\boldsymbol{\theta}}

\newcommand{\calB}{\mathcal{B}}

\newcommand{\calD}{\mathcal{D}}

\newcommand{\calK}{\mathcal{K}}

\newcommand{\calS}{\mathcal{S}}

\newcommand{\bE}{\mathbf{E}}

\newcommand{\bI}{\mathbf{I}}

\newcommand{\bfs}{\mathbf{s}}

\newcommand{\bR}{\mathbf{R}}
\newcommand{\bv}{\mathbf{v}}
\newcommand{\bV}{\mathbf{V}}

\newcommand{\bw}{\mathbf{w}}
\newcommand{\bW}{\mathbf{W}}
\newcommand{\bx}{\mathbf{x}}

\newcommand{\bz}{\mathbf{z}}

\newcommand{\bzero}{\mathbf{0}}

\newcommand{\mathR}{\mathbb{R}}

\numberwithin{equation}{section}

\begin{document}
\singlespacing

\if1\blind
{
	\title{\bf A unified framework for fitting Bayesian semiparametric models to arbitrarily censored survival data, including spatially-referenced data}
	\author{Haiming Zhou \\
		Division of Statistics, Northern Illinois University\\
		and \\
		Timothy Hanson\\
		Department of Statistics, University of South Carolina}
	\maketitle \thispagestyle{empty}
} \fi

\if0\blind
{
	\bigskip
	\bigskip
	\bigskip
	\begin{center}
		{\bf A unified framework for fitting Bayesian semiparametric models to arbitrarily censored survival data, including spatially-referenced data}
	\end{center}
	\medskip
	\thispagestyle{empty}
} \fi

\noindent {\large (\textit{To appear in Journal of the American Statistical Association})}

\bigskip
\begin{abstract}
	A comprehensive, unified approach to modeling arbitrarily censored spatial survival data is presented for the three most commonly-used semiparametric models: proportional hazards, proportional odds, and accelerated failure time. Unlike many other approaches, all manner of censored survival times are simultaneously accommodated including uncensored, interval censored, current-status, left and right censored, and mixtures of these. Left-truncated data are also accommodated leading to models for time-dependent covariates.  Both georeferenced (location exactly observed) and areally observed (location known up to a geographic unit such as a county) spatial locations are handled; formal variable selection makes model selection especially easy.  Model fit is assessed with conditional Cox-Snell residual plots, and model choice is carried out via LPML and DIC.  Baseline survival is modeled with a novel transformed Bernstein polynomial prior. All models are fit via a new function which calls efficient compiled C++ in the R package \texttt{spBayesSurv}. The methodology is broadly illustrated with simulations and real data applications.  An important finding is that proportional odds and accelerated failure time models often fit significantly better than the commonly-used proportional hazards model. Supplementary materials are available online.
\end{abstract}

\noindent%
\vfill

\clearpage
\setcounter{page}{1}
\doublespacing
\section{Introduction}

Spatial location often plays a key role in prediction, serving as a proxy for unmeasured regional characteristics such as socioeconomic status, access and quality of healthcare, pollution, etc.  Spatial models use location both as a means for blocking, leading to more precisely estimated non-spatial risk factors, but also as a focal point of inference in its own right, for example to delineate regional ``hot spots'' or outbreaks that merit closer attention or increased resources.  Literature on the spatial analysis of time-to-event data related to human health has flourished over the last decade, including data on leukemia survival \citep{Henderson.etal2002}, infant/childhood mortality \citep{Banerjee.etal2003,Kneib2006}, coronary artery bypass grafting \citep{Hennerfeind.etal2006}, asthma \citep{Li.Ryan2002,Li.Lin2006}, breast cancer \citep{Zhao.Hanson2011,Hanson.etal2012,Zhou.etal2015}, mortality due to air pollution \citep{Jerrett.etal2013}, colorectal cancer survival \citep{Liu.etal2014b}, smoking cessation \citep{Pan.etal2014}, HIV/AIDS patients \citep{Martins.etal2016}, time to tooth loss \citep{Schnell.etal2015}, and many others.  Spatial survival models have also been used in other important areas such as the study of political event processes \citep{Darmofal2009}, agricultural mildew outbreaks \citep{Ojiambo.Kang2013}, forest fires \citep{Morin2014}, pine trees \citep{Li.etal2015}, health and pharmaceutical firms \citep{Arbia.etal2017}, and emergency service response times \citep{Taylor2017} to name a few.  

All twenty papers referenced above use Cox's \citep{Cox1972} proportional hazards (PH) model for inference; competing models are not considered.  There are a few papers using models alternative to PH in a spatial context, e.g. \cite{Diva.etal2008}, \cite{Zhao.etal2009}, \cite{Wang.etal2012}, and \cite{Li.etal2015b} but they tend to be limited in scope.  For example, all four of these latter papers only consider areal data, do not employ variable selection or allow time-dependent covariates, and are developed for right censored data only.  In general the literature is fragmented in that a method for variable selection in the PH model for non-spatial right censored data will comprise one paper; another paper may extend the PH model for fitting current status data to general interval censored data, but does not include variable selection, spatial frailties, etc.  In this paper, a broadly inclusive, comprehensive treatment of three competing, highly interpretable semiparametric survival models, PH, proportional odds (PO), and accelerated failure time (AFT), is developed and illustrated on several real and simulated data sets.  This work represents the culmination of a great deal of effort tying together many disparate ideas and methodologies in the literature as well as developing an efficient approach to mixed interval censored, exactly observed, and truncated data that can be widely applied to different semiparametric models.  

Model formulation consists of a parametric portion giving relative risk (PH), odds (PO), or acceleration factors (AFT) coupled with georeferenced or areal spatial frailties, and a nonparametric baseline survival function modeled as a transformed Bernstein polynomial (TBP) \citep{Chen.etal2014} centered at a standard parametric family: one of log-logistic, log-normal, or Weibull.  Centering the baseline allows prior mass to be roughly guided by the parametric family.  The resulting distribution behaves similarly to a B-spline but where knot locations are guided by the centering family, blending the merits of both parametric and nonparametric approaches.  Unlike mixtures of Dirichlet processes \citep{Antoniak1974}, gamma processes \citep{Kalbfleisch1978}, and mixtures of Polya trees \citep{Lavine1992}, the TBP is smooth and has a finite number of parameters, yielding an explicit likelihood that allows for many immediate generalizations and efficient block-adaptive Markov chain Monte Carlo (MCMC) inference.  With judicious choice of blocking and careful use of existing parametric survival model fits, the block-adaptive approach cultivated here is robust, fully automated (e.g. no ``tuning'' is required), is relatively fast, and has been applied to data sets of up to a million observations.  Furthermore, all of the machinery developed is available in a powerful, freely-available R function \texttt{survregbayes} calling compiled C++ in the \texttt{spBayesSurv} package \citep{Zhou.Hanson2016} for R.  Note that although the methodology is developed for both areal and georeferenced spatial time-to-event data, non-spatial data are also accommodated.  All manner of exactly observed, right censored, interval censored, and left-truncated data are accommodated, as well as mixtures of these. Left-truncation allows for the inclusion of time-dependent covariates and spike-and-slab variable selection is also implemented.  Finally, a separate function that computes a variation on the Cox-Snell residual plot allows for gross assessment of model fit.  The ready availability of software to easily fit the models developed herein allows researchers to empirically compare various competing semiparametric models on their own survival data, as well as allowing comparison to other survival models in the literature.

Section 2 describes the models including the TBP, georeferenced and areal frailties, MCMC, diagnostics and model selection criteria.  Three illustrative data analyses comprise Section 3.  Section 4 offers simulations illustrating the quality of estimation as well as comparing to the R packages \texttt{ICBayes} \citep{Pan.etal2015}, \texttt{bayesSurv} \citep{Komarek.Lesaffre2008} and \texttt{R2BayesX} \citep{Umlauf.etal2015,Belitz.etal2015}.  The paper is concluded in Section 5.  Tests for parametric baseline, stochastic search variable selection, left-truncation, time-dependent covariates, partially linear predictors, more simulations, as well as many other details are further discussed in the online supplementary material accompanying this paper.  

\section{The Models}
Subjects are observed at $m$ distinct spatial locations $\bfs_1,\dots,\bfs_m$.  For areally-observed outcomes, e.g. county-level, there is typically replication at each location; for georeferenced data there may or may not be replication. Let $t_{ij}$ be the (possibly censored) survival time for subject $j$ at location $\bfs_i$ and $\bx_{ij}$ be the corresponding $p$-dimensional vector of covariates, $i=1, \ldots, m, j=1,\ldots,n_i$; let $n=\sum_{i=1}^{m}n_i$ be the total number of subjects. Assume the survival time $t_{ij}$ lies in the interval $(a_{ij},b_{ij})$, $0\leq a_{ij}\leq b_{ij}\leq \infty$. Here left censored data are of the form $(0,b_{ij})$, right censored $(a_{ij},\infty)$, interval censored $(a_{ij},b_{ij})$ and uncensored values simply have $a_{ij}=b_{ij}$, i.e. define $(x,x)=\{x\}$. The event and censoring times are assumed to be independent given the observed covariates. Note that both current status data and case 2 interval censored data \citep{Sun2006}, arise as particular special cases. 

Spatial dependence often arises among survival outcomes due to region-specific similarities in ecological and/or social environments that are typically not measurable or omitted due to confidentiality concerns. To incorporate such spatial dependence, a traditional way is to introduce random effects (frailties) $v_1,\dots,v_m$ into the linear predictor of survival models. In this paper we consider three commonly-used semiparametric models: AFT, PH, and PO.  Given spatially-varying frailties $v_1,\dots,v_m$, regression effects $\bbeta=(\beta_1, \ldots, \beta_p)'$, and baseline survival $S_0(\cdot)$ with density $f_0(\cdot)$ corresponding to $\bx_{ij}=\bzero$ and $v_i=0$, the AFT model has survival and density functions 
\begin{equation} \label{aft}
	S_{\bx_{ij}}(t)=S_0(e^{\bx_{ij}'\bbeta+v_i} t), \ f_{\bx_{ij}}(t)=e^{\bx_{ij}'\bbeta+v_i} f_0(e^{\bx_{ij}'\bbeta+v_i}t), 
\end{equation}
while the PH model has survival and density functions
\begin{equation} \label{ph} 
	S_{\bx_{ij}}(t)=S_0(t)^{e^{\bx_{ij}'\bbeta+v_i}}, \ f_{\bx_{ij}}(t)=e^{\bx_{ij}'\bbeta+v_i} S_0(t)^{e^{\bx_{ij}'\bbeta+v_i}-1} f_0(t), 
\end{equation}  
and the PO model has survival and density functions
\begin{equation} \label{po}
	S_{\bx_{ij}}(t)=\frac{e^{-\bx_{ij}'\bbeta-v_i}S_0(t)}{1+(e^{-\bx_{ij}'\bbeta-v_i}-1)S_0(t)}, \ f_{\bx_{ij}}(t)=\frac{e^{-\bx_{ij}'\bbeta-v_i}f_0(t)}{[1+(e^{-\bx_{ij}'\bbeta-v_i}-1)S_0(t)]^2}.
\end{equation}
In semiparametric survival analysis, a wide variety of Bayesian nonparametric priors can be used to model $S_0(\cdot)$;
see \cite{Mueller.etal2015} and \cite{Zhou.Hanson2015} for reviews. In this paper we consider the models (\ref{aft}), (\ref{ph}) and (\ref{po}) with a TBP prior on $S_0(\cdot)$. 

\subsection{Transformed Bernstein Polynomial Prior}
For a given positive integer $J \ge 1$, define the Bernstein polynomial of degree $J-1$ as a particular B-spline over $(0,1)$ given by
\begin{equation}\label{bp:pdf}
	d(x|J,\bw_J) = \sum_{j=1}^{J} w_{j} \delta_{j,J}(x) \equiv \sum_{j=1}^{J} w_{j} \frac{\Gamma(J+1)}{\Gamma(j)\Gamma(J-j+1)}x^{j-1}(1-x)^{J-j},
\end{equation}
where $\bw_J=(w_{1}, \ldots, w_{J})'$ is a vector of positive weights satisfying $\sum_{j=1}^{J}\bw_{j}=1$ and $\delta_{j,J}(x)$ denotes a beta density with parameters $(j,J-j+1)$. Smooth densities with support $(0,1)$ can be well approximated by a Bernstein polynomial \citep{Ghosal2001}: if $f(x)$ is any continuously differentiable density with support $(0,1)$ and bounded second derivative, $\bw_J$ can be chosen such that
\[ \sup_{0<x<1} |f(x) - d(x|J, \bw_J)| = O(J^{-1}).
\]
Integrating (\ref{bp:pdf}) gives the corresponding cumulative distribution function (cdf)
\begin{equation}\label{bp:cdf}
	D(x|J,\bw_J) = \sum_{j=1}^{J} w_{j} \Delta_{j,J}(x),
\end{equation}
where $\Delta_{j,J}(x)$ is the cdf associated with $\delta_{j,J}(x)$. One can calculate the summands in (\ref{bp:cdf}) recursively as
\[ \Delta_{j+1,J}(x)=\Delta_{j,J}(x)- \frac{\Gamma(J+1)}{\Gamma(j+1)\Gamma(J-j+1)}x^j (1-x)^{J-j}.
\] 

By assigning a joint prior distribution to $(J, \bw_J)$, the random $d(x|J,\bw_J)$ in (\ref{bp:pdf}) is said to have the Bernstein polynomial (BP) prior. \cite{Petrone1999b} showed that if the prior on $(J, \bw_J)$ has full support, the BP prior has positive support on all continuous density functions on $(0,1)$. However, for practical reasons, the degree $J$ is often truncated to a large value, say $\calK$, so that the prior has support $\calB_{\calK}=\{d(x|J, \bw_J): J\leq \calK\}$. Under mild conditions, \cite{Petrone.Wasserman2002} showed that, for a fixed $\calK$, the posterior density almost surely converges (as $n\rightarrow \infty$) to a density that minimizes the Kullback-Leibler divergence of the true density against $d(x)\in\calB_{\calK}$. 
BP priors of lesser degree have the same support as $\calB_{\calK}$ because any Bernstein polynomial can be written in terms of Bernstein polynomials of higher degree through the relationship
\[\delta_{j,J-1}(x)   = \frac{J-j}{J}  \delta_{j,J}(x)  + \frac{j}{J}  \delta_{j+1,J}(x).\]
It follows that $d(x|J-1, \bw_{J-1})$ can be written as $d(x|J, \bw^*_{J})$ with a suitable
choice of $\bw^*_{J}$, so every $d(x|J, \bw_{J})$ with $J\leq \calK$ belongs to $\{ d(x|\calK, \bw_{\calK}) \}$.  Therefore, following \cite{Chen.etal2014} we fix $J$ throughout, with $J=15$ being the software's default.  This is roughly equivalent to having 15 knots in a B-spline representation of the transformed baseline survival function, described next.

Let $\{S_{\btheta}(\cdot):\btheta \in \boldsymbol{\Theta}\}$ denote a parametric family of survival functions with support on positive reals $\mathR^+$. The log-logistic family $S_{\btheta}(t)=\{1+(e^{\theta_1} t)^{\exp(\theta_2)}\}^{-1}$ is considered in the examples in Section 3, where $\btheta=(\theta_1, \theta_2)'$; Weibull and log-normal families are also included in the R package. Note that $S_{\btheta}(t)$ always lies in the interval $(0,1)$ for $0<t<\infty$, so a natural prior on $S_0(\cdot)$, termed the TBP prior, is simply 
\begin{equation}\label{bp:S0}
	S_0(t) = D(S_{\btheta}(t)|J, \bw_J) \text{ with density } f_0(t) = d(S_{\btheta}(t)|J, \bw_J)f_{\btheta}(t),
\end{equation}
where $f_{\btheta}$ is the density associated with $S_{\btheta}(\cdot)$.  Clearly, the random distribution $S_0(\cdot)$ is centered at $S_{\btheta}(\cdot)$, that is, $E[S_0(t)]=S_{\btheta}(t)$ and $E[f_0(t)]=f_{\btheta}(t)$. The weight parameters $\bw_J$ adjust the shape of the baseline survival $S_0(\cdot)$ relative to the centering distribution $S_{\btheta}(\cdot)$. This adaptability makes the TBP prior attractive in its flexibility, but also anchors the random $S_0(\cdot)$ firmly about $S_{\btheta}(\cdot)$: $w_j=1/J$ for $j=1,\dots,J$ implies $S_0(t)=S_{\btheta}(t)$ for $t \ge 0$.  Moreover, unlike the mixture of Polya trees \citep{Lavine1992} or mixture of Dirichlet process \citep{Antoniak1974} priors, the TBP prior selects smooth densities, leading to efficient posterior sampling.  Initially we had implemented the models using mixtures of Polya trees (and this function is still available in the \texttt{spBayesSurv} R package with limited functionality); these models suffered poor mixing, especially for the AFT model.  The TBP provided essentially the same posterior inference or better (as measured by LPML) with vastly improved MCMC mixing; see supplemental Appendix J.4.

Regarding the prior for $\bw_J$, we consider a Dirichlet distribution, $\bw_J|J \sim \textrm{Dirichlet}(\alpha, \ldots, \alpha)$, 
where $\alpha>0$ acts like the precision in a Dirichlet process \citep{Ferguson1973}, controlling how stochastically ``pliable'' $S_0(\cdot)$ is relative to $S_{\btheta}(\cdot)$. Large values of $\alpha$ indicate a strong belief that $S_0(\cdot)$ is close to $S_{\btheta}(\cdot)$: as $\alpha \rightarrow \infty$, $S_0(t) \rightarrow S_{\btheta}(t)$ with probability 1. Smaller values of $\alpha$ allow more pronounced deviations of $S_0(\cdot)$ from $S_{\btheta}(\cdot)$. 
A gamma prior on $\alpha$ is considered: $\alpha\sim\Gamma(a_{\alpha},b_{\alpha})$ where $E(\alpha)=a_{\alpha}/b_{\alpha}$.

\subsection{Spatial Frailty Modeling}
Spatial frailty models are usually grouped into two general settings according to their underlying data structure \citep{Banerjee.etal2014}: \textit{georeferenced} data, where $\bfs_i$ varies continuously throughout a fixed study region $\calS$ (e.g.,  $\bfs_i$ is recorded as longitude and latitude); and \textit{areal} data, where $\calS$ is partitioned into a finite number of areal units with well-defined boundaries (e.g., $\bfs_i$ represents a county).

\subsubsection{Areal Data Modeling} \label{sec:adm}
We consider an intrinsic conditionally autoregressive (ICAR) prior \citep{Besag1974} on $\bv=(v_1, \ldots, v_m)'$. Let $e_{ij}$ be $1$ if regions $i$ and $j$ are neighbors (which can be defined in various ways) and 0 otherwise; set $e_{ii}=0$. Then the $m\times m$ matrix $\bE=[e_{ij}]$ is called the adjacency matrix for the $m$ regions. The ICAR prior on $\bv$ is defined through the set of the conditional distributions
\begin{equation}\label{eq:areal-prior}
v_i | \{v_j\}_{j \neq i} \sim N\left(\sum_{j=1}^m e_{ij}v_j/e_{i+}, ~\tau^2/e_{i+} \right), ~ i=1,\ldots,m,
\end{equation}
where $e_{i+} = \sum_{j=1}^m e_{ij}$, $\tau$ is a scale parameter, and $\tau^2/e_{i+}$ is the conditional variance.  The induced prior on $\bv$ under ICAR is improper; the constraint $\sum_{j=1}^m v_j=0$ is used for identifiability \citep{Banerjee.etal2014}. Note that we assume that every region has at least one neighbor, so the proportionality constant for the improper density of $\bv$ is $(\tau^{-2})^{(m-1)/2}$ \citep{Lavine.Hodges2012}.  

A referee has asked about the inclusion of the proper CAR prior, which is not currently supported by the software accompanying this article.  The ICAR has been generally favored over its proper counterpart, the CAR prior, for several reasons.  
The CAR prior includes a parameter $\kappa$ which shrinks the frailties toward zero: $v_i | \kappa,\{v_j\}_{j \neq i} \sim N\left(\kappa\sum_{j=1}^m e_{ij}v_j/e_{i+}, \tau^2/e_{i+} \right)$.  As $\kappa \rightarrow 1^-$ the improper ICAR results.  \cite{Paciorek2009} suggests that the shrinkage from a proper CAR is ``generally unappealing'' in the spatial setting; similarly \cite{Banerjee.etal2014} question the sensibility of smoothing a spatial effect toward a proportion $\kappa<1$ of the mean of its neighbors. \cite{Paciorek2009} further notes that posterior estimates of $\kappa$ are often close to one, essentially yielding the ICAR.  This is likely due to the rather modest amount of correlation the proper CAR provides unless $\kappa \approx 1$; see \cite{Assuncao.Krainski2009} and \cite{Banerjee.etal2014}.   

When spatial smoothing is not of interest, we consider independent and identically distributed (IID) Gaussian frailties, $v_1, \ldots, v_m \overset{iid}{\sim}N(0,\tau^2)$.  This is a special case of the proper CAR prior where $\kappa=0$, so the software allows for either no correlation or the maximal limiting correlation of the proper CAR prior.

\subsubsection{Georeferenced Data Modeling}
For georeferenced data, it is commonly assumed that $v_i=v(\bfs_i)$ arises from a Gaussian random field (GRF) $\{v(\bfs), \bfs\in \calS\}$ such that $\bv=(v_1,\ldots, v_m)$ follows a multivariate Gaussian distribution as $
\bv\sim N_m(\bzero, \tau^2\bR)$, where $\tau^2$ measures the amount of spatial variation across locations and the $(i,j)$ element of $\bR$ is modeled as $\bR[i,j]=\rho(\bfs_i, \bfs_j)$, where $\rho(\cdot,\cdot)$ is a correlation function controlling the spatial dependence of $v(\bfs)$. In this paper, we consider the powered exponential correlation function $\rho(\bfs, \bfs')=\rho(\bfs, \bfs';\phi) = \exp\{ -(\phi \|\bfs-\bfs'\|)^\nu\}$, where $\phi>0$ is a range parameter controlling the spatial decay over distance, $\nu\in(0,2]$ is a shape parameter, and $\|\bfs-\bfs'\|$ is the distance (e.g., Euclidean, great-circle) between $\bfs$ and $\bfs'$.  Note that $\phi \rightarrow \infty$ gives the IID Gaussian prior described in Section \ref{sec:adm} as a special case.  Similar to ICAR, the conditional prior is given by
\begin{equation}\label{eq:point-prior}
v_i | \{v_j\}_{j \neq i} \sim N\left(-\sum_{\{j:j\neq i\}} p_{ij}v_j/p_{ii}, ~\tau^2/p_{ii} \right), ~ i=1,\ldots,m,
\end{equation}
where $p_{ij}$ is the $(i,j)$ element of $\bR^{-1}$. 

As $m$ increases evaluating $\bR^{-1}$ from $\bR$ becomes computationally impractical.  Various approaches have been developed to overcome this computational issue such as process convolutions \citep{Higdon2002}, fixed rank kriging \citep{Cressie.Johannesson2008}, predictive process models \citep{Banerjee.etal2008}, sparse approximations \citep{Kaufman.etal2008}, and the full-scale approximation \citep{Sang.Huang2012}. We consider the full-scale approximation (FSA) as an option when fitting models in the accompanying R software due to its capability of capturing both large- and small-scale spatial dependence; see supplementary Appendix B for a brief introduction.  

The FSA was arrived at after a number of lengthy, unsuccessful attempts at using other approximations.  BayesX \citep{Belitz.etal2015} uses what have been termed ``Mat\'ern splines,'' first introduced in an applied context by \cite{Kammann.Wand2003}.  Several authors have used this approach including \cite{Kneib2006}, \cite{Hennerfeind.etal2006}, and \cite{Kneib.Fahrmeir2007}.  This approximation was termed a ``predictive process'' and given a more formal treatment by \cite{Banerjee.etal2008}.  In our experience, the predictive process tends to give biased regression effects and prediction when the rank (i.e. the number of knots) was chosen too low; the problem worsened with no replication and/or when spatial correlation was high.  The FSA fixes the predictive process by adding tapering to the ``residual'' process.  We have been able to successfully analyze data with several thousand georeferenced spatial locations via MCMC using the FSA option with little to no artificial bias in parameter estimation and consistently good predictive ability. 

Choices alternative to the powered exponential such as the spherical and Mat\'ern correlation functions may be of interest, however the current package only supports the powered exponential correlation with pre-specified $\nu$.  This choice has proven to be adequate across many simulated data sets and is partially borne from our experience with very weakly identified parameters in ``too flexible'' correlation structures.  Initially we used the Mat\'ern but moved to the powered exponential when Mat\'ern parameters would invariably have very poor MCMC mixing and biased estimates in all but very large samples.  Note that the exponential correlation function, also a special case of Mat\'ern, is given by $\nu=1$ yielding continuous but not differentiable sample paths.  Gaussian correlation, a limiting case of Mat\'ern, is given by $\nu=2$, providing (infinitely) differentiable sample paths.

BayesX and R-INLA \citep{Martins.etal2013} both include the Mat\'ern correlation function $\rho(\bfs, \bfs';\phi) = [2^{\nu-1}\Gamma(\nu)]^{-1}(\phi \|\bfs-\bfs'\|)^{\nu}\mathcal{K}_{\nu}(\phi \|\bfs-\bfs'\|)$ where $\mathcal{K}_{\nu}$ is the modified Bessel function of order $\nu$.  Noting weakly identified Mat\'ern parameters, BayesX requires the user to fix one of $\nu \in \{0.5,1.5,2.5,3.5\}$ ($\nu=0.5$ gives exponential correlation; $\nu \rightarrow \infty$ gives Gaussian) and fixes $\phi^{-1}$ at $\hat{\phi}^{-1}=\max_{i,j}||\bfs_i-\bfs_j||/c$ with $c$ chosen so the correlation between the two farthest locations is small, e.g. 0.001.  R-INLA requires the user to fix one of $\nu \in \{1.5,2.5,3.5\}$ but allows the range parameter $\phi$ to be estimated from the data.  So the approach of BayesX fixes both parameters of the Mat\'ern correlation and R-INLA allows the range $\phi$ to be random (as do we).

\subsection{Likelihood Construction and MCMC}\label{sec:mcmc}
Let $\mathcal{D}=\{(a_{ij},b_{ij},\bx_{ij},\bfs_i); i=1,\ldots, m, j=1,\ldots,n_i\}$ be the observed data. Assume $t_{ij} \sim S_{\bx_{ij}}(t)$ following one of (\ref{aft}), (\ref{ph}), or (\ref{po}) with the TBP prior on $S_0(t)$ defined in (\ref{bp:S0}), and $\bv$ following (\ref{eq:areal-prior}) or (\ref{eq:point-prior}). The likelihood for $(\bw_J, \btheta, \bbeta, \bv)$ is 
\begin{equation} 
	L(\bw_J, \btheta, \bbeta, \bv)=\prod_{i=1}^m \prod_{j=1}^{n_i} \left[S_{\bx_{ij}}(a_{ij})-S_{\bx_{ij}}(b_{ij})\right]^{I\{a_{ij}<b_{ij}\}} f_{\bx_{ij}}(a_{ij})^{I\{a_{ij}=b_{ij}\}}. \label{like} 
\end{equation}
MCMC is carried out through an empirical Bayes approach coupled with adaptive Metropolis samplers \citep{Haario.etal2001}.  The posterior density is 
\[ p(\bw_J, \btheta, \bbeta, \bv, \alpha, \tau^2, \phi|\calD) \propto L(\bw_J, \btheta, \bbeta, \bv) p(\bw_J|\alpha)p(\alpha)p(\btheta)p(\bbeta)p(\bv|\tau^2)p(\tau^2)p(\phi), \]
where each $p(\cdot)$ represents a prior density, and $p(\phi)$ is only included for georeferenced data. Assume $\btheta\sim N_2(\btheta_0, \bV_0)$, $\bbeta\sim N_p(\bbeta_0, \bW_0)$, $\alpha\sim \Gamma(a_{\alpha}, b_{\alpha})$, $\tau^{-2}\sim \Gamma(a_\tau, b_\tau)$ and $\phi\sim \Gamma(a_\phi, b_\phi)$. 

Recall that $w_{j}=1/J$ implies the underlying parametric model with $S_0(t)=S_{\btheta}(t)$.  Thus, the parametric model provides good starting values for the TBP survival model. Let $\hat{\btheta}$ and $\hat{\bbeta}$ denote the parametric estimates of $\btheta$ and $\bbeta$, and let $\hat{\bV}$ and $\hat{\bW}$ denote their estimated covariance matrices, respectively. Set $\bz_{J-1}=(z_1,\ldots,z_{J-1})'$ with $z_j=\log(w_{j})-\log(w_{J})$. The $\bbeta$, $\btheta$, $\bz_{J-1}$, $\alpha$ and $\phi$ are all updated using adaptive Metropolis samplers, where the initial proposal variance is $\hat{\bW}$ for $\bbeta$, $\hat{\bV}$ for $\btheta$, $0.16\bI_{J-1}$ for $\bz_{J-1}$ and $0.16$ for $\alpha$ and $\phi$. Each frailty term $v_i$ is updated via Metropolis-Hastings, with proposal variance as the conditional prior variance of $v_i | \{v_j\}_{j \neq i}$; $\tau^{-2}$ is updated via a Gibbs step from its full conditional. A complete description and derivation of the updating steps are in supplementary Appendix A. To determine the running length of an MCMC run, one may first run a short chain without thinning, then use R packages such as \texttt{coda} \citep{Plummer.etal2006} and \texttt{mcmcse} \citep{Flegal.etal2016} for convergence diagnosis and effective sample size calculations. 

Regarding the default choice for the hyperparameters, when variable selection is not implemented (see Appendix E in the online supplementary material) we set $\bbeta_0=\bzero$, $\bW_0=10^{10}\bI_p$, $\btheta_0=\hat{\btheta}$, $\bV_0=10\hat{\bV}$, $a_{\alpha}=b_{\alpha}=1$, and $a_\tau=b_\tau=.001$. Note here we assume a somewhat informative prior on $\btheta$ to obviate confounding between $\btheta$ and $\bw_J$.  For georeferenced data, we set $a_{\phi}=2$ and $b_\phi=(a_{\phi}-1)/\phi_0$ so that the prior of $\phi$ has mode at $\phi_0$, where $\phi_0$ satisfies $\rho(\bfs',\bfs'';\phi_0)=0.001$ with $\|\bfs'-\bfs''\|=\max_{i,j}\|\bfs_i-\bfs_j\|$. Note that \cite{Kneib.Fahrmeir2007} simply fix $\phi$ at $\phi_0$, while we allow $\phi$ to be random around $\phi_0$. 

\subsection{Model Diagnostics and Comparison}\label{sec:diagnoistics}

For model diagnostics, a general residual defined in \cite{Cox.Snell1968} has been widely used in a variety of regression settings. Define $r(t_{ij}) = - \log S_{\bx_{ij}}(t_{ij})$ for $i=1,\ldots,m$, $j=1,\ldots, n_i$, then $r(t_{ij})$, given $S_{\bx_{ij}}(\cdot)$, has a standard exponential distribution. Therefore, if the model is ``correct,'' and under arbitrary censoring, the pairs $\{ r(a_{ij}), r(b_{ij}) \}$ are approximately a random arbitrarily censored sample from an $\textrm{Exp}(1)$ distribution, and the estimated \citep{Turnbull1974} cumulative hazard plot should be approximately straight with slope 1. Uncertainty in the plot is assessed through several cumulative hazards based on a random sample from $[\bw_J, \btheta, \bbeta, \bv|\calD]$. This is in contrast to typical Cox-Snell plots which only use point estimates.

Several researchers have pointed out that Cox-Snell residuals are conservative in that they may be straight even under quite large departures from the model \citep{Baltazar-Aban.Pena1995,OQuigley.Xu2005}. In this case, model criteria will be more informative. We consider two popular model choice criteria: the deviance information criterion (DIC) \citep{Spiegelhalter.etal2002} and the log pseudo marginal likelihood (LPML) \citep{Geisser.Eddy1979}, where DIC (smaller is better) places emphasis on the relative quality of model fitting and LPML (larger is better) focuses on the predictive performance. Both criteria are readily computed from the MCMC output; see supplemental Appendix C for more details.

\section{Real Data Applications}

\subsection{Loblolly Pine Survival Data}
Loblolly pine is the most commercially important timber species in the Southeastern United States; estimating loblolly survival is crucial to forestry research. The dataset used in this section consists of 45,525 loblolly pine trees at $m=168$ distinct sites, which were established in 1980-1981 and monitored annually until 2001-2002. During the 21-year follow-up, 5,379 trees died; the rest  survived until the last follow-up and are treated as right censored. It is of interest to investigate the association between  loblolly pine survival and several important risk factors after adjusting for spatial dependence among different sites. The risk factors considered include two time-independent variables, treatment and physiographic region, and three time-dependent variables---diameter at breast height, tree height and crown class---which were repeatedly measured every 3 years. Supplementary Table S6 presents some baseline characteristics for the trees. 

\begin{figure}
	\centering
	\subfigure[]{ \includegraphics[width=0.25\textwidth]{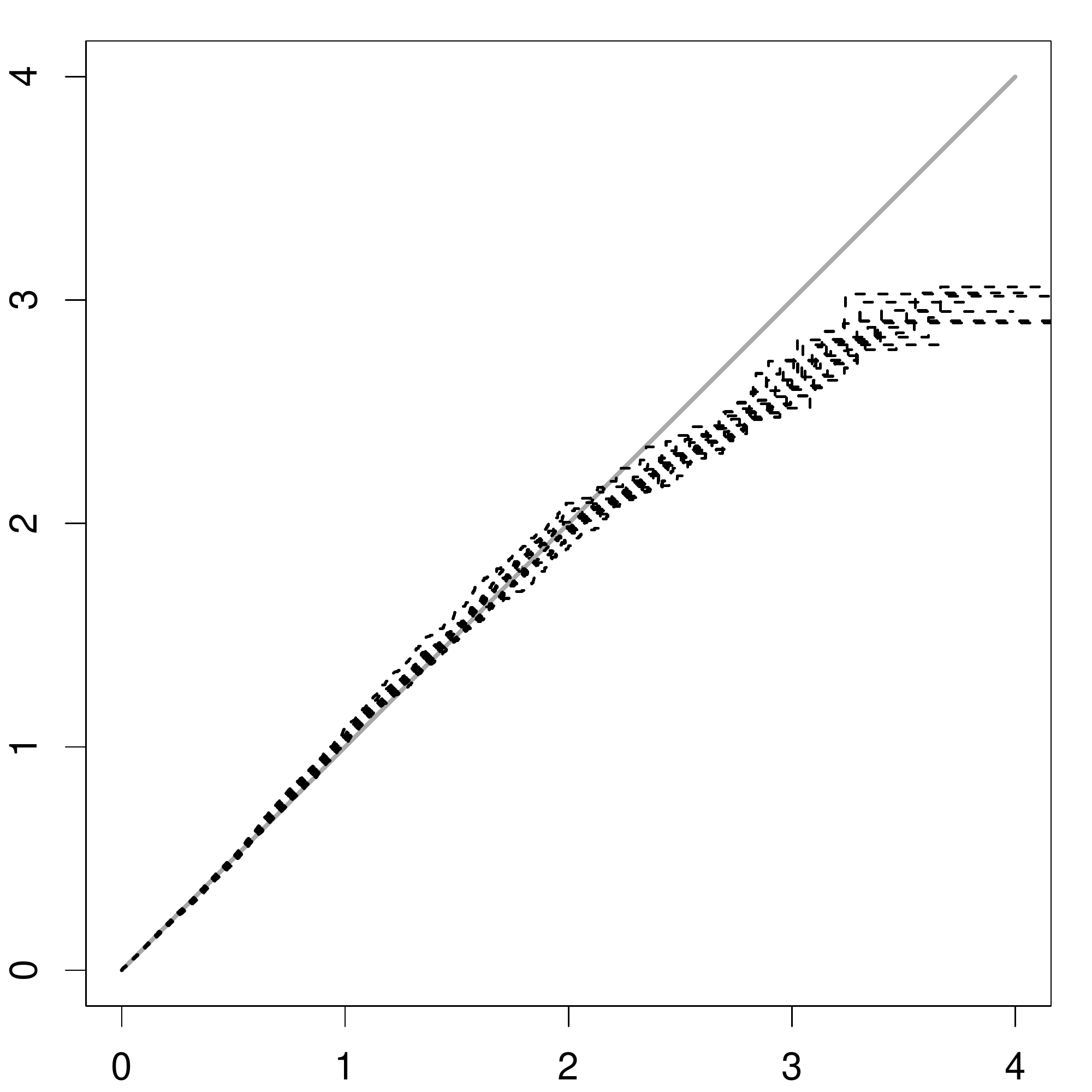} }
	\subfigure[]{ \includegraphics[width=0.25\textwidth]{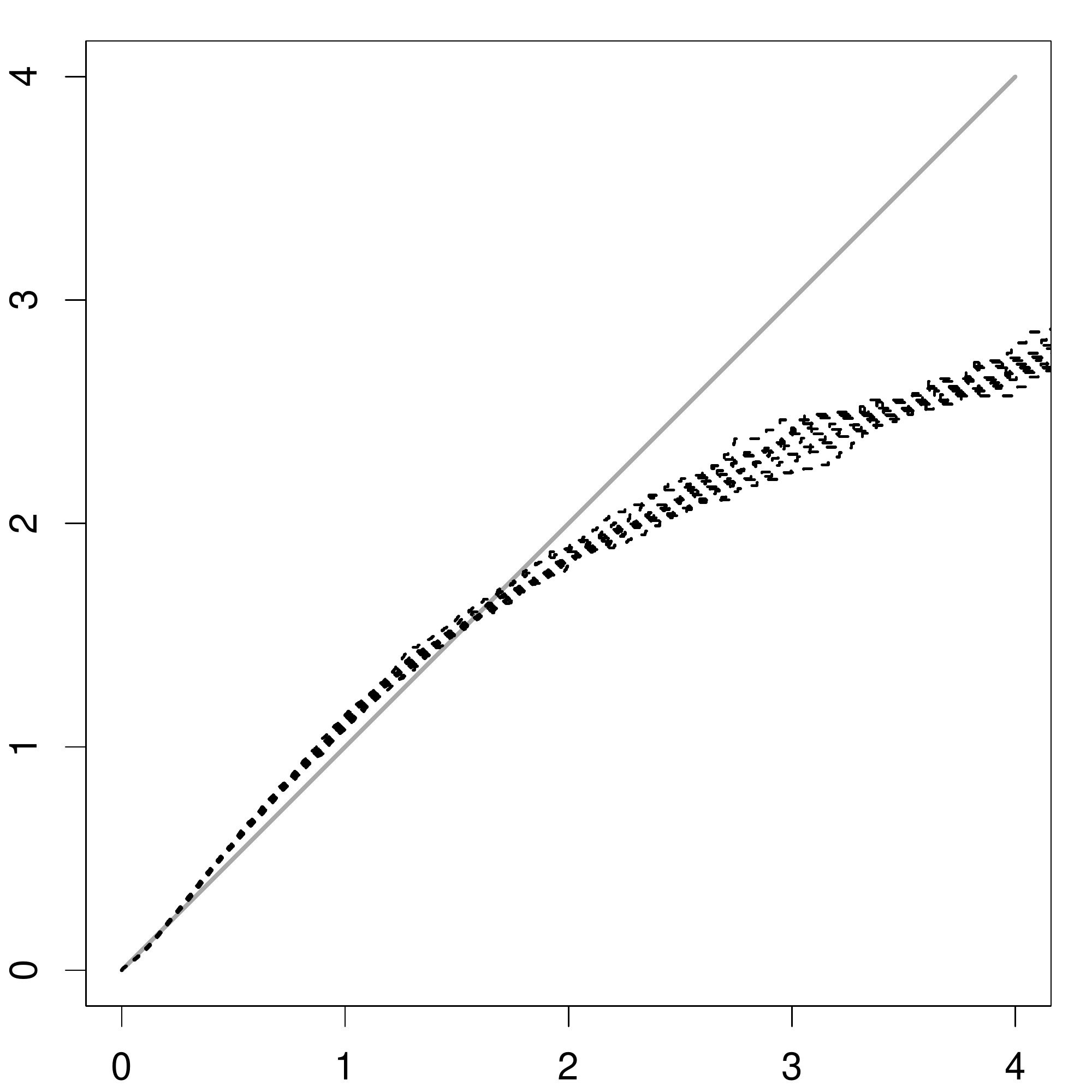} }
	\subfigure[]{ \includegraphics[width=0.25\textwidth]{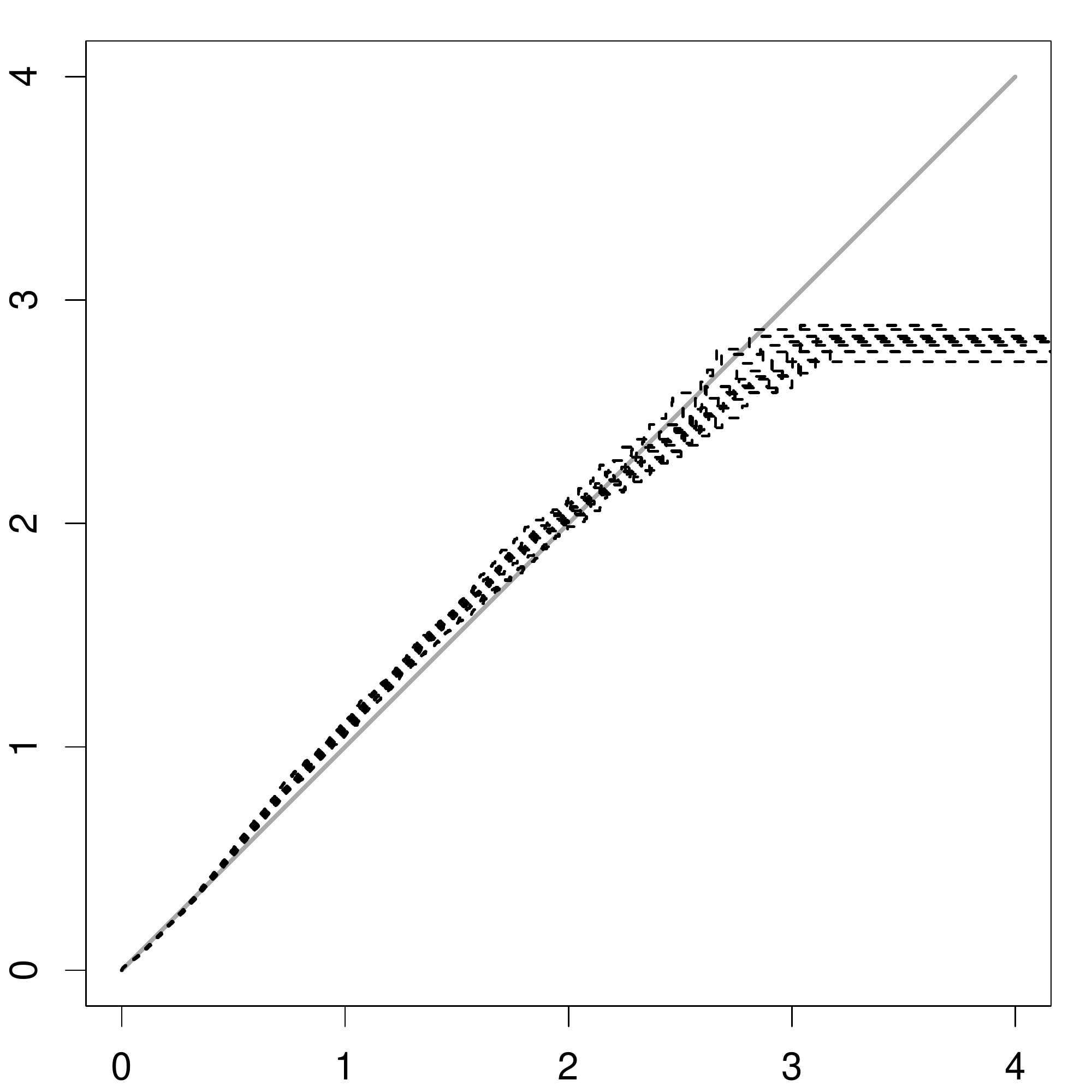} }
	\caption{Loblolly pine data. Cox-Snell residual plots under AFT (panel a), PH (panel b) and PO (panel c) with GRF frailties.}
	\label{loblolly:residual}
\end{figure}

\cite{Li.etal2015} fitted a semiparametric PH model with several georeferenced spatial frailty specifications. However, they showed that the PH assumption does not hold very well for treatment and physiographic region. To investigate whether the AFT or PO provides better fit, we fit each of the AFT, PH and PO models with GRF frailties to the data using the same covariates as those in \cite{Li.etal2015}; the log-logistic centering distribution was used throughout. Here $\nu=1$ was fixed giving the exponential correlation function. To better investigate the spatial frailty effect, we also fit each model with IID Gaussian frailties and non-frailty models. For each MCMC run we retained $2,000$ scans thinned from $50,000$ after a burn-in period of $10,000$ iterations. Figure~\ref{loblolly:residual} reports the Cox-Snell residual plots under the three GRF frailty models, where we see that the PH model severely deviates from the 45 degree line, and the AFT fits the data much better than the PH and PO. Table~\ref{loblolly:comparison} compares all fitted models using the LPML and DIC criteria, where we can see that the AFT always outperforms the PH and PO regardless of the frailty assumptions. Under all three models, incorporating IID frailties significantly improves the goodness of fit over the non-frailty model. Some goodness of fit improvement for the GRF over the IID occurs under the AFT, but this does not happen under the PH and PO, indicating that adding spatially structured frailties can deteriorate the model if the overarching model assumption is violated. 

\begin{table}
	\caption{Loblolly pine data. Model comparison.} 
	\label{loblolly:comparison}
	\centering
	{
		\begin{tabular}{lc|ccc}
			\hline\noalign{\smallskip}
			~			 	& ~					& AFT 			& PH   			& PO		\\
			\noalign{\smallskip}\hline\noalign{\smallskip}
			GRF frailty	& LPML			& {-23,812} & -23,991	&  -23,882 	\\
			~				& DIC			 & {~47,611} & ~47,971	 & ~47,767 	\\
			IID	frailty	 & LPML			& -23,832 & -23,966	&  -23,865 	\\
			~				& DIC			 & ~47,648 & ~47,897	 & ~47,731 	\\
			Non-frailty	& LPML			& -25,447 & -25,508	&  -25,549 	\\
			~				& DIC			 & ~50,893 & ~51,015	 & ~51,099	\\
			\noalign{\smallskip}\hline
		\end{tabular}
	}
\end{table}

Table~\ref{loblolly:coeff} gives covariate effects under the AFT models.  Coefficient estimates under the model with frailties (IID or GRF) have changed significantly from those under the non-frailty model. For example, the effect of total tree height on tree survival is reversed when the model is fit with frailties. This indicates that shorter trees are associated with longer survival rates when averaged over spatial location; however taller trees have better survival rates than shorter adjusting for location.  Thus a type of ``Simpson's paradox'' occurs with space being confounded with tree height in some fashion.

\begin{table}
	\caption{Loblolly pine data. Posterior means (95\% credible intervals) of fixed effects $\bbeta$ from fitting the AFT model with different frailty settings. }
	\label{loblolly:coeff}
	\centering
	{
		\begin{tabular}{lccc}
			\hline\noalign{\smallskip}
			~	     	  &  Non-frailty	 		  & IID	frailty             	& GRF frailty	\\
			\noalign{\smallskip}\hline\noalign{\smallskip}
			DBH & -0.233(-0.255,-0.212) & -0.127(-0.144,-0.111) & -0.126(-0.142,-0.110) \\ 
			TH & 0.027(0.024,0.030) & -0.012(-0.015,-0.010) & -0.011(-0.014,-0.009) \\ 
			treat2 & -0.521(-0.583,-0.466) & -0.381(-0.423,-0.339) & -0.388(-0.431,-0.349) \\ 
			treat3 & -0.719(-0.798,-0.641) & -0.529(-0.589,-0.473) & -0.544(-0.601,-0.495) \\ 
			PhyReg2 & -0.302(-0.367,-0.241) & -0.362(-0.514,-0.198) & -0.390(-0.594,-0.201) \\ 
			PhyReg3 & -0.007(-0.110,0.099) & -0.254(-0.530,0.050) & -0.260(-0.511,0.014) \\ 
			C2 & 0.097(0.028,0.169) & 0.031(-0.023,0.078) & 0.044(-0.002,0.097) \\ 
			C3 & 0.835(0.747,0.923) & 0.401(0.331,0.465) & 0.430(0.375,0.491) \\ 
			C4 & 1.919(1.799,2.029) & 1.040(0.951,1.128) & 1.101(1.018,1.194) \\ 
			treat2:PhyReg2 & 0.146(0.041,0.244) & 0.118(0.053,0.190) & 0.105(0.046,0.168) \\ 
			treat3:PhyReg2 & 0.372(0.246,0.503) & 0.255(0.169,0.349) & 0.246(0.162,0.332) \\ 
			treat2:PhyReg3 & -0.404(-0.625,-0.197) & -0.220(-0.374,-0.072) & -0.216(-0.368,-0.064) \\ 
			treat3:PhyReg3 & 0.109(-0.129,0.325) & 0.110(-0.067,0.267) & 0.125(-0.037,0.286) \\ 
			$\tau^2$ 		  & ~ & 0.318(0.248,0.402) & 0.350(0.270,0.457)\\
			$\phi$				& ~ & ~ & 0.274(0.165,0.434)\\
			\noalign{\smallskip}\hline
		\end{tabular}
	}
\end{table}

\begin{figure}
	\centering
	\subfigure[]{ \includegraphics[width=0.35\textwidth]{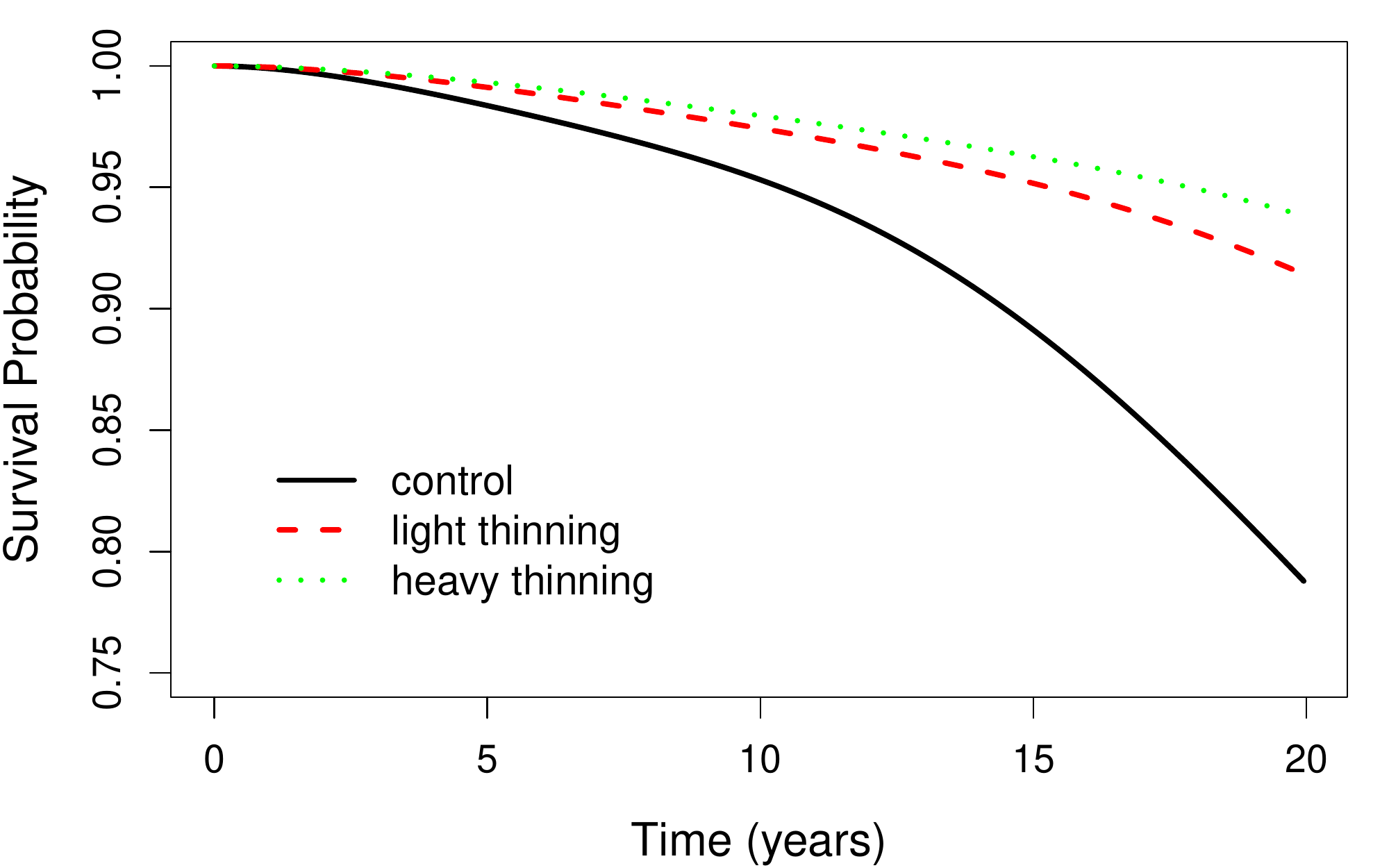} }
	\subfigure[]{ \includegraphics[width=0.35\textwidth]{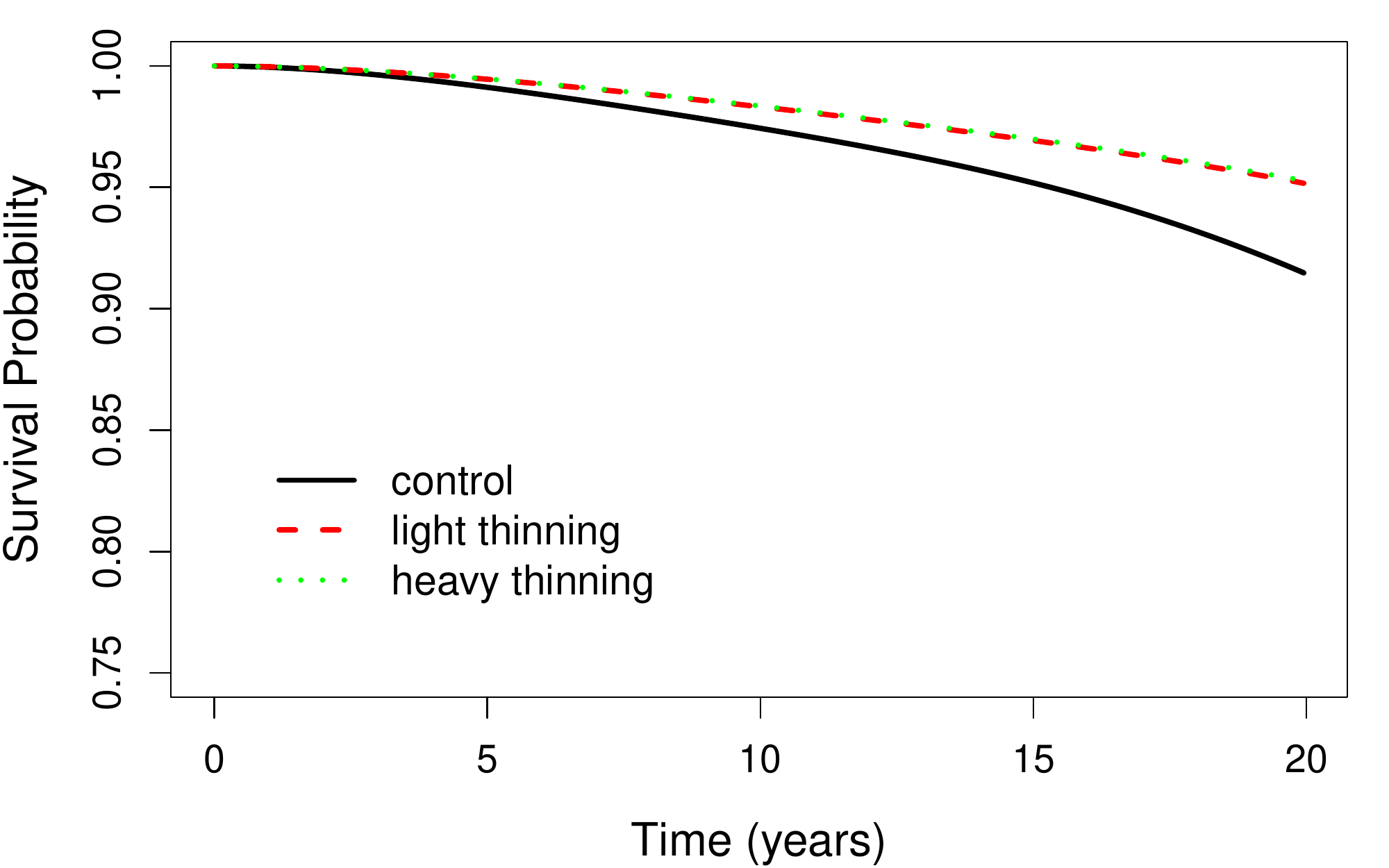} }\\
	\subfigure[]{ \includegraphics[width=0.35\textwidth]{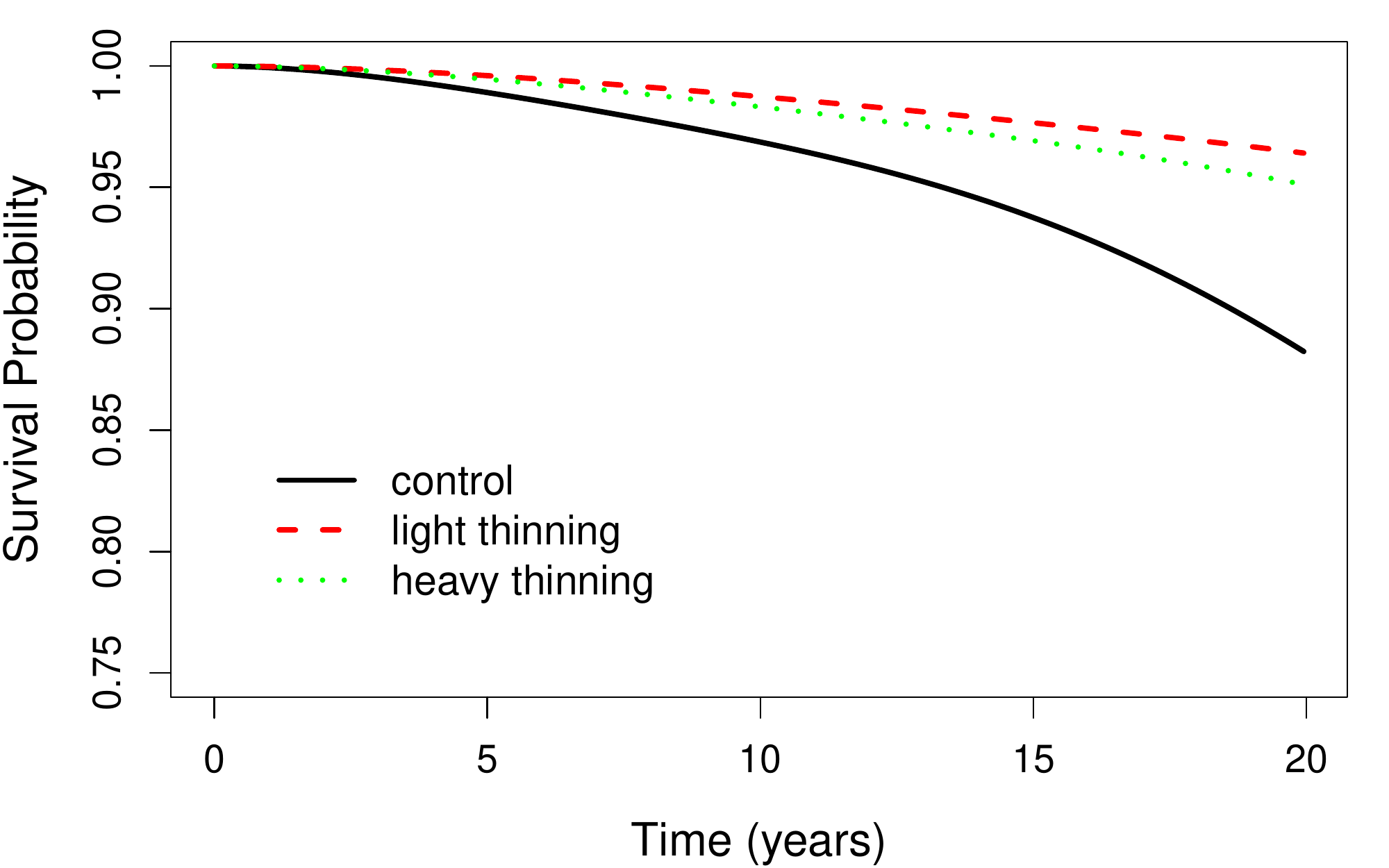} }
	\subfigure[]{ \includegraphics[width=0.35\textwidth]{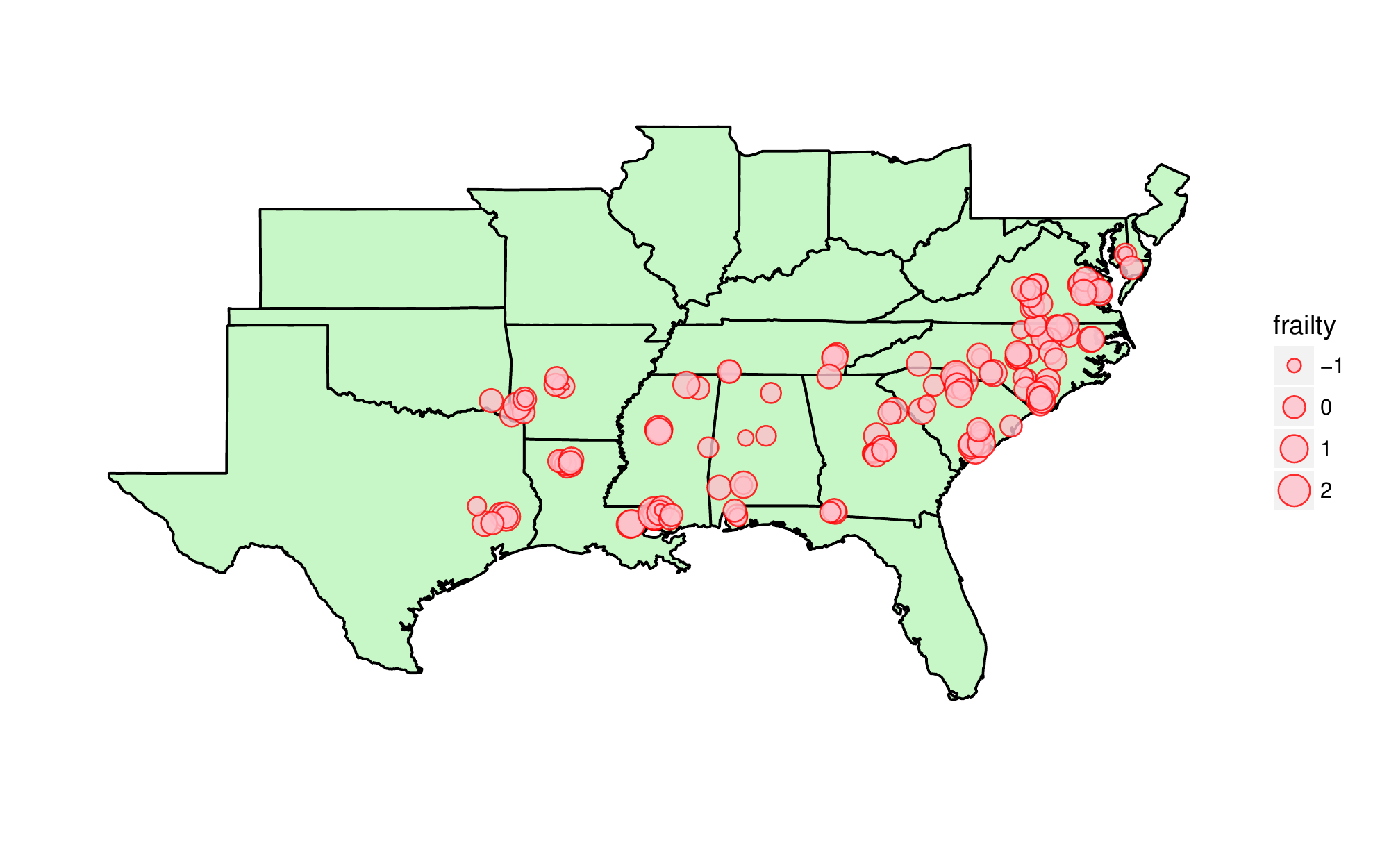} }
	\caption{Loblolly pine data. Survival curves for treatment effect under coastal (panel a), Piedmont (panel b), and other (panel c) regions. The posterior means of spatial frailties for each location are mapped in panel d.}
	\label{loblolly:curves}
\end{figure}

We next interpret the results under the AFT model for time-dependent covariates \citep{Prentice.Kalbfleisch1979} with GRF frailties, as it outperforms all other models. Table~\ref{loblolly:coeff} shows all covariates are significant risk factors for loblolly pine survival. For example, the mean or median survival time will increase by a factor $e^{0.126}=1.134$ for every 1 cm increase in diameter at breast height, holding other covariates and the frailty constant. \cite{Hanson.etal2009} note that this interpretation holds for mean or median \emph{residual life} as well, which is of greater interest in medical contexts when choosing a course of treatment.  Significant interaction effects are present between treatment and physiographic region; Figure~\ref{loblolly:curves} presents the survival curves for the treatment effect under different physiographic regions. The thinning treatment has the largest effect on survival rates in coastal regions, while for Piedmont regions, heavy thinning is required to improve survival rates. The posterior means of spatial frailties for each location are also mapped in Figure~\ref{loblolly:curves}, where we do not see a clear spatial pattern, implying that the spatial dependence may not be very strong for these data. To further confirm this, the posterior mean of the spatial range parameter $\phi$ is $0.274$ km. Note that $99\%$ of the pairwise distances among the 168 locations are greater than 9.93 km, that is, $99\%$ of the pairwise correlations are lower than $1-e^{-0.274\times 9.93}\approx 0.066$. 

\subsection{The Signal Tandmobiel Study}
Oral health data collected from the Signal Tandmobiel study are next considered, available in the R package \texttt{bayesSurv}; a description of the data can be found in \cite{Komarek.Lesaffre2009}. It is of interest to investigate the impact of gender ($1=$ girl, $0=$ boy), dmf (1, if the predecessor of the permanent first premolar was decayed, missing due to caries or filled, 0, if the predecessor was sound), and tooth location ($1=$ mandibular, $0=$ maxillary) on the emergence time of each of the four permanent first premolar (teeth 14, 24, 34, 44). The data set consists of $4,430$ children with four tooth emergence times recorded for each child, yielding a sample size of  $n=17,594$. The emergence times are interval censored due to annual examinations. \cite{Komarek.Lesaffre2009} fit an IID frailty AFT model for taking into account the dependency between emergence times within each child.  Although their models allow for a nonparametric frailty density, their Figure 6 shows a remarkably Gaussian-shaped estimate.  Each of the AFT, PH and PO models with IID Gaussian frailties were fit to the data using the same covariates as the Model S in \cite[][Table 4]{Komarek.Lesaffre2009}, giving LPML values -15125, -15124, and -15503, respectively. Thus, the PH and AFT models perform similarly and both outperform the PO in terms of predictive performance; however, the AFT model fits these data better than the PH according to DIC. Cox-Snell residual plots (not shown) also slightly favor AFT.  Table~\ref{dental:coeff} reports the fixed effects and frailty variances under each model, from which we see that the effect of dmf is different for boys and girls, and it is also different for mandibular and maxillary teeth.  

\begin{table}
	\caption{The Signal Tandmobiel study. Posterior means (95\% credible intervals) of fixed effects from fitting the AFT, PH and PO models with IID frailties. The LPML and DIC are also shown for each model.}
	\label{dental:coeff}
	\centering
	{
		\begin{tabular}{lccc}
			\hline\noalign{\smallskip}
			~	     	  &  AFT 		  & PH             	& PO	\\
			~			  & {\small (LPML: -15125)}  & {\small(LPML: -15124)}	& {\small(LPML: -15503)}  \\
			~			  & {\small (DIC: 29067)}  & {\small(DIC: 30045)}	& {\small(DIC: 30596)}  \\
			\noalign{\smallskip}\hline\noalign{\smallskip}
			gender & 0.045(0.038,0.053) & 1.034(0.884,1.191) & 1.433(1.206,1.667) \\ 
			dmf & 0.031(0.026,0.036) & 0.799(0.690,0.908) & 1.424(1.264,1.573) \\ 
			tooth & 0.020(0.017,0.022) & 0.408(0.341,0.472) & 0.594(0.496,0.694) \\ 
			dmf:tooth & -0.018(-0.022,-0.014) & -0.478(-0.578,-0.378) & -0.770(-0.921,-0.625) \\ 
			gender:dmf & -0.009(-0.015,-0.003) & -0.258(-0.401,-0.126) & -0.470(-0.674,-0.260) \\ 
			$\tau^2$	  & 0.010(0.009, 0.010)   & 4.619(4.326, 4.918)	 & 9.261(8.679, 9.885)\\
			\noalign{\smallskip}\hline
		\end{tabular}
	}
\end{table}

\subsection{Leukemia Data}
Finally, a dataset on the survival of acute myeloid leukemia in $n=1,043$ patients is considered, available in the package \texttt{spBayesSurv}. It is of interest to investigate possible spatial variation in survival after accounting for known subject-specific prognostic factors, which include age, sex, white blood cell count (wbc) at diagnosis, and the Townsend score (tpi) for which higher values indicates less affluent areas. There are $m=24$ administrative districts and each one forms a spatial cluster \citep[][Figure 1]{Henderson.etal2002}. \cite{Henderson.etal2002} fitted a multivariate gamma frailty PH model with linear predictors. Here we fit each of the PH, AFT and PO models with ICAR frailties to see whether the AFT or PO model provides better fit. To allow for non-linear effects of continuous predictors, we consider partially linear age, wbc and tpi as described in supplementary Appendix G. 

\begin{figure}
	\centering
	\subfigure[]{ \includegraphics[width=0.25\textwidth]{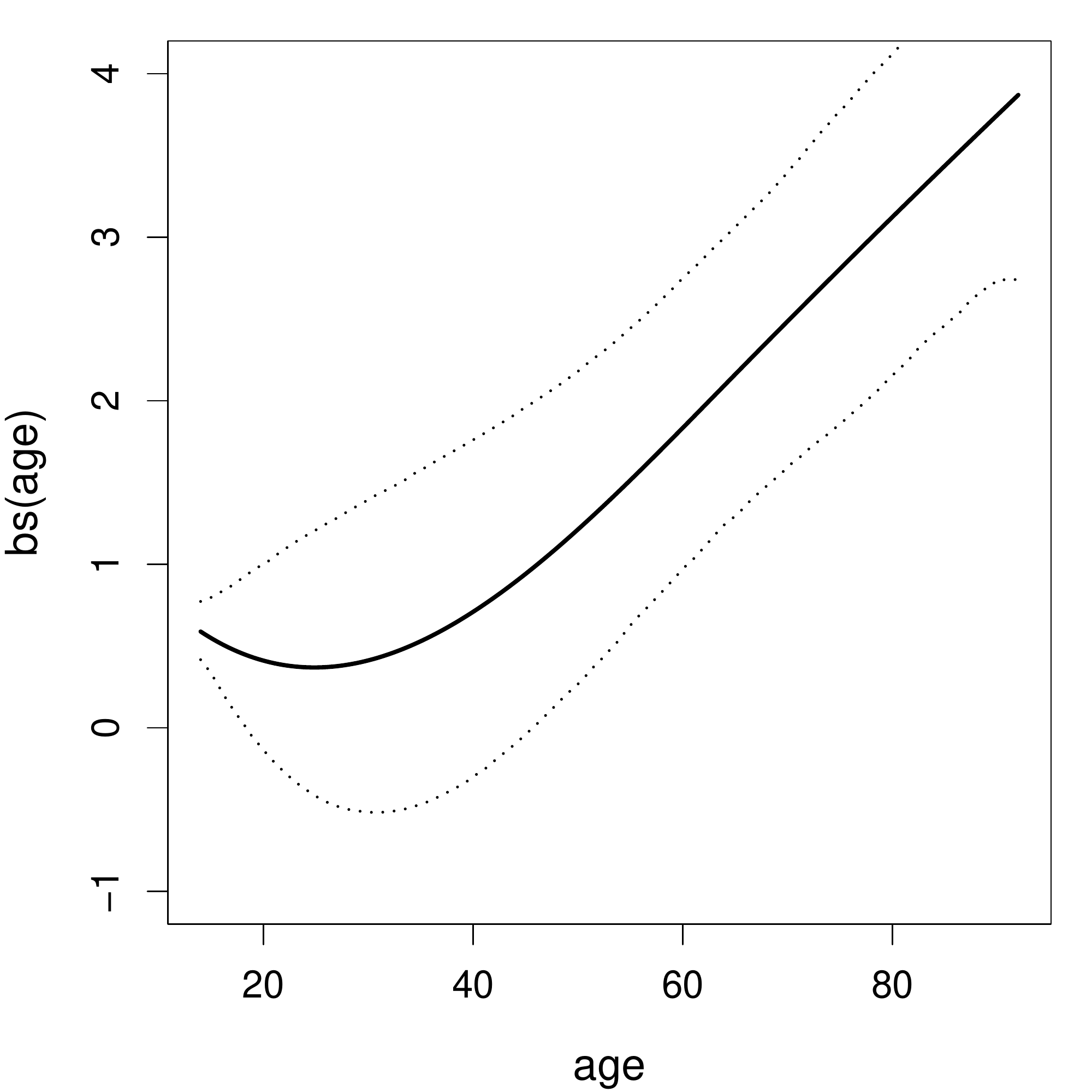} }
	\subfigure[]{ \includegraphics[width=0.25\textwidth]{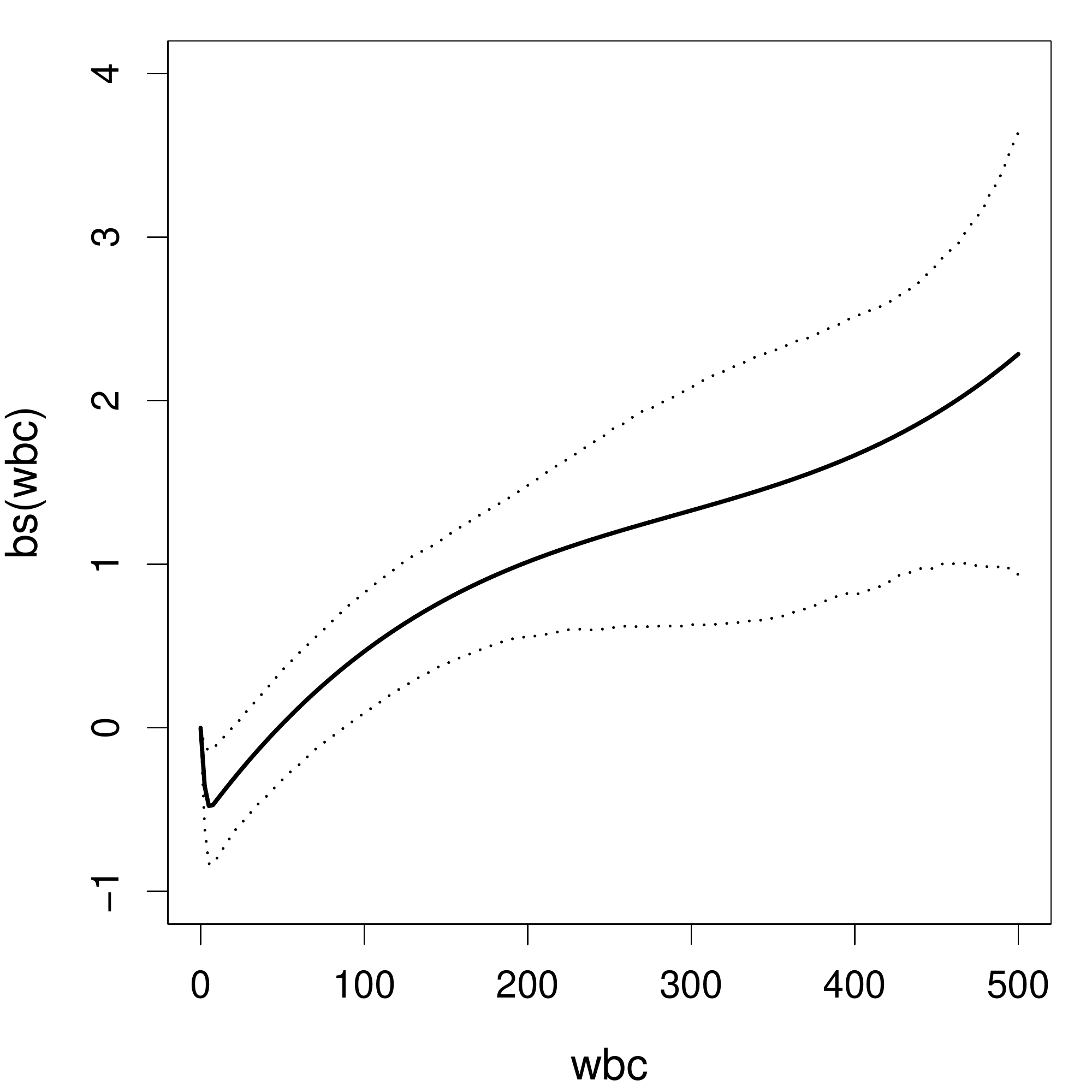} }
	\subfigure[]{ \includegraphics[width=0.25\textwidth]{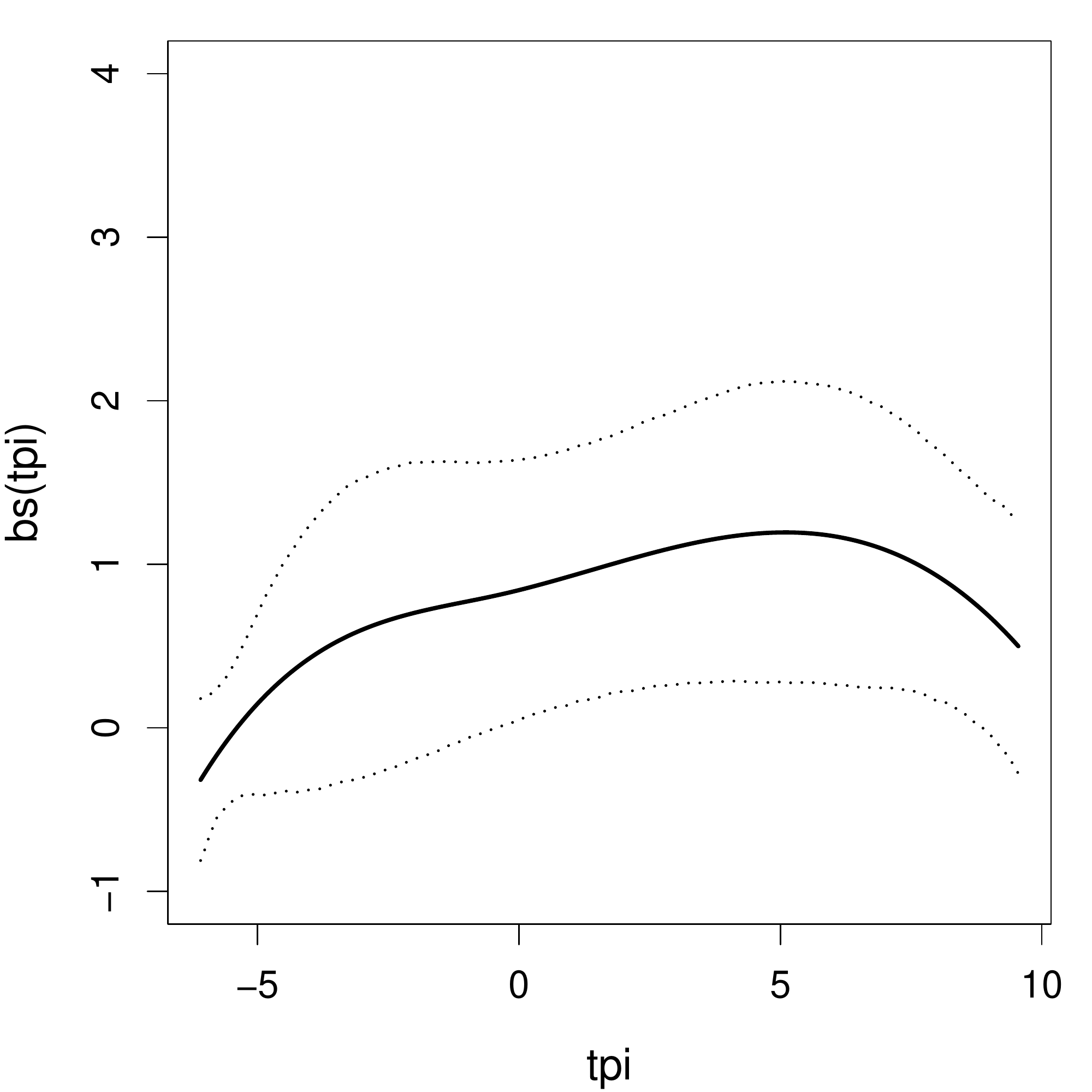} }
	\caption{Leukemia data. Posterior mean estimates (solid lines) for the effects of age (panel a), wbc (panel b) and tpi (panel c), together with 95\% credible intervals (dotted lines).}
	\label{leukemia:curves}
\end{figure}

The LPML values for the PH, AFT and PO models are -5946, -5945, and -5919, respectively.  The PO model significantly outperforms others from a predictive point of view with a pseudo Bayes factor on the order of $10^{10}$ relative to PH and AFT. Figure~\ref{leukemia:curves} presents the estimated effects of age, wbc and tpi under the PO model. The Bayes Factors for testing the linearity of age, wbc and tpi are 0.13, 0.04 and 0.01, respectively; non-linear effects are not needed. 

\section{Simulations}\label{sec:sim}
Extensive simulations were carried out to evaluate the proposed MCMC algorithms (implemented in \texttt{spBayesSurv}) under the three survival models with arbitrarily censored spatial data. The proposed methodology is then compared to monotone splines \citep{Lin.etal2015} as implemented in \texttt{ICBayes} for PH and PO, Bayesian G-splines \citep{Komarek2006} implemented in \texttt{bayesSurv} for AFT, and geoadditive PH modeling \citep{Hennerfeind.etal2006} implemented in \texttt{R2BayesX}. Additional simulations for georeferenced data and variable selection are available in supplementary Appendixes J.2 and J.3. 

\subsection{Simulation I: Areal Data}
For each of AFT, PH and PO, data were generated from (\ref{aft}), (\ref{ph}) and (\ref{po}), respectively, where covariates $\bx_{ij}=(x_{ij1},x_{ij2})'$ were sampled $x_{ij1}\overset{iid}{\sim} \mathrm{Bernoulli}(0.5)$ independent of $x_{ij2}\overset{iid}{\sim} N(0,1)$ for $i=1\ldots,37$ and $j=1,\ldots,20$ ($n=740$),  regression effects set to $\bbeta=(\beta_1, \beta_2)'=(1,1)'$, the baseline survival distribution $S_0(t)=1-0.5[\Phi\left(2(\log t+1)\right)+\Phi\left(2(\log t-1)\right)]$ was bimodal, and $\bv$ followed the ICAR model (\ref{eq:areal-prior}) with $\tau^2=1$ and $\bE$ equaling the Nigeria adjacency matrix used in the childhood mortality data analysis of \cite{Kneib2006}.  Half the data were right censored (including uncensored survival times) and half interval censored.  The times at which survival was right censored were independently simulated from a $\mathrm{Uniform}(2,6)$ distribution. For interval censoring, each subject was assumed to have $N$ observation times, $O_1, O_2, \ldots, O_N$, where $(N-1)\sim \mathrm{Poisson}(2)$ and $(O_{k}-O_{k-1})|N\overset{iid}{\sim} \mathrm{Exp}(1)$ with $O_0=0$, $k=1, \ldots, N$. The censoring interval endpoints are the two adjacent observation times among $\{0,O_1,\dots,O_N,\infty\}$ that include the true survival time. The final data yield around 20\% right censored, 40\% uncensored, 25\% left censored and 15\% interval censored. For each model, 500 Monte Carlo (MC) replicate data sets were generated. Models were fit using the default priors introduced in Section~\ref{sec:mcmc}. For each MCMC run, $5,000$ scans were thinned from $50,000$ after a burn-in period of $10,000$ iterations; convergence diagnostics deemed this more than adequate. 

Table~\ref{sim:coeff} summarizes the results for regression parameters $\bbeta$ and the ICAR variance $\tau^2$, including the averaged bias (BIAS) and posterior standard deviation (PSD) of each point estimate (posterior mean for $\bbeta$ and median for $\tau^2$), the standard deviation (across 500 MC replicates) of the point estimate (SD-Est), the coverage probability (CP) of the $95\%$ credible interval, and average effective sample size (ESS) out of 5,000 \citep{Sargent.etal2000} for each point estimate. The results show that the point estimates of ${\bbeta}$ and $\tau^2$ are unbiased under all three models, SD-Est values are close to the corresponding PSDs, the CP values are close to the nominal $95\%$ level, and ESS values are promising. Supplementary Figure S1 presents the average across the 500 MC replicates of the fitted baseline survival functions revealing that the proposed model is capable to capture complex baseline survival curves very well. 

\begin{table}
	\caption{Simulation I: average bias (BIAS) and posterior standard deviation (PSD) of each point estimate, standard deviation (across 500 MC replicates) of the point estimate (SD-Est), coverage probability (CP) for the $95\%$ credible interval, and effective sample size (ESS) out of 5,000 with thinning=10 for each point estimate. }
	\label{sim:coeff}
	\centering
	{
		\begin{tabular}{ccrrccc}
			\hline
			\hline\noalign{\smallskip}
			Model & Parameter   & BIAS 		& PSD        & SD-Est      & CP	    & ESS \\
			\noalign{\smallskip}\hline\noalign{\smallskip}
			AFT	& $\beta_1=1$    & 0.000	& 0.067      & 0.062       & 0.960	& 3194 \\
			~	& $\beta_2=1$    & 0.002	& 0.036      & 0.034       & 0.960	& 2992 \\
			~ 	& $\tau^2=1$ 	 & 0.015	& 0.311 	 & 0.298	   & 0.942	& 4561 \\
			\noalign{\smallskip}
			PH	& $\beta_1=1$    & -0.021	& 0.100      & 0.099       & 0.936	& 3024 \\
			~	& $\beta_2=1$    & -0.014	& 0.061      & 0.060       & 0.944	& 2095 \\
			~ 	& $\tau^2=1$ 	 & -0.039	& 0.352		 & 0.319	   & 0.950	& 3478 \\
			\noalign{\smallskip}
			PO	& $\beta_1=1$    & 0.012	& 0.151	     & 0.152       & 0.940  & 3672 \\
			~	& $\beta_2=1$    & 0.008	& 0.083      & 0.086       & 0.944	& 2822\\
			~ 	& $\tau^2=1$ 	 & -0.008	& 0.465		 & 0.448	   & 0.924	& 2451 \\
			\noalign{\smallskip}\hline
			
		\end{tabular}
	}
\end{table}

\subsection{Simulation II: Comparison}
In this section we compare our R function \texttt{survregbayes} in the package \texttt{spBayesSurv} with R functions \texttt{ICBayes} in the package \texttt{ICBayes}, \texttt{bayessurvreg2} in the package \texttt{bayesSurv} and \texttt{bayesx} in the package \texttt{R2BayesX} in terms of mixing and computing speed. Our \texttt{survregbayes} function can fit spatial or non-spatial PH, AFT and PO models (either no, IID, ICAR or GRF frailties) for all common types of survival (uncensored, interval censored, current status, right censored, etc.) and/or left-truncated data. In contrast, \texttt{ICBayes} fits only non-frailty PH and PO models to interval censored data (uncensored data is not supported), \texttt{bayessurvreg2} fits the AFT model (either no or IID frailties) to general interval censored data, and \texttt{bayesx} fits the PH model to right-censored data using MCMC. Data are generated from three cases: (C1) the non-frailty PH model with interval censoring, (C2) the non-frailty AFT model with arbitrary censoring, and (C3) the non-frailty PH model with right censoring, where $\bbeta$ and $S_0(\cdot)$ are the same as those used in \textbf{Simulation I}. Under each setting we generate 500 MC replicates, each with sample size $n=500$. For each MCMC run, we retain $10,000$ scans without thinning after a burn-in period of $10,000$ iterations.

Table~\ref{sim:comparison} reports the comparison results; the ESS from \texttt{survregbayes} range from 3 to 20 times as large as those using \texttt{ICBayes} and \texttt{bayessurvreg2}, indicating that the proposed MCMC algorithms are more efficient in terms of mixing. In addition, \texttt{survregbayes} is about 5 times faster than \texttt{ICBayes}, although it is much slower than \texttt{bayessurvreg2} and \texttt{bayesx}. Note that \texttt{bayessurvreg2} is almost 9 times faster than our \texttt{survregbayes}, but its ESS for $\beta_1$ is nearly 19 times smaller than ours. That is, \texttt{bayessurvreg2} needs to take 19 times more iterates than ours to obtain the same ESS, thus tempering superior speed with much poorer MCMC mixing. In comparison with \texttt{bayesx} for right censored data, our \texttt{survregbayes} performs slightly worse in terms of both speed and mixing. This is not surprising, because our MCMC is designed for all AFT, PH and PO with all manner of censoring schemes while the MCMC in \texttt{bayesx} is tailored to PH with right censored data. 

\begin{table}
	\caption{Simulation II: average bias (BIAS) and posterior standard deviation (PSD) of each point estimate, standard deviation (across 500 MC replicates) of the point estimate (SD-Est), coverage probability (CP) for the $95\%$ credible interval, average effective sample size (ESS) out of 10,000 without thinning for each point estimate, and average computing time in seconds.}
	\label{sim:comparison}
	\centering
	{
		\begin{tabular}{ccccrrccc}
			\hline
			\hline\noalign{\smallskip}
			 Case & R function & Time & Parameter   & BIAS 	& PSD        & SD-Est      & CP	& ESS \\
			\noalign{\smallskip}\hline\noalign{\smallskip}
			C1  & \texttt{survregbayes}	& 63  	 & $\beta_1=1$    & -0.018	& 0.134      & 0.134       & 0.940	& 1139 \\
			~ & ~		& ~   					& $\beta_2=1$    & -0.015	& 0.086      & 0.087       & 0.940	& 934 \\
			\noalign{\smallskip}
			{\smallskip}
			~ & \texttt{ICBayes}	& 310 & $\beta_1=1$    & -0.036	& 0.133      & 0.132       & 0.938	& 346 \\
			~& ~	& ~       		& $\beta_2=1$    & -0.019	& 0.084      & 0.085       & 0.938	& 292 \\
			\noalign{\smallskip}\hline\noalign{\smallskip}
			C2  & \texttt{survregbayes}	& 54  & $\beta_1=1$    & 0.000	& 0.075      & 0.074       & 0.950	& 1207 \\
			~ & ~					& ~      & $\beta_2=1$    & 0.002	& 0.041      & 0.040       & 0.952	& 1186 \\
			\noalign{\smallskip}
			{\smallskip}
			~ & \texttt{bayessurvreg2}	& 6 & $\beta_1=1$    & -0.002	& 0.072      & 0.075       & 0.936	& 64 \\
			~& ~					& ~     & $\beta_2=1$    & -0.001	& 0.039      & 0.040       & 0.934	& 109 \\
			\noalign{\smallskip}\hline\noalign{\smallskip}
			C3  & \texttt{survregbayes}	& 85  & $\beta_1=1$    & -0.009	& 0.103      & 0.061       & 0.954	& 1028 \\
			~ & ~		& ~   					& $\beta_2=1$    & -0.010	& 0.107      & 0.061       & 0.950	& 673 \\
			\noalign{\smallskip}
			{\smallskip}
			~ & \texttt{bayesx}	& 44 	& $\beta_1=1$    & 0.017	& 0.103      & 0.062       & 0.940	& 989 \\
			~& ~	& ~       		& $\beta_2=1$    & 0.014	& 0.111      & 0.064       & 0.952	& 2817 \\
			\noalign{\smallskip}\hline
		\end{tabular}
	}
\end{table}

\texttt{ICBayes} stems from a burgeoning literature on fitting PO and PH models to interval censored data using a monotone spline, essentially an integrated B-spline, as the baseline cumulative hazard function.  In particular \cite{Cai.etal2011} consider a PH model for current status data; \cite{Lin.Wang2011} consider the PO model with current status data; case 2 interval censoring is considered for PO by \cite{Wang.Lin2011}.  \cite{Pan.etal2014} consider the PH model for general interval censored data with spatial CAR frailties and \cite{Lin.etal2015} consider the same PH model with general interval censored data without frailties.  \cite{Wang.etal2016} take the E-M approach to the model of \cite{Lin.etal2015}.  Among these papers \cite{Lin.etal2015} state concerning the \texttt{ICBayes} function that ``\emph{...our developed Gibbs sampler is easy to execute
with only four steps and efficient because it does not require imputing any unobserved
failure times or contain any complicated Metropolis-Hastings steps.}''  They further say that their approach is efficient in the sense that their approach is easy to program.  However, although the \cite{Lin.etal2015} approach does not require imputing failure times, it does require imputing as many latent Poisson variates as there are spline basis functions per observation, i.e. much more latent data than simply imputing the survival times. Furthermore, they update one regression coefficient $\beta_j$ at a time for $j=1,\ldots,p$ via the ARS method \citep{Gilks.Wild1992}, and it is well-known that any component-at-a-time sampler will suffer from poor mixing if there is strong correlation in the posterior.  The simplest way to improve mixing is through the consideration correlated proposals such as the adaptive version developed here or the iteratively weighted least squares proposal used in \cite{Hennerfeind.etal2006}.  This simulation shows that adaptive block updates can dramatically outperform a Gibbs sampler with latent data, both in terms of speed and mixing. Furthermore, the monotone spline approach models the baseline cumulative hazard over the range of the observed data only; however, the AFT model maps observed data onto ``baseline support'' through a scale factor.  The TBP prior used here includes a scale and so is applicable to all of PH, PO, and AFT, as well as accelerated hazards (which also needs a scale term), additive hazards, proportional mean residual life, and others.  

\section{Discussion}

The recent surge in literature analyzing spatially correlated time-to-event data focusing on PH highlights the need for flexible, robust methodology and software to enable the ready use of other competing yet easily-interpretable survival models, spatial or not.  This article provides a unified framework for considering competing models to PH, i.e. AFT and PO, including variable selection, areal or georeferenced frailties, and additive structure.  The freely available \texttt{survregbayes} function allows for easy fitting of these competing models.  An important aspect associated with the Bayesian nonparametric TBP -- or more accurately richly parametric -- formulation of the AFT, PH, and PO models presented here is that the assumption of the same flexible prior on $S_0(\cdot)$ places these models common ground. Differences in fit and/or predictive performance can therefore be attributed to the semiparametric models only, rather than to additional possible differences in quite different nonparametric priors (e.g. gamma process vs. beta process vs. Dirichlet process mixture) or estimation methods (e.g. NPMLE vs. partial likelihood vs. sieves). 

In two of the three data analyses in Section 3 a model alternative to PH was favored by the predictive LPML measure; for the dental data in Section 3.2, the AFT and PH predictively perform about equally well and outperform PO. However, PH can be certainly be the best model choice from a predictive point of view; this is, of course, data-dependent. In additional analyses not presented here, the proposed AFT, PH and PO models were applied to spatial smoking cessation data set from the Lung Health Study carried out by \cite{Murray.etal1998} and analyzed by \cite{Pan.etal2014}. The LPML values were -211, -206, and -206, respectively, indicating that PH and PO models predict equally well and outperform AFT; the pseudo Bayes factor comparing PH to AFT (or PO to AFT) is approximately $e^{-206-(-211)} \approx 150$.  We also analyzed the current status lung cancer data set available in the R package \texttt{ICBayes}, analyzed by \cite{Sun2006}.  The LPML values under the AFT, PH and PO are all essentially -83, respectively, indicating that there is essentially no difference on the predictive performance among the three models.  This makes sense as the only covariate (treatment) was not significant: $\bbeta=\mathbf{0}$ implies PH, PO, and AFT reduce to $S_0(\cdot)$.

\section*{Supplementary Materials}
\begin{description}
	\item[Appendices:] Appendix A: MCMC sampling algorithms; Appendix B: brief introduction to the full-scale approximation; Appendix C: definitions of the DIC and LPML criteria; Appendix D: testing for parametric $S_0(\cdot)$; Appendix E: stochastic search variable selection; Appendix F: left-truncation and time-dependent covariates; Appendix G: partially linear predictors; Appendix H: introduction to the R function \texttt{survregbayes}; Appendix I: additional results for real data applications; Appendix J: additional simulation results. (pdf file)
	\item[Data and Code:] Data and R code for the three analyses in Section 3. (zip file)
\end{description}

\bibliographystyle{apalike}
\bibliography{bibliography}
\end{document}


\title{\textbf{Supplementary Materials for \emph{``A unified framework for fitting Bayesian semiparametric models to arbitrarily censored survival data, including spatially-referenced data''} by Haiming Zhou and Timothy Hanson}}
\date{}
\author{}

\maketitle



\section*{Appendix 0 \quad Notation and Prior Tables}
Table~\ref{table:notation} presents the notation symbols used in the main paper and their definitions. Table~\ref{table:priors} lists the priors for all parameters and the reasons of choosing them, where $\TBP_J(\alpha,S_{\btheta})$ is the TBP prior, $\ICAR(\tau^2)$ is the ICAR prior, $\GRF(\tau^2, \phi)$ is the GRF prior, and $\IID(\tau^2)$ is the IID prior. 

\begin{longtable}{|p{3.3cm}|p{12cm}|}
	\caption{List of Notations.}\label{table:notation}\\
	\hline
	\hline\noalign{\smallskip}
	Notation & Definition  \\
	\noalign{\smallskip}\hline\noalign{\smallskip}
	\endhead
	\hline
	\endfoot
	$\alpha$ & precision parameter of the TBP prior\\
	$\bbeta=(\beta_1,\ldots,\beta_p)'$ & $p$-vector of regression coefficients for survival models\\
	$\bbeta_0$ & mean of the normal $N_p(\bbeta_0, \bW_0)$ prior on $\bbeta$\\
	$\hat{\bbeta}$ & estimate of $\bbeta$ under the parametric survival model with $S_0=S_{\btheta}$ \\
	$\delta_{j,J}(\cdot)$ & beta density function with parameters $(j, J-1+1)$\\
	$\Delta_{j,J}(\cdot)$ & beta cumulative distribution function with parameters $(j, J-1+1)$\\
	$\bgamma=(\gamma_1,\ldots,\gamma_p)'$ & latent binary variable with $\gamma_\ell=1$ indicating the presence of the $\ell$th covariate, $\ell=1,\ldots,p$ \\
	$\Gamma(a,b)$ & gamma distribution with shape parameter $a$ and rate parameter $b$\\
	$\btheta=(\theta_1,\theta_2)'$ & parameters of the centering distribution families $S_{\btheta}$ \\
	$\btheta_0$ & mean of the normal $N_2(\btheta_0, \bV_0)$ prior on $\btheta$\\
	$\hat{\btheta}$ & estimate of $\btheta$ under the parametric survival model with $S_0=S_{\btheta}$ \\
	$\nu$ &  powered exponential correlation function shape parameter, $\nu\in(0,2]$\\
	$\bxi_\ell=(\xi_{\ell1},\ldots, \xi_{\ell K})'$ &  coefficients of the cubic B-spline basis functions for the $\ell$th covariate, $\ell=1,\ldots,p$\\
	$\rho(\cdot, \cdot)$ & correlation function; arguments are two spatial locations \\
	$\rho(\cdot, \cdot; \phi)$ &  correlation function indexed by the range parameter $\phi$\\
	$\kappa$ & shrinkage parameter used under the proper CAR\\
	$\tau^2$ &  scale parameter under the ICAR or GRF or IID frailty prior\\
	$\phi$ & range parameter in the powered exponential correlation function\\
	$a_{ij}, b_{ij}$  &  endpoints of the interval $(a_{ij}, b_{ij})$ that contains the survival time $t_{ij}$, $i=1,\ldots,m$, $j=1,\ldots,n_i$\\
	$a_\alpha, b_\alpha$ & shape and rate parameters of the $\Gamma(a_{\alpha},b_{\alpha})$ prior on $\alpha$\\
	$a_\tau, b_\tau$ & shape and rate parameters of the $\Gamma(a_\tau, b_\tau)$ prior on $\tau^{-2}$\\
	$a_\phi, b_\phi$ & shape and rate parameters of the $\Gamma(a_\phi, b_\phi)$ prior on $\phi$\\
	$A$ & number of knots used in the FSA\\
	$B$ & number of blocks used in the FSA\\
	$\bC$ & precision matrix of the vector of frailties $\bv=(v_1,\ldots,v_m)'$\\
	$d(\cdot|J,\bw_J)$ & density function of Bernstein polynomial \\
	$D(\cdot|J,\bw_J)$ & cdf associated with density $d(\cdot|J,\bw_J)$\\
	$\bF_e$ & $m\times m$ diagonal matrix with the $i$ diagonal element being $e_{i+}$\\
	$e_{ij}$ & equals 1 if regions $i$ and $j$ share a common boundary and 0 otherwise, $i,j=1,\ldots,m$; set $e_{ii}=0$\\ 
	$e_{i+}$ & number of adjacent regions for region $i$, i.e. $e_{i+}=\sum_{j=1}^{m}e_{ij}$\\
	$\bE$ & $m\times m$ adjacency matrix with the $ij$th element equal to $e_{ij}$\\
	$f_{\bx_{ij}}(\cdot)$ & density function of the survival time $t_{ij}$ given the covariate $\bx_{ij}$\\
	$f_0(\cdot)$ & baseline density function in the survival models\\
	$g$ & parameter in the $g$-prior for variable selection\\
	$G$ & distribution of covariate vectors $\bx$ with support on $\calX\subseteq\mathR^p$\\
	$h_{\bx_{ij}}(\cdot)$ & hazard function of the survival time $t_{ij}$ given the covariate $\bx_{ij}$\\
	$h_0(\cdot)$ & baseline hazard function in the survival models\\
	$\bI_p$ & $p \times p$ identity matrix\\
	$I(\cdot)$ & the usual indicator function\\
	$J$ & number of Bernstein polynomials used for defining $d(\cdot|J,\bw_J)$\\
	$K$ & number of basis functions used for modeling the nonlinear function $u_\ell(\cdot)$\\
	$L(\bw_J, \btheta, \bbeta, \bv)$ & likelihood function for $(\bw_J, \btheta, \bbeta, \bv)$\\
	$m$ & number of distinct spatial locations\\
	$M$ & used to determine the $g$ in the $g$-prior for variable selection \\
	$n_i$ & number of subjects within the $i$th spatial location, $i=1\ldots,m$\\
	$n$ & total number of subjects in the data, i.e. $n=\sum_{i=1}^{m}n_i$\\
	$N(a,b^2)$ & normal distribution with mean $a$ and variable $b^2$ \\
	$N_k(\bmu, \bSigma)$ & $k$-variate normal distribution with mean $\bmu$ and covariance $\bSigma$\\
	$o_{ij}$ & number of observations for the time-dependent covariate vector $\bx_{ij}(t)$, $i=1,\ldots,m$, $j=1,\ldots,n_i$\\
	$p$ & dimension of the covariate vector $\bx_{ij}$\\
	$p(\cdot)$ & generic symbol for prior and posterior density functions\\
	$q$ & used to determine the $g$ in the $g$-prior for variable selection \\ 
	$r(t_{ij})$ & Cox-Snell residual equal to $-\log\{S_{\bx_{ij}}(t_{ij})\}$\\
	$\bR$ & correlation matrix of $\bv$ under the GRF prior\\
	$S_{\bx_{ij}}(\cdot)$ & survival function of the survival time $t_{ij}$ given the covariate $\bx_{ij}$\\
	$S_0(\cdot)$ & baseline survival function in the survival models\\
	$u_\ell(\cdot)$ & nonlinear function for the $\ell$th covariate, $\ell=1,\ldots,p$\\
	$\bv=(v_1,\ldots,v_m)'$ & vector of frailties\\
	$\bV_0$ & covariance matrix of the normal $N_2(\btheta_0, \bV_0)$ prior on $\btheta$\\
	$\hat{\bV}$ & estimate of the covariance of $\hat{\btheta}$ under the parametric survival model\\
	$\bw_J=(w_1,\ldots,w_J)'$ & $J$-vector of positive weights used in the TBP prior\\
	$\bW_0$ & covariance matrix of the normal $N_p(\bbeta_0, \bW_0)$ prior on $\bbeta$\\
	$\hat{\bW}$ & estimate of the covariance of $\hat{\bbeta}$ under the parametric survival model\\
	$\bx_{ij}$ & $p$-vector of covariates for subject $ij$, $i=1,\ldots,m$, $j=1,\ldots,n_i$\\
	$x_{ij\ell}$ & $\ell$th element of the $\bx_{ij}$, $\ell=1,\ldots,p$\\
	$\bx$ & generic symbol for $p$-vector of covariates\\
	$x_{\ell}$ & $\ell$th element of the $\bx$, $\ell=1,\ldots,p$\\
	$\bX$ & design matrix associated with $\{\bx_{ij}\}$ with mean-centered columns\\
	$\bX_\ell$ & design matrix associated with $u_\ell(x_\ell)$ with mean-centered columns $\ell=1,\ldots,p$\\
	$z_j$ & equals $\log(w_j)-\log(w_J)$\\
	$\bz_{J-1}$ & equals $(z_1,\ldots,z_{J-1})'$\\
	\noalign{\smallskip}\hline
\end{longtable}

\begin{longtable}{|p{2cm}|p{2cm}|p{11cm}|}
	\caption{List of priors.}\label{table:priors}\\
	\hline
	\hline\noalign{\smallskip}
	Parameter & Prior & Justification  \\
	\noalign{\smallskip}\hline\noalign{\smallskip}
	\endhead
	\hline
	\endfoot
	$S_0(\cdot)$ & $\TBP(\alpha, S_{\btheta})$ & Selects smooth densities and can be centered at a standard parametric family: one of log-logistic, log-normal, and Weibull. \\
	$\alpha$ & $\Gamma(a_{\alpha}, b_{\alpha})$ & $\alpha>0$ acts like the precision in a Dirichlet process controlling how stochastically pliable $S_0$ is close to $S_{\btheta}$. A gamma prior has been widely used for Dirichlet processes. Defaults: $a_{\alpha}=b_{\alpha}=1$.\\
	$\bbeta$ & $N_p(\bbeta_0, \bW_0)$ & Gaussian is common for regression effects. Defaults: $\bbeta_0=\bzero$, $\bW_0=10^{10}\bI_p$ or $\bW_0=gn(\bX'\bX)^{-1}$ when the SSVS is performed. \\
	$\btheta$ & $N_2(\btheta_0, \bV_0)$ & Centering distribution $S_{\btheta}$ is parameterized so that $\btheta$ is defined on $\mathR^2$, so a Gaussian prior is appropriate. Defaults: $\btheta_0=\hat{\btheta}$, $\bV_0=10\hat{\bV}$. Note here we assume a somewhat informative prior on $\btheta$ to obviate confounding between $\btheta$ and $\bw_J$. \\
	$\bv$ & $\ICAR(\tau^2)$ & When clusters are formed by spatial regions and spatial smoothing is of interest, the ICAR prior is commonly used for modeling the frailties in survival models. \\
	$\bv$ & $\GRF(\tau^2, \phi)$ & Very common prior for georeferenced data.\\
	$\bv$ & $\IID(\tau^2)$ & When spatial dependence among clusters is not of interest, the IID Gaussian frailties are commonly assumed. \\
	$\tau^{-2}$ & $\Gamma(a_\tau, b_\tau)$ &  The gamma distribution is a conjugate prior on $\tau^{-2}$. Defaults: $a_\tau=b_\tau=0.001$. \\
	$\phi$ & $\Gamma(a_\phi, b_\phi)$ & The range parameter $\phi$ is positive and the gamma prior is a natural choice. Defaults: $a_{\phi}=2$ and $b_\phi=(a_{\phi}-1)/\phi_0$ so that the prior of $\phi$ has mode at $\phi_0$, where $\phi_0$ satisfies $\rho(\bfs',\bfs'';\phi_0)=0.001$ with $\|\bfs',\bfs''\|=\max_{i,j}\|\bfs_i-\bfs_j\|$. \\
	$\bgamma$ & $\ds\prod_{\ell=1}^p\textrm{Bern}(q_\ell)$ & Commonly used for Bayesian variable selection \citep[e.g.][]{Kuo.Mallick1998}. Defaults: $q_\ell=0.5$, $\ell=1,\ldots,p$.\\
	$\bxi_\ell$ & $N_K(\bzero, \bS_{\bxi})$ & Here $\bS_{\bxi}=g n (\bX_\ell'\bX_\ell)^{-1}$ was chosen following the idea of informative $g$-prior introduced Appendix E.  \\
	\noalign{\smallskip}\hline
\end{longtable}

\section{MCMC Sampling}

The joint posterior distribution for all parameters is given by
\begin{equation}
\begin{aligned}
\calL(\bbeta, \btheta, \bw_J, \alpha,  \bv, \tau^2, \phi) &\propto L(\bw_J, \btheta, \bbeta, \bv) \\
& \times \exp \left\{ -\frac{1}{2} (\bbeta - \bbeta_0)' \bW_0^{-1} (\bbeta - \bbeta_0) \right\}\\
& \times \exp \left\{ -\frac{1}{2} (\btheta - \btheta_0)' \bV_0^{-1} (\btheta - \btheta_0) \right\}\\
& \times \frac{\Gamma(\alpha J)}{\Gamma(\alpha)^J}\prod_{j=1}^{J}(w_{j})^{\alpha-1} \times \alpha^{a_{\alpha} -1} \exp\{-b_{\alpha} \alpha\}\\
& \times \left(\tau^{-2}\right)^{\frac{\mathrm{rank}(\bC)}{2}} \exp\left\{ -\frac{1}{2\tau^2} \bv'\bC \bv \right\} \times \left(\tau^{-2}\right)^{a_\tau-1} \exp\left\{-b_\tau \tau^{-2} \right\}\\
& \times p(\phi) |\bC|^{1/2}\\
\end{aligned}
\end{equation}
For the GRF prior, $\bC=\bR^{-1}$ and $p(\phi)=\phi^{a_\phi-1} \exp\left\{-b_\phi \phi \right\}$. The ICAR prior does not need $p(\phi) |\bC|^{1/2}$, and $\bC=\bF_e - \bE$, where $\bF_e$ is an $m\times m$ diagonal matrix with $\bF_e[i,i]=e_{i+}$. For the IID prior, $p(\phi) |\bC|^{1/2}$ is also not needed and $\bC=\bI_m$ is an identity matrix. 

Note that when $w_{j}=1/J$ the underlying parametric model with $S_0(t)=S_{\btheta}(t)$ is obtained, so a fit from a standard parametric survival model can provide starting values for the TBP survival model. Let $\hat{\btheta}$ and $\hat{\bbeta}$ denote the parametric point estimates of $\btheta$ and $\bbeta$, and let $\hat{\bV}$ and $\hat{\bW}$ denote their asymptotic covariance matrices, respectively. These estimates can be easily obtained by running the proposed MCMC below with $w_{j}\equiv 1/J$ and relatively vague priors on $(\btheta, \bbeta)$. 

\part{1} Update $\bw_J$.\\
Set $\bz_{J-1}=(z_1,\ldots,z_{J-1})'$ with $z_j=\log(w_{j})-\log(w_{J})$.  The full conditional distribution for $\bz_{J-1}$ is 
\begin{equation*}
\begin{aligned}
p(\bz_{J-1} | \mathrm{else}) &\propto L(\bw_J, \btheta, \bbeta, \bv) \times \prod_{j=1}^{J} \left[ \frac{e^{z_j}}{\sum_{k=1}^{J}e^{z_j}} \right]^\alpha,
\end{aligned}
\end{equation*}
where $z_J=0$. The vector $\bz_{J-1}$ can be updated using adaptive Metropolis samplers \citep{Haario.etal2001}. Suppose we are currently in iteration $l$ and have sampled the states $\bz_{J-1}^{(1)}, \ldots, \bz_{J-1}^{(l-1)}$. We select an index $l_0$ (e.g., $l_0=5000$) for the length of an initial period and define
\[
\bSigma_l =
\begin{cases}
\bSigma_0,      & l \leq l_0 \\
\frac{(2.4)^2}{d}(\calC_l+10^{-10}\bI_{d}) & l> l_0.
\end{cases}
\]
Here $\calC_l$ is the sample variance of $\bz_{J-1}^{(1)}, \ldots, \bz_{J-1}^{(l-1)}$, $d=J-1$ is the dimension of $\bz_{J-1}$, and $\bSigma_0$ is an initial diagonal covariance matrix of $\bz$, defined so that the variance of $z_{j}$ is $0.16$. The choice of $0.16$ is based on extensive simulation studies; other choices (as long as it is not too small or large) will have little impact on posterior inferences. We generate $\bz_{J-1}^*=(z_1^*, \ldots, z_{J-1}^*)'$ from $N_{J-1}(\bz_{J-1}^{(l-1)}, \bSigma_l)$ and accept it with probability  
$$\min\left\{1, \frac{p(\bz_{J-1}^*|\mathrm{else})}{p(\bz_{J-1}^{(l-1)}|\mathrm{else})}\right\}.$$

\part{2} Update $\btheta$.\\
The full conditional distribution for $\btheta$ is
\begin{equation*}
\begin{aligned}
p(\btheta| \mathrm{else}) &\propto L(\bw_J, \btheta, \bbeta, \bv) \times \exp \left\{ -\frac{1}{2} (\btheta - \btheta_0)' \bV_0^{-1} (\btheta - \btheta_0) \right\}.
\end{aligned}
\end{equation*}
The centering distribution parameters $\btheta$ are updated via adaptive Metropolis samplers. At iteration $l$, each candidate is sampled as $\btheta^*\sim N_2(\btheta^{(l-1)}, \bSigma_l)$ and accepted with probability
\[\min\left\{1, \frac{p(\btheta^*|\mathrm{else})}{p(\btheta^{(l-1)}|\mathrm{else})}\right\}.
\]
where   $\bSigma_l$ is defined similarly as above, but with $\bSigma_0$ set to be $\hat{\bV}$. 

\part{3} Update $\bbeta$.\\
The full conditional distribution for $\bbeta$ is
\begin{equation*}
\begin{aligned}
p(\btheta| \mathrm{else}) &\propto L(\bw_J, \btheta, \bbeta, \bv)\times \exp \left\{ -\frac{1}{2} (\bbeta - \bbeta_0)' \bW_0^{-1} (\bbeta - \bbeta_0) \right\}.
\end{aligned}
\end{equation*}
The survival model coefficients $\bbeta$ are updated via adaptive Metropolis samplers as well with proposal $\bbeta^*\sim N_p(\bbeta^{(l-1)}, \bSigma_l)$ and acceptance probability 
\[\min\left\{1, \frac{p(\bbeta^*|\mathrm{else})}{p(\bbeta^{(l-1)}|\mathrm{else})}\right\}.
\]
where $\bSigma_l$ is defined similarly as above with $\bSigma_0=\hat{\bW}$.

\part{4} Update $\alpha$.\\
The full conditional distribution for $\alpha$ is
\begin{equation*}
\begin{aligned}
p(\alpha | \mathrm{else}) & \propto \frac{\Gamma(\alpha J)}{\Gamma(\alpha)^J}\prod_{j=1}^{J}(w_{j})^{\alpha-1} \times \alpha^{a_{\alpha} -1} \exp\{-b_{\alpha} \alpha\}.
\end{aligned}
\end{equation*}
The precision parameter $\alpha$ is updated via adaptive Metropolis samplers with normal proposal $\alpha^*\sim N_1(\alpha^{(l-1)}, \bSigma_l)$ with $\bSigma_l$ is defined similarly as above with $\bSigma_0=0.16$, and the acceptance probability is 
$$\min\left\{1, \frac{p(\alpha^*|\mathrm{else})}{p(\alpha^{(l-1)}|\mathrm{else})}\right\}.$$

\part{5} Update $\bv$.\\
Let $L(\bw_J, \btheta, \bbeta, \bv)=\prod_{i=1}^m \prod_{j=1}^{n_i} L_{ij}(\bw_J, \btheta, \bbeta, \bv)$. For the ICAR prior, the full conditional distribution for $v_i$, $i=1,\ldots, m$, is 
\begin{equation*}
\begin{aligned}
p(v_i|\text{else}) &\propto \prod_{j=1}^{n_i} L_{ij}(\bw_J, \btheta, \bbeta, \bv) \exp\left\{ -\frac{e_{i+}}{2\tau^2}\left(v_i - \sum_{j=1}^m e_{ij}v_j/e_{i+} \right)^2 \right\}.
\end{aligned}
\end{equation*}
The $v_j$ is updated via Metropolis-Hastings sampling steps with proposal $v_j^*\sim N(v_j^{(l-1)}, \tau^2/e_{j+})$. The candidate $v_j^*$ is accepted with probability 
\[\min\left\{1, \frac{p(v_j^*|\mathrm{else})}{p(v_j^{(l-1)}|\mathrm{else})}\right\}.\]  
After each individual frailty update, the vector of $\bv$ is updated to have sample mean zero through $\bv \leftarrow \bv-\tfrac{1}{m} \mathbf{1}_m'\bv$.  Although \emph{ad hoc}, this approach to enforcing the sum-to-zero constraint on $v_1,\dots,v_m$ has negligible effect on the posterior and has been advocated by many authors, e.g. \cite{Banerjee.etal2014} and \cite{Lang.Brezger2004}.

For the IID prior, the full conditional distribution for $v_i$, $i=1,\ldots, m$, is 
\begin{equation*}
\begin{aligned}
p(v_i|\text{else}) &\propto \prod_{j=1}^{n_i} L_{ij}(\bw_J, \btheta, \bbeta, \bv) \exp\left\{ -\frac{1 }{2\tau^2}v_i^2 \right\}.
\end{aligned}
\end{equation*}
The $v_j$ is updated via Metropolis-Hastings sampling steps with proposal $v_j^*\sim N(v_j^{(l-1)}, \tau^2)$. The candidate $v_j^*$ is accepted with probability 
\[\min\left\{1, \frac{p(v_j^*|\mathrm{else})}{p(v_j^{(l-1)}|\mathrm{else})}\right\}.
\]
For the GRF prior, the full conditional distribution for $v_i$, $i=1,\ldots, m$, is 
\begin{equation*}
\begin{aligned}
p(v_i|\text{else}) &\propto \prod_{j=1}^{n_i} L_{ij}(\bw_J, \btheta, \bbeta, \bv) \exp\left\{ -\frac{p_{ii} }{2\tau^2}\left(v_i +\sum_{\{j:j\neq i\}} p_{ij}v_j/p_{ii} \right)^2 \right\},
\end{aligned}
\end{equation*}
where $p_{ij}$ is the $(i,j)$ element of $\bR^{-1}$.  The $v_j$ is updated via Metropolis-Hastings sampling steps with proposal $v_j^*\sim N(v_j^{(l-1)}, \tau^2/p_{ii})$. The candidate $v_j^*$ is accepted with probability 
\[\min\left\{1, \frac{p(v_j^*|\mathrm{else})}{p(v_j^{(l-1)}|\mathrm{else})}\right\}.
\]

\part{6} Update $\tau^2$.\\
The full conditional distribution for $\tau^{-2}$ is
\begin{equation*}
p(\tau^{-2}|\text{else})\propto \left(\tau^{-2}\right)^{a_\tau + \frac{\mathrm{rank}(\bC)}{2}-1} \exp\left\{-\left[b_\tau + \frac{1}{2}\bv'\bC \bv \right] \tau^{-2} \right\}.
\end{equation*}
Thus $\tau^{-2}$ is sampled from $\Gamma(a_\tau^*, b_\tau^*)$, where $a_\tau^*=a_\tau + \frac{\mathrm{rank}(\bC)}{2}-1$ and $b_\tau^*=b_\tau + \frac{1}{2}\bv'\bC \bv$.

\part{7} Update $\phi$ for georeferenced data.\\
The full conditional distribution for $\phi$ is
\begin{equation*}
\begin{aligned}
p(\phi | \mathrm{else}) & \propto |\bR|^{-1/2} \exp\left\{ -\frac{1}{2\tau^2} \bv' \bR^{-1}\bv \right\} \phi^{a_\tau-1} \exp\left\{-b_\phi \phi \right\} 
\end{aligned}
\end{equation*}
The range parameter $\phi$ is updated via adaptive Metropolis samplers with normal proposal $\phi^*\sim N_1(\phi^{(l-1)}, \bSigma_l)$ with $\bSigma_l$ is defined similarly as above with $\bSigma_0=0.16$, and the acceptance probability is 
$$\min\left\{1, \frac{p(\phi^*|\mathrm{else})}{p(\phi^{(l-1)}|\mathrm{else})}\right\}.$$

\part{8} Update $\bgamma$ when variable selection is performed.\\
When variable selection is performed, all $\bbeta$s in steps 1-7 need to be replaced by $\bgamma\odot\bbeta$, where $\odot$ denotes componentwise multiplication. Then each $\gamma_j$ is generated from its full conditional, i.e. a Bernoulli distribution with the success probability
\[ \frac{q_j}{q_j+(1-q_j)L(\bw_J, \btheta, \bgamma_{j0}\odot\bbeta, \bv)/L(\bw_J, \btheta, \bgamma_{j1}\odot\bbeta, \bv)},
\]
where the vector $\bgamma_{j0}$ $(\bgamma_{j1})$ is obtained from $\bgamma$ with the $j$th element replaced by $0$ (1). 

\section{The Full Scale Approximation}

For georeferenced data, a computational bottleneck of the MCMC sampling scheme is inverting the $m\times m$ matrix $\bR$, which typically has computational cost $O(m^3)$. In this section, we introduce a full scale approximation (FSA) approach proposed by \cite{Sang.Huang2012}, which provides a high quality approximation to the correlation function $\rho$ at both the large and the small spatial scales, such that the inverse of $\bR$ can be substantially sped up for large value of $m$, e.g., $m\geq 500$. 

Consider a fixed set of ``knots'' $\calS^* = \{\bfs^*_1, \ldots, \bfs^*_A\}$ chosen from the study region. These knots can be chosen using the function \texttt{cover.design} within the R package \texttt{fields}, which computes space-filling coverage designs using the swapping algorithm \citep{Johnson.etal1990}. Let $\rho(\bfs,\bfs')$ be the correlation between locations $\bfs$ and $\bfs'$. The FSA approach approximates the correlation function $\rho(\bfs,\bfs')$ with
\begin{equation}\label{FSA}
	\rho^\dag (\bfs, \bfs') = \rho_l(\bfs, \bfs') + \rho_s(\bfs, \bfs').
\end{equation}
The $\rho_l(\bfs,\bfs')$ in (\ref{FSA}) is the reduced-rank part capturing the long-scale spatial dependence, defined as $ \rho_l(\bfs, \bfs') = \rho'(\bfs, \calS^*)\rho_{AA}^{-1}(\calS^*, \calS^*)\rho(\bfs', \calS^*)$,
where $\rho(\bfs, \calS^*) = [\rho(\bfs, \bfs_i^*)]_{i=1}^A$ is an $A\times 1$ vector, and $\rho_{AA}(\calS^*, \calS^*)=[\rho(\bfs_i^*, \bfs_j^*)]_{i,j=1}^A$ is an $A\times A$ correlation matrix at knots $\calS^*$. However, $\rho_l(\bfs,\bfs')$ cannot well capture the short-scale dependence due to the fact that it discards entirely the residual part $\rho(\bfs,\bfs')-\rho_l(\bfs,\bfs')$. The idea of FSA is to add a small-scale part $\rho_s(\bfs,\bfs')$ as a sparse approximate of the residual part, defined by $\rho_s(\bfs, \bfs') = \left\{ \rho(\bfs,\bfs')-\rho_l(\bfs,\bfs') \right\}\Delta(\bfs,\bfs')$, where $\Delta(\bfs,\bfs')$ is a modulating function, which is specified so that the $\rho_s(\bfs,\bfs')$ can well capture the local residual spatial dependence while still permits efficient computations. Motivated by \cite{Konomi.etal2014}, we first partition the total input space into $B$ disjoint blocks, and then specify $\Delta(\bfs,\bfs')$ in a way such that the residuals are independent across input blocks, but the original residual dependence structure within each block is retained. Specifically, the function $\Delta(\bfs,\bfs')$ is taken to be $1$ if $\bfs$ and $\bfs'$ belong to the same block and $0$ otherwise. The approximated correlation function $\rho^\dag (\bfs, \bfs')$ in (\ref{FSA}) provides an exact recovery of the true correlation within each block, and the approximation errors are $\rho(\bfs,\bfs')-\rho_l(\bfs,\bfs')$ for locations $\bfs$ and $\bfs'$ in different blocks. Those errors are expected to be small for most entries because most of these location pairs are farther apart. To determine the blocks, we first use the R function \texttt{cover.design} to choose $B\leq m$ locations among the $m$ locations forming $B$ blocks, then assign each $\bfs_i$ to the block that is closest to $\bfs_i$. Here $B$ does not need to be equal to $A$. When $B=1$, no approximation is applied to the correlation $\rho$. When $B=m$, it reduces to the approach of \cite{Finley.etal2009}, so the local residual spatial dependence may not be well captured. 

Applying the above FSA approach to approximate the correlation function $\rho(\bfs,\bfs')$, we can approximate the correlation matrix $\bR$ with
\begin{equation}\label{FSA:rho}
	\begin{aligned}
		\brho_{mm}^\dag &= \brho_l + \brho_s = \brho_{mA}\brho_{AA}^{-1}\brho_{mA}' + \left(\brho_{mm} - \brho_{mA}\brho_{AA}^{-1}\brho_{mA}' \right) \circ \bDelta,
	\end{aligned}
\end{equation}
where $\brho_{mA} = [\rho(\bfs_i, \bfs_j^*)]_{i=1:m,j=1:A}$, $\brho_{AA}=[\rho(\bfs_i^*, \bfs_j^*)]_{i,j=1}^A$, and $\bDelta=[\Delta(\bfs_i, \bfs_j)]_{i,j=1}^m$. Here, the notation ``$\circ$'' represents the element-wise matrix multiplication. To avoid numerical instability, we add a small nugget effect $\epsilon=10^{-10}$ when define $\bR$, that is, $\bR=(1-\epsilon)\brho_{mm}+\epsilon \bI_m$.  It follows from equation (\ref{FSA:rho}) that $\bR$ can be approximated by
$$\bR^\dag = (1-\epsilon)\brho_{mm}^\dag+\epsilon\bI_m = (1-\epsilon)\brho_{mA}\brho_{AA}^{-1}\brho_{mA}'  + \bR_s,$$
where $\bR_s=(1-\epsilon)\left(\brho_{mm} - \brho_{mA}\brho_{AA}^{-1}\brho_{mA}' \right) \circ \bDelta + \epsilon\bI_m$. Applying the Sherman-Woodbury-Morrison formula for inverse matrices, we can approximate $\bR^{-1}$ by 
\begin{equation}\label{FSA:inv}
	\begin{aligned}
		\left(\bR^\dag\right)^{-1}=\bR_s^{-1} - (1-\epsilon)\bR_s^{-1}\brho_{mA} \left[\brho_{AA}+(1-\epsilon)\brho_{mA}'\bR_s^{-1}\brho_{mA}\right]^{-1} \brho_{mA}' \bR_s^{-1}.
		\normalsize
	\end{aligned}
\end{equation}
In addition, the determinant of $\bR$ can be approximated by 
\begin{equation}\label{FSA:det}
	\det\left(\bR^\dag\right) = \det\left\{\brho_{AA}+(1-\epsilon)\brho_{mA}'\bR_s^{-1}\brho_{mA}\right\} \det(\brho_{AA})^{-1}\det(\bR_s).
\end{equation}
Since the $m\times m$ matrix $\bR_s$ is a block matrix, the right-hand sides of equations (\ref{FSA:inv}) and (\ref{FSA:det}) involve only inverses and determinants of $A\times A$ low-rank matrices and $m\times m$ block diagonal matrices. Thus the computational complexity can be greatly reduced relative to the expensive computational cost of using original correlation function for large value of $m$.

\section{The DIC and LPML Criteria}
To set notation, denote by $\calD$ the observed dataset, by $\calD_i$ the $i$th data point, and by $\calD_{-i}$ the dataset with $\calD_i$ removed, $i=1,\ldots,n$. Let $\Omega$ denote the entire collection of model parameters under a particular model, $L(\calD|\Omega)$ be the likelihood function based on observed data $\calD$, and $L_i(\cdot|\Omega)$ be the likelihood contribution based on $\calD_i$. Suppose $\{\Omega^{(1)}, \ldots, \Omega^{(\calL)}\}$ are random draws from the full posterior $p_{post}(\Omega|\calD)$. Let $\hat{\Omega}=\sum_{l=1}^{\calL}\Omega^{(l)}/\calL$ be the posterior mean estimate for $\Omega$.

The DIC, a generalization of the Akaike information criterion (AIC), is commonly used for comparing complex hierarchical models for which the asymptotic justification of AIC is not appropriate. The DIC is defined as
\begin{equation}\label{DIC}
	\DIC = -2\log L(\calD|\hat{\Omega}) + 2p_D,
\end{equation} 
where 
\[ p_D=2\left(\log L(\calD|\hat{\Omega})-\frac{1}{\calL}\sum_{l=1}^{\calL}\log L(\calD|\Omega^{(l)}) \right)
\]
is referred to as the effective number of parameters measuring the model complexity. Similar to AIC, a smaller value of DIC indicates a better fit model. 

The definition of LPML is based on the conditional predictive ordinate (CPO) statistic. The CPO for data point $\calD_i$ is given by
\begin{equation*}\label{CPO}
\mbox{CPO}_{i} = f(\mathcal{D}_i|\mathcal{D}_{-i}) = \int L_i(\calD_i|\Omega) p_{post}(\Omega|\calD_{-i}) d\Omega,
\end{equation*}  
where $p_{post}(\cdot|\calD_{-i})$ is the posterior density of $\Omega$ give $\calD_{-i}$. Let $\CPO_{i,1}$ and $\CPO_{i,2}$ denote the CPO for the $i$th data point under models 1 and 2, respectively. The ratio $\mbox{CPO}_{i,1}/\mbox{CPO}_{i,2}$ measures how well model 1 supports the data point $\mathcal{D}_i$ relative to model 2, based on the remaining data $\mathcal{D}_{-i}$. The product of the CPO ratios gives an overall aggregate summary of how well supported the data are by model 1 relative to model 2 and is called the pseudo Bayes factor (PBF),
\[ B_{12} = \prod_{i=1}^n
\frac{\mbox{CPO}_{i,1}}{\mbox{CPO}_{i,2}}.
\]
It is well known that Bayes factors \citep{Kass.Raftery1995,Han.Carlin2001} are usually difficult to obtain in practice. The PBF is a surrogate for the more traditional Bayes factor and can be interpreted similarly, but is more analytically tractable, much less sensitive to prior assumptions, and does not suffer from Lindley's paradox. 

As noted by \cite{Gelfand.Dey1994}, one can use importance sampling to estimate $\CPO_i$ by 
\begin{equation*}
\left\{ \frac{1}{\calL}\sum_{l=1}^{\calL}\frac{1}{L_i(\calD_i|\Omega^{(l)})} \right\}^{-1}.
\end{equation*}
However, these estimates may be unstable since the weights $\omega_{i,l}=1/L_i(\calD_i|\Omega^{(l)})$ can have infinite variance \citep{Epifani.etal2008}, depending on the tail behavior of $p_{post}(\Omega|\calD_{-i})$ relative to $L_i(\calD_i|\Omega)$ as a function of $\Omega$. To stabilize the weights, \cite{Vehtari.Gelman2014} suggest replacing $\omega_{i,l}$ with $\tilde{\omega}_{i,l}=\min\{ \omega_{i,l},\sqrt{\calL} \bar{\omega}_{i}\}$, where $\bar{\omega}_{i}=\sum_{l=1}^{\calL}\omega_{i,l}/\calL$. Therefore, the stabilized estimate of the CPO statistic is 
\begin{equation*}\label{CPO:est}
\widehat{\CPO}_i = \frac{\sum_{l=1}^\calL L_i(\calD_i|\Omega^{(l)})\tilde{\omega}_{i,l}}{\sum_{l=1}^{\calL} \tilde{\omega}_{i,l}}.
\end{equation*}
Finally, the LPML is defined as 
\begin{equation}\label{LPML}
\mbox{LPML}=\sum_{i=1}^n \log \widehat{\CPO}_i.
\end{equation}
A further improved estimate was recently proposed by \cite{Vehtari.etal2017} using Pareto-smoothed importance sampling; this version will be implemented in later versions of the R package.

The LPML can be viewed as a predictive measure that generalizes leave-one-out cross-validated prediction error to more heavily penalize ``bad predictions.''  Consider the frequentist LPML proposed by \cite{Geisser.Eddy1979} for normal-errors regression data.  Let $y_i \stackrel{ind.}{\sim} N(\bx_i'\bbeta, \sigma^2)$; then \[ \mbox{CPO}_i = \frac{1}{\sqrt{2\pi} \hat{\sigma}_i} \exp\left\{ -\frac{(y_i-\bx_i'\hat{\bbeta}_i)^2}{2 \hat{\sigma}_i^2}\right\},\] where $(\hat{\bbeta}_i,\hat{\sigma}_i)$ is the MLE of $(\bbeta,\sigma)$ leaving out $(\bx_i,y_i)$.  Then \[ -\mbox{LPML}=\underbrace{\sum_{i=1}^n \frac{1}{2\hat{\sigma}_i}(y_i-\hat{y}_{-i})^2}_{\mbox{squared bias}}+\underbrace{\sum_{i=1}^n \log \hat{\sigma}_i}_{\mbox{variance}} + \mbox{ constant}.\] This generalizes to any location-scale family, e.g. parametric survival models $\log t_i = \bx_i'\bbeta +\epsilon_i$, where $\epsilon_i$ has a scaled standard extreme value distribution, scaled log-logistic distribution, or scaled normal distribution yielding common Weibull, log-logistic, and log-normal regression models.  Note that unlike the usual predicted residual error sum of squares (PRESS) statistic the bias terms are weighted by the variability of the prediction: ``bad'' predictions with less variability (more precision) provide much more discrepancy than ``bad'' predictions with large variability.  Having both bias and variance pieces, the LPML is of similar form to the L-measure \citep{Ibrahim.etal2001b}, but more naturally generalizes to survival data; note that \cite{Ibrahim.etal2001b} advocating taking the log of the survival time and require a different L-measure for each family of distributions.

A Bayesian might view the frequentist CPO$_i$ using the MLE above as overoptimistic.  The MLE is the posterior mode under a flat prior and sampling variability is not taken into account.  Instead, one might want to average the CPO$_i$ statistic over the (perhaps asymptotic) estimated sampling distribution of $\Omega_i$, e.g. $\Omega_i \stackrel{\bullet}{\sim} N(\hat{\Omega}_i, \mathbf{V}_i)$.  Equivalently, and more precisely, the Bayesian approach averages the predictive density for a new observation with covariates $\bx_i$ over the leave-$i$-out posterior $[\Omega|\mathcal{D}_{-i}]$.  Thus the Bayesian LPML used here can be viewed as a measure similar to PRESS or prediction error, but a properly pessimistic one that averages over the (non-asymptotic) sampling distribution of the parameters.  The more sampling variability there is (reflecting smaller sampler sizes $n$), the more heavily each CPO$_i$ is penalized. 

In addition to DIC and LPML, the Watanabe-Akaike information criterion (WAIC) \citep{Watanabe2010} has also gained popularity in recent years due to its stability compared to DIC \citep{Gelman.etal2014,Vehtari.Gelman2014}. The WAIC is defined as 
\begin{equation}\label{WAIC}
\WAIC = -2\sum_{i=1}^{n}\log \left(\frac{1}{\calL}\sum_{l=1}^{\calL} L_i(\calD_i|\Omega^{(l)})\right) + 2p_W,
\end{equation}
where 
\[ p_W=\sum_{i=1}^{n} \left[\frac{1}{\calL-1}\sum_{l=1}^{\calL} \left\{ \log L_i(\calD_i|\Omega^{(l)})-\frac{1}{\calL}\sum_{k=1}^{\calL} \log L_i(\calD_i|\Omega^{(k)}) \right\}^2 \right]
\]
is the effective number of parameters. A smaller value of WAIC indicates a better predictive model. WAIC can be viewed as an approximation to $-2\sum_{i=1}^{n}\log\CPO_i$ \citep{Gelman.etal2014}, so WAIC is also used to compare models' predictive performance. The WAIC has been implemented in the function \texttt{survregbayes} and saved in its returned object. 

\section{Parametric vs. Nonparametric $S_0(\cdot)$}
Many authors have found parametric models to fit as well or better than competing semiparametric models (\citealp[p. 123]{Cox.Oakes1984}; \citealp{Nardi.Schemper2003}). Here, testing for the adequacy of the simpler underlying parametric model is developed.  
The proposed semiparametric models have their baseline survival functions centered at a parametric family $S_{\btheta}(t)$. Note that $\bz_{J-1}=\bzero$ implies $S_0(t)=S_{\btheta}(t)$. Therefore, testing $H_0:\bz_{J-1}=\bzero$ versus $H_1:\bz_{J-1}\neq\bzero$ leads to the comparison of the semiparametric model with the underlying parametric model. Let $BF_{10}$ be the Bayes factor between $H_1$ and $H_0$. \cite{Zhou.etal2017} proposed to estimate $BF_{10}$ by a large-sample approximation to the generalized Savage-Dickey density ratio \citep{Verdinelli.Wasserman1995}. Adapting their approach $BF_{10}$ is estimated
\begin{equation}
\widehat{BF}_{10} = \frac{p(\bzero|\hat{\alpha})}{N_{J-1}(\bzero; \hat{\bfm}, \hat{\bSigma})},
\end{equation}
where $p(\bzero|\alpha)=\Gamma(\alpha J)/[J^{\alpha}\Gamma(\alpha) ]^J$ is the prior density of $\bz_{J-1}$ evaluated at $\bz_{J-1}=\bzero$, $\hat{\alpha}$ is the posterior mean of $\alpha$, $N_{p}(\cdot; \bfm, \bSigma)$ denotes a $p$-variable normal density with mean $\bfm$ and covariance $\bSigma$, and $\hat{\bfm}$ and $\hat{\bSigma}$ are posterior mean and covariance of $\bz_{J-1}$. 

\section{Variable Selection}\label{sec:selection}
There is a large amount of literature on Bayesian variable selection methods; see \cite{OHara.Sillanpaeae2009} for a comprehensive review. Let $\bx=(x_1,\ldots,x_p)'$ denote the $p$-vector of covariates in general. The most direct approach is to multiply $\beta_\ell$ by a latent Bernoulli variable $\gamma_\ell$ for $\ell=1,\ldots,p$, where $\gamma_\ell=1$ indicates the presence of $x_{\ell}$ in the model, and then assume an appropriate prior on $(\bbeta, \bgamma)$, where $\bgamma=(\gamma_1, \ldots, \gamma_p)'$. \cite{Kuo.Mallick1998} considered an independent prior $p(\bbeta, \bgamma)=N_p(\bzero, \bW_0)\times \prod_{\ell=1}^p\textrm{Bern}(q_\ell)$, where $\bW_0$ was taken as a diagonal matrix yielding a diffuse prior on $\bbeta$, and $q_\ell$ is a prior probability of including $x_{\ell}$ in the model. The resulting MCMC algorithm does not require any tuning, but mixing can be poor if the prior on $\bbeta$ is too diffuse \citep{OHara.Sillanpaeae2009}. The $g$-prior of \cite{Zellner1983} and its various extensions \citep{Bove.etal2011,Hanson.etal2014} have been widely used for variable selection. We consider one such prior adapted for use in the semiparametric survival models considered here. Specifically, the same prior as \cite{Kuo.Mallick1998} is considered, but with 
\begin{equation} \label{eq:gprior}
	\bbeta \sim N_p(\mathbf{0}, g n (\bX'\bX)^{-1}),  
\end{equation}
where $\bX$ is the usual design matrix, but with mean-centered covariates, i.e. 
$\mathbf{1}_n' \bX = \mathbf{0}_p'$. 
Assume that the covariate vectors $\bx_{ij}$ arise from a distribution $G$ with support on ${\calX}\subseteq \mathR^p$, and are independent of $\bbeta$.  Following \cite{Hanson.etal2014} $g$ is set equal to a constant based on prior information on $e^{\bx'\bbeta}$, i.e. the relative risks (PH), acceleration factors (AFT), or odds factors (PO) of random subjects $\bx$ relative to their mean $\int_{\mathcal{X}} \bx G(d\bx)$.  Under the prior (\ref{eq:gprior}), \cite{Hanson.etal2014} showed that $\bx'\bbeta$ has an approximately normal distribution with mean 0 and variance $ng$. Thus, a simple method of choosing $g$ is to pick a number $M$ such that a random $e^{\bx'\bbeta}$ is less than $M$ with probability $q$. It follows that $g = \left[ {\log M}/{\Phi^{-1}(q)} \right]^2/{p}$. Here, $M=10$ and $q=0.9$ are fixed.  The MCMC procedure is described in supplementary Appendix A.  Posterior output includes a list of sub-models with their posterior probabilities, i.e. a ranking of models much like the best subsets $C_p$ statistic.

This variable selection method was originally termed ``stochastic search variable selection'' (SSVS) by \cite{George.McCulloch1993} who instead of using Bernoulli point masses for each regression effect used highly concentrated normal distributions centered at zero.  This approach has also been called ``spike and slab'' variable selection by many authors.  A recent review and extensive simulation study by \cite{Pavlou.etal2016} suggests that SVSS can routinely outperform other variable selection approaches.  They found that SSVS performed overall the best across many realistic data scenarios for variable selection among methods that also include versions of the LASSO (regular, adaptive, and Bayesian), SCAD, and the elastic net.  All methods grossly outperformed backwards elimination; see Table 4 in \cite{Pavlou.etal2016}.

\section{Left-Truncation and Time-Dependent Covariates}
To avoid an explosion of subscripts, drop the $ij$ from $t_{ij}$, etc.  
The survival time $t$ is left-truncated at $u\geq 0$ if $u$ is the time when the subject under consideration is first observed. Left-truncation often occurs when age is used as the time scale. Given the observed left-truncated data $\{ (u, a, b, \bx, \bfs)\}$, where $a\geq u$, the likelihood contribution is
\begin{equation*}\label{like-truncation}
	L=\left[S_{\bx}(a)-S_{\bx}(b)\right]^{I\{a<b\}} f_{\bx}(a)^{I\{a=b\}}/S_{\bx}(u).
\end{equation*}
Note that the left censored data under left-truncation are of the form $(u, b)$. 

We next discuss how to extend the semiparametric AFT, PH and PO models to handle time-dependent covariates. Let $\{(u, a, b, \bx(t), \bfs): u \leq t\leq a\}$ be the observed data with time-dependent covariates and possible left-truncation. Suppose we observe $\bx(t)$ at $o$ ordered times $t=t_{1}, \ldots, t_{o}$, denoted as $\bx_{1},\ldots,\bx_{o}$, respectively, where $t_{1}=u$ and $t_{o}\leq a$.  Following \cite{Kneib2006} and \cite{Hanson.etal2009}, we assume that $\bx(t)$ is a step function given by  
\[ \bx(t) = \sum_{k=1}^{o} \bx_{k} I(t_{k}\leq t < t_{k+1}),
\] where $t_{o+1}=\infty$.   Assuming the AFT, PH or PO holds conditionally on each interval, the survival function at time $a$ is
  \begin{align*} 
 P(t>a) 
 & = P(t>a|t>t_{o}) \prod_{k=1}^{o-1}P(t>t_{k+1}|t>t_{k}) \\
 & = \frac{S_{\bx_{o}}(a)}{S_{\bx_{o}}(t_{o})}
 \prod_{k=1}^{o-1} \frac{S_{\bx_{k}}(t_{k+1})}{S_{\bx_{k}}(t_{k})}.
 \end{align*} 
This leads to the usual PH model for time-dependent covariates \citep{Cox1972}, the AFT model first proposed by \cite{Prentice.Kalbfleisch1979}, and a particular piecewise PO model.  

Returning to the use of subscripts for the $ij$th subject, for time-dependent covariates replace $(u_{ij}, a_{ij}, b_{ij}, \bx_{ij}(t), \bfs_i)$ by a set of new $o_{ij}$ observations $(t_{ij,1}, t_{ij,2}, \infty, \bx_{ij,1}, \bfs_i)$, $(t_{ij,2}, t_{ij,3}, \infty, \bx_{ij,2}, \bfs_i)$, $\ldots$, $(t_{ij,o_{ij}}, a_{ij}, b_{ij}, \bx_{ij,o_{ij}}, \bfs_i)$ yielding an augmented left-truncated data set of size $\sum_{i=1}^{m}\sum_{j=1}^{n_i} o_{ij}$.  Then the likelihood function becomes
\begin{align*}
	L(\bw_J, \btheta, \bbeta, \bv) = & \prod_{i=1}^m \prod_{j=1}^{n_i} \bigg\{ \left[S_{\bx_{ij,o_{ij}}}(a_{ij})-S_{\bx_{ij,o_{ij}}}(b_{ij})\right]^{I\{a_{ij}<b_{ij}\}}  f_{\bx_{ij,o_{ij}}}(a_{ij})^{I\{a_{ij}=b_{ij}\}}/S_{\bx_{ij,o_{ij}}}(t_{ij,o_{ij}})  \\
	&\times \prod_{k=1}^{o_{ij}-1} \frac{S_{\bx_{ij,k}}(t_{ij,k+1})}{S_{\bx_{ij,k}}(t_{ij,k})}\bigg\} . 
\end{align*}
Note that the derivations above still hold for time-dependent covariates without left-truncation (i.e. $u_{ij}=0$ for all $i$ and $j$).

\section{Partially linear predictors}\label{sec:additive}
An additive PH model was first considered by \cite{Gray1992} as
\[ h_{\bx_{ij}}(t) = h_0(t)\exp\{\bx_{ij}'\bbeta + \sum_{\ell=1}^p u_\ell(x_{ij\ell})\},\] where the nonlinear functions $u_1(\cdot),\dots,u_p(\cdot)$ are modeled via penalized B-splines with the linear portion removed.  Setting some of the $u_\ell(\cdot) \equiv 0$ gives the so-called ``partially linear PH model'' that has been given a great deal of attention in recent literature.  This model has been extended to spatial versions by \cite{Kneib2006} and \cite{Hennerfeind.etal2006} for PH and can be easily fit in \texttt{R2BayesX}. 

Additive partially linear predictors can be implemented in the proposed AFT, PH and PO models by simply adding a linear basis expansion for any continuous covariate; cubic B-splines are considered here and illustrated in Section 3.3. Specifically, $u_\ell(\cdot)$ is parameterized as 
\[u_\ell(\cdot)=\sum_{k=1}^{K}\xi_{\ell k}B_{\ell k}(\cdot),\] 
where $\{B_{\ell k}(\cdot): k=0,\ldots,K+1\}$ are the standard cubic B-spline basis functions with knots determined by the data; the first and last basis functions have been dropped to ensure a full-rank model (the linear term is already included). Independent normal priors are considered for $\bbeta$ and $\bxi_\ell=(\xi_{\ell1},\ldots,\xi_{\ell K})'$:
\[ \bbeta\sim N_p(\bzero, \bW_0), \quad \bxi_\ell \sim N_K(\bzero, g n (\bX_\ell'\bX_\ell)^{-1}), \ell=1,\ldots,p
\]
where $\bW_0=10^{10}\bI_p$, $\bX_\ell$ is the design matrix for the $u_\ell(\cdot)$ term, and $g=\left[ {\log 10}/{\Phi^{-1}(0.9)} \right]^2/{K}$. This approach can be viewed as a simplified version of the Bayesian P-splines \citep{Lang.Brezger2004} with fewer basis functions and a g-prior ``penalty'' instead of a random-walk penalty. Note that posterior updating could be inefficient if a large number of basis functions is considered, as $(\bbeta,\bxi_1,\ldots,\bxi_p)$ is currently updated in one large block via adaptive Metropolis. For this reason, the full Bayesian P-spline approach may be a better choice, but requires updating high-dimensional vectors of spline coefficient parameters, and their suggested iteratively weighted least squares proposals would need to be modified to handle our survival models. We hope to include this in future updates of the R package

Bayes factors can be used to test the linearity of $x_{ij\ell}$ through the hypothesis $H_0:\bxi_\ell=\bzero$ versus $H_1:\bxi_\ell\neq \bzero$. Let $BF_{10}$ be the Bayes factor between $H_1$ and $H_0$. We estimate $BF_{10}$ by a large-sample approximation to the Savage-Dickey density ratio \citep{Dickey1971} 
\begin{equation}
\widehat{BF}_{10} = \frac{N_K(\bzero; \bzero, g n (\bX_\ell'\bX_\ell)^{-1}) }{N_{K}(\bzero; \hat{\bfm}_\ell, \hat{\bSigma}_\ell)},
\end{equation}
where $\hat{\bfm}_\ell$ and $\hat{\bSigma}_\ell$ are posterior mean and covariance of $\bxi_\ell$. 

\section{Implementation Using R}
An illustrative use of the \texttt{R} function \texttt{survregbayes} in the package \texttt{spBayesSurv} is presented to fit AFT, PH and PO frailty models with the TBP prior on baseline survival functions using simulated data. We take {Example 2} of the variable selection simulation (see \textbf{Simulation IV} below) as an example. The following code is used to generate data:

{\footnotesize
\begin{verbatim}
##-------------Load libraries-------------------##
rm(list=ls())
library(coda)
library(survival)
library(spBayesSurv)
library(BayesX)

##-------------Set the true models--------------##
betaT = c(1,1,0,0,0); 
## Baseline Survival
f0oft = function(t) 0.5*dlnorm(t, -1, 0.5)+0.5*dlnorm(t,1,0.5);
S0oft = function(t) 0.5*plnorm(t, -1, 0.5, lower.tail=FALSE)+
		0.5*plnorm(t, 1, 0.5, lower.tail=FALSE)
## The Survival function:
Sioft = function(t,x,v=0)  exp( log(S0oft(t))*exp(sum(x*betaT)+v) ) ;
Fioft = function(t,x,v=0) 1-Sioft(t,x,v);
## The inverse for Fioft
Finv = function(u, x,v=0) uniroot(function (t) Fioft(t,x,v)-u, lower=1e-100, upper=1e100,
extendInt ="yes")$root

##-------------Generate data-------------------##
## read the adjacency matrix of Nigeria for the 37 states
nigeria=read.bnd(system.file("otherdata/nigeria.bnd", 
package="spBayesSurv"));
adj.mat=bnd2gra(nigeria)
W = diag(diag(adj.mat)) - as.matrix(adj.mat); m=nrow(W);
tau2T = 1;
covT = tau2T*solve(diag(rowSums(W))-W+diag(rep(1e-10, m)));
v0 = MASS::mvrnorm(n=1, mu=rep(0,m), Sigma=covT); 
v = v0-mean(v0);
mis = rep(20, m); n = sum(mis);
vn = rep(v, mis);
id = rep(1:m, mis); 
## generate x 
x1 = rbinom(n, 1, 0.5); x2 = rnorm(n, 0, 1);
x3 = x2+0.15*rnorm(n); x4 = rnorm(n, 0, 1); x5 = rnorm(n, 0, 1);
X = cbind(x1, x2, x3, x4, x5);
colnames(X) = c("x1", "x2", "x3", "x4", "x5");
## generate survival times
u = runif(n);
tT = rep(0, n);
for (i in 1:n){
tT[i] = Finv(u[i], X[i,], vn[i]);
}
## generate partly interval-censored data
t1=rep(NA, n);t2=rep(NA, n); delta=rep(NA, n); 
n1 = floor(0.5*n); ## right-censored part
n2 = n-n1; ## interval-censored part
# right-censored part
rcen = sample(1:n, n1);
t1_r=tT[rcen];t2_r=tT[rcen];
Centime = runif(n1, 2, 6);
delta_r = (tT[rcen]<=Centime) +0 ; length(which(delta_r==0))/n1;
t1_r[which(delta_r==0)] = Centime[which(delta_r==0)];
t2_r[which(delta_r==0)] = NA;
t1[rcen]=t1_r; t2[rcen]=t2_r; delta[rcen] = delta_r;
# interval-censored part
intcen = (1:n)[-rcen];
t1_int=rep(NA, n2);t2_int=rep(NA, n2); delta_int=rep(NA, n2);
npois = rpois(n2, 2)+1;
for(i in 1:n2){
gaptime = cumsum(rexp(npois[i], 1)); 
pp = Fioft(gaptime, X[intcen[i],], vn[intcen[i]]);
ind = sum(u[intcen[i]]>pp); 
if (ind==0){
delta_int[i] = 2;
t2_int[i] = gaptime[1];
}else if (ind==npois[i]){
delta_int[i] = 0;
t1_int[i] = gaptime[ind];
}else{
delta_int[i] = 3;
t1_int[i] = gaptime[ind];
t2_int[i] = gaptime[ind+1];
}
}
t1[intcen]=t1_int; t2[intcen]=t2_int; delta[intcen] = delta_int;
## make a data frame
d = data.frame(t1=t1, t2=t2, X, delta=delta, tT=tT, ID=id, frail=vn); table(d$delta)/n;

##------- Fit the PH model with variable selection -----------##
# MCMC parameters
nburn=10000; nsave=2000; nskip=4; niter = nburn+nsave
mcmc=list(nburn=nburn, nsave=nsave, nskip=nskip, ndisplay=500);
prior = list(maxL=15, a0=1, b0=1);
state <- list(cpar=1);
ptm<-proc.time()
res2 = survregbayes(formula = Surv(t1, t2, type="interval2")~x1+x2+x3+
x4+x5+frailtyprior("car", ID), 
data=d, survmodel="PH", selection=TRUE, prior=prior, mcmc=mcmc, state=state, 
dist="loglogistic", Proximity = W);
sfit2=summary(res2); sfit2;
systime2=proc.time()-ptm; systime2;
\end{verbatim}}
\normalsize

Note that the data have to be sorted by region ID before model fitting. The argument \texttt{mcmc} above specifies that the chain is subsampled every 5 iterates to get a total of $2,000$ scans after a burn-in period of $10,000$ iterations. The argument \texttt{prior} is used set all the priors; if nothing is specified, the default priors in the paper are used. The output is given below:

{\footnotesize
	\begin{verbatim}
	Posterior inference of regression coefficients
	(Adaptive M-H acceptance rate: 0.105):
	     Mean      Median    Std. Dev.  95%CI-Low  95%CI-Upp
	x1   1.00002   1.00201   0.09491    0.82401    1.18337 
	x2   0.93568   0.97427   0.16605    0.44710    1.15790 
	x3  -0.68349  -0.66875   0.83818   -2.26155    0.56054 
	x4   0.03566   0.06164   0.75003   -1.42845    1.47316 
	x5  -0.02822   0.01343   0.72955   -1.43050    1.28246 
	
	Posterior inference of precision parameter
	(Adaptive M-H acceptance rate: 0.2652):
	       Mean     Median   Std. Dev.  95%CI-Low  95%CI-Upp
	alpha  0.3843  0.3642  0.1509     0.1541     0.7288   
	
	Posterior inference of conditional CAR frailty variance
	          Mean    Median  Std. Dev.  95%CI-Low  95%CI-Upp
	variance  0.6576  0.6162  0.2504     0.2994     1.2456   
	
	Variable selection:
	       x1,x2   x1,x2,x3  x1,x2,x4  x1,x2,x5  x1,x2,x3,x5  x1,x2,x3,x4  x1,x2,x4,x5
	prop.  0.6490  0.2245    0.0505    0.0485    0.0155       0.0075       0.0045     
	
	Log pseudo marginal likelihood: LPML=-417.0232
	Deviance Information Criterion: DIC=833.0498
	Number of subjects: n=740
\end{verbatim}
}
\normalsize

\textbf{Remarks:} The function \texttt{survregbayes} can also fit a semiparametric survival model (AFT, PH, or PO) with independent Gaussian frailties by setting \texttt{frailtyprior("iid", ID)}, with Gaussian random field frailties by setting \texttt{frailtyprior("grf", ID)}, a model without frailties by removing \texttt{frailtyprior()} in the formula, and a parametric (loglogistic, lognormal or weibull) survival model by specifying \texttt{a0} at a negative value and adding an argument \texttt{state=list(cpar=Inf)}. If FSA is used for GRF frailty models, the number of knots $A$ and the number of blocks $B$ are specified via \texttt{prior=list(nknots=$A$, nblock=$B$)}.

\section{Additional Results for Real Data Applications}

\subsection{Loblolly Pine Survival Data}
Table~\ref{loblolly:summary} presents some baseline characteristics for the trees. 

\begin{table}
	\caption{Loblolly pine data. Baseline characteristics of the 45,525 trees.}
	\label{loblolly:summary}
	\centering
	{
		\begin{tabular}{lcc}
			\hline\noalign{\smallskip}
			\textbf{Categorical variables}    & \textbf{Level}     & \textbf{Proportion (\%)} \\
			\noalign{\smallskip}\hline\noalign{\smallskip}
			Censoring status         & uncensored         & 12.65 \\
			& right censored  & 87.35 \\
			Treatment (treat)	  	  & 1--control			  		& 24.78 \\
			~									& 2--light thinning		  & 40.32 \\
			~									& 3--heavy thinning			& 34.90 \\
			Physiographic region (PhyReg)	  & 1--coastal			  		& 55.53 \\
			~									& 2--piedmont		  					& 37.01 \\
			~									& 3--other									& 7.46 \\
			Crown class (C)		 		& 1--dominant			  & 28.21 \\
			~									& 2--codominant		  & 52.22 \\
			~									& 3--intermediate			& 15.50 \\
			~									& 4--suppressed			& 4.07 \\
			\noalign{\smallskip} \noalign{\smallskip} 
			\textbf{Continuous variables}     & \textbf{Mean} 	& \textbf{Std. Dev.} \\
			\noalign{\smallskip}\hline\noalign{\smallskip}
			Total height of tree in meters (TH)  & 38.47 & 11.77 \\
			Diameter at breast height in cm (DBH)            & 5.88 & 1.77 \\
			\noalign{\smallskip}\hline
		\end{tabular}
	}
\end{table}

\section{Additional Results for Simulations}
\subsection{Simulation I: Areal Data}
\begin{figure}
	\centering
	\subfigure[]{ \includegraphics[width=0.3\textwidth]{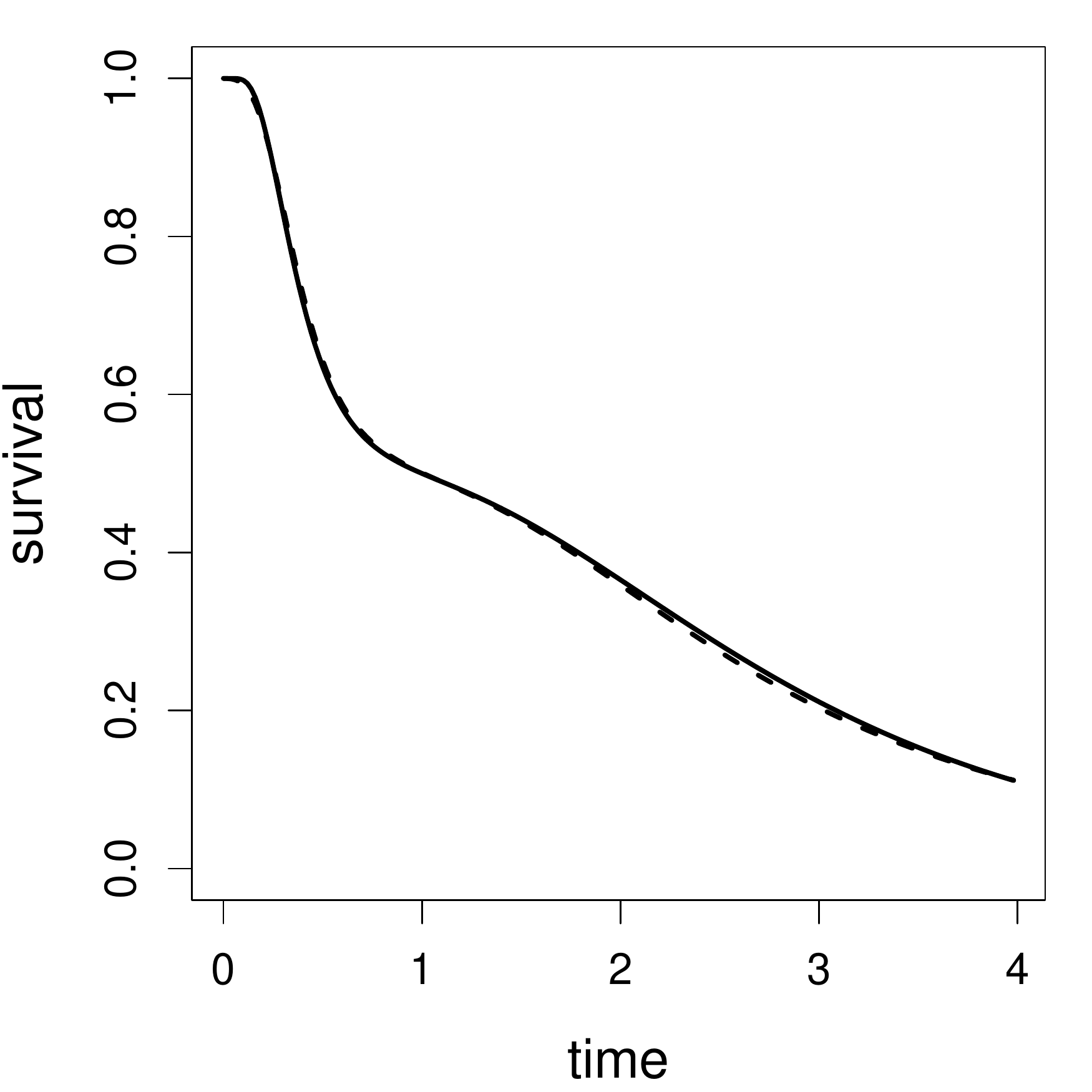} }
	\subfigure[]{ \includegraphics[width=0.3\textwidth]{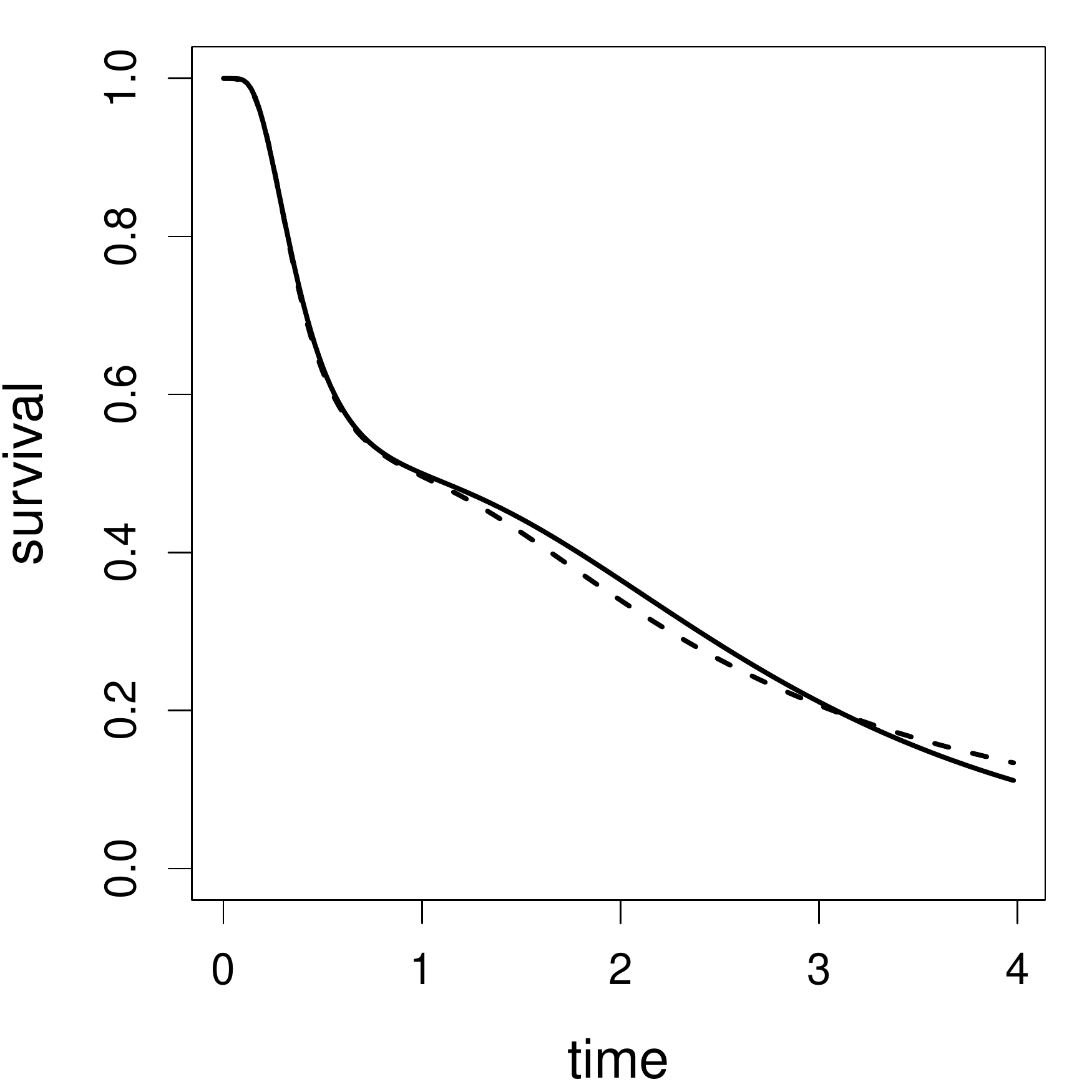} }
	\subfigure[]{ \includegraphics[width=0.3\textwidth]{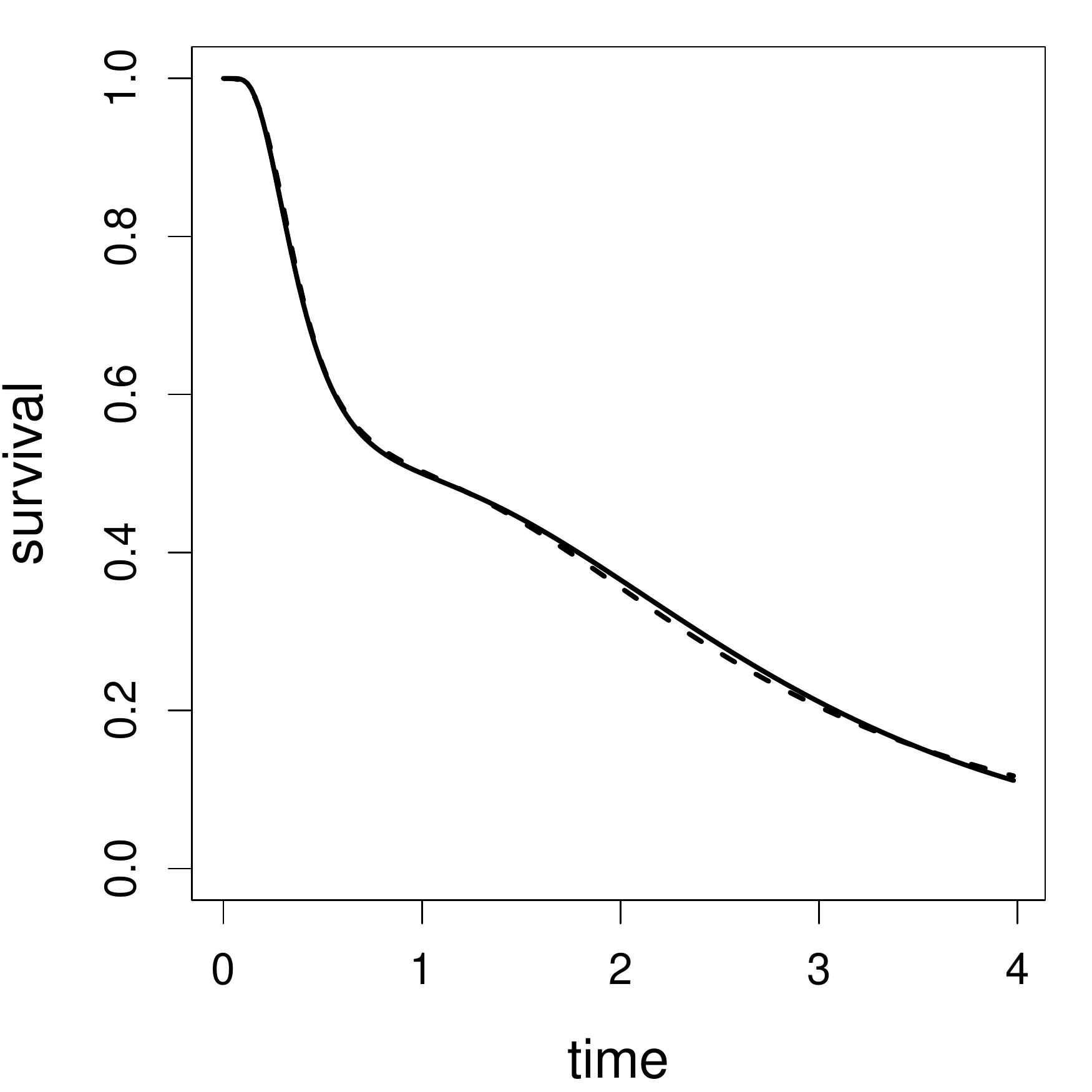} }
	\caption{Simulated I. Mean, across the 500 MC replicates, of the posterior mean of the baseline survival functions under AFT (panel a), PH (panel b) and PO (panel c). The true curves are represented by continuous lines and the fitted curves are represented by dashed lines.}
	\label{sim:curv}
\end{figure}

Figure~\ref{sim:curv} presents the average, across the 500 MC replicates, of fitted (posterior means over a grid of time points) baseline survival functions; the proposed method capably captures complex (here bimodal) baseline survival curves. 

\subsection{Simulation III: Georeferenced Data}
We generated the data using the same settings as \textbf{Simulation I} except that $i=1,\ldots,150, j=1,\ldots,5$, and $v_i$ follows the GRF prior with $\tau^2=1$, $\nu=1$ and $\phi=1$. The locations $\{\bfs_i\}_{i=1}^{150}$ were generated from $[0,10]^2$ uniformly. Table~\ref{sim2:coeff} summaries the results, where we see that the point estimates of ${\bbeta}$ are unbiased under all three models, SD-Est values are close to the corresponding PSDs, and the CP values are close to the nominal $95\%$ level. We also observe that  $\phi$ tends to be overestimated and the standard deviations for $\tau^2$ and $\phi$ are underestimated (because SD-Est is smaller than PSD). Even though, the CP values are still close to $95\%$. The ESS values for $\bbeta$ are much smaller than these obtained for areal data, indicating that the georeferenced spatial dependency makes the posterior samples more correlated. 

\begin{table}
	\caption{Simulation III. Averaged bias (BIAS) and posterior standard deviation (PSD) of each point estimate, standard deviation (across 500 MC replicates) of the point estimate (SD-Est), coverage probability (CP) for the $95\%$ credible interval, and effective sample size (ESS) for each point estimate. }
	\label{sim2:coeff}
	\centering
	{
		\begin{tabular}{ccrrccc}
			\hline
			\hline\noalign{\smallskip}
			Model & Parameter   & BIAS 		& PSD        & SD-Est      & CP	    & ESS \\
			\noalign{\smallskip}\hline\noalign{\smallskip}
			AFT	& $\beta_1=1$    & -0.002	& 0.085      & 0.089       & 0.946	& 1933 \\
			~	& $\beta_2=1$    & -0.000	& 0.045      & 0.042       & 0.964	& 1815 \\
			~ 	& $\tau^2=1$ 	 & 0.000	& 0.329 	 & 0.220	   & 0.948	& 548 \\
			~ 	& $\phi=1$ 	 & 0.082	& 0.388 	 & 0.357	   & 0.962	& 471 \\
			\noalign{\smallskip}
			PH	& $\beta_1=1$    & -0.016	& 0.112      & 0.116       & 0.934	& 1943 \\
			~	& $\beta_2=1$    & -0.015	& 0.068      & 0.068       & 0.942	& 1110 \\
			~ 	& $\tau^2=1$ 	 & 0.042	& 0.451		 & 0.316	   & 0.938	& 366 \\
			~ 	& $\phi=1$ 	 & 0.066	& 0.471 	 & 0.420	   & 0.918	& 351 \\
			\noalign{\smallskip}
			PO	& $\beta_1=1$    & -0.001	& 0.157	     & 0.159       & 0.952  & 3006 \\
			~	& $\beta_2=1$    & 0.003	& 0.087      & 0.088       & 0.944	& 1960\\
			~ 	& $\tau^2=1$ 	 & 0.034	& 0.410		 & 0.341	   & 0.954	& 502\\
			~ 	& $\phi=1$ 	 & 0.313	& 1.361 	 & 0.768	   & 0.912	& 353 \\
			\noalign{\smallskip}\hline
			
		\end{tabular}
	}
\end{table}

\subsection{Simulation IV: Variable Selection}
We next assess the performance of our variable selection method via three simulated examples. For each example, one data set was generated from the PH model with $S_0(t)$ and ICAR as in \textbf{Simulation I}. Under Example 1, we set $\bx_{ij}=(x_{ij1}, \ldots, x_{ij5})$ with $x_{ij1}{\sim} \mathrm{Bernoulli}(0.5)$ and $x_{ij2},\ldots, x_{ij5}\overset{iid}{\sim} N(0,1)$, and $\bbeta=(1,1,0,0,0)'$. Example 2 is identical to Example 1 except that $x_{ij3}=x_{ij2}+0.15z$ where $z\sim N(0,1)$, yielding a $0.989$ correlation between $x_2$ and $x_3$. For Example 3, we set $\bx_{ij}=(x_{ij1}, \ldots, x_{ij10})$ with $\bbeta=(1,1,1,1,1,0,0,0,0,0)$ and $x_{ijk}|z\overset{iid}{\sim}N(z,1)$ where $z\sim N(0,1)$, which induces pairwise correlations of about 0.5. We applied our method to the three simulated datasets using all default priors designed for variable selection. A sample of $10,000$ scans was thinned from $50,000$ after a burn-in period of $10,000$ iterations. Table~\ref{sim:selection} lists the proportions for the four highest frequency models under each example. The results reveal that our method predicts the right model very well even in the presence of extreme collinearity. 

\begin{table}
	\caption{Simulated IV. High frequency models with selected variables.}
	\label{sim:selection}
	\centering
	\scalebox{1}{
		\begin{tabular}{lcclcclc}
			\hline \hline
			\multicolumn{2}{c}{Example 1} & & \multicolumn{2}{c}{Example 2}	& & \multicolumn{2}{c}{Example 3}\\  \cline{1-2}\cline{4-5}\cline{7-8}
			Variables	& Proportions	& & Variables	& Proportions	& & Variables	& Proportions\\
			\hline	
			1 2 		& 0.80			& & 1 2			& 0.49			& & 1-5			& 0.63 \\
			1 2 3 		& 0.08			& & 1 2 3		& 0.22			& & 1-5, 10		& 0.15 \\
			1 2 5 		& 0.05			& & 1 3 		& 0.17			& & 1-5, 7		& 0.09 \\
			1 2 4	 	& 0.05			& & 1 2 5		& 0.04			& & 1-5, 8		& 0.05 \\			
			\hline
		\end{tabular}}
	\end{table}

\subsection{Comparing with Polya Trees}
\cite{Zhao.etal2009} considered the AFT, PH and PO models for right censored areal data, and used the mixture of Polya trees (MPT) prior on the baseline survival function. In their MCMC scheme, most parameters were updated using simple random walk Metropolis-Hastings steps, so a careful tuning of the proposal distribution was required to achieve desirable acceptance rate. We instead used adaptive Metropolis samplers \citep{Haario.etal2001} on most parameters and implemented the three MPT models into an R function \texttt{survregbayes2}; this function can also fit arbitrarily censored data.  We generated data using the same settings as \textbf{Simulation I}, then fitted each model with finite Polya tree level equal to $4$, a $\Gamma(5,1)$ prior on the Polya tree precision parameter, and priors on other parameters similar to Section 2.3 in the main paper. For each MCMC algorithm, 5,000 scans were thinned from 50,000 after a burn-in period of 10,000 iterations.

Table~\ref{simMPT:coeff} summarizes the results for regression parameters $\bbeta$ and the ICAR variance $\tau^2$, including the averaged bias (BIAS) and posterior standard deviation (PSD) of each point estimate (posterior mean for $\bbeta$ and median for $\tau^2$), the standard deviation (across 500 MC replicates) of the point estimate (SD-Est), the coverage probability (CP) of the $95\%$ creditable interval, and effective sample size (ESS) out of 5,000 \citep{Sargent.etal2000} for each point estimate. We can see that effective sample sizes for $\beta_1$ and $\beta_2$ under the MPT AFT are 2 times smaller than those under the TBP AFT. In addition, the MPT PH model provides more biased estimates than the TBP PH. 

Due to the non-smoothness of Polya tree densities, the MPT AFT often suffers poor mixing when the true baseline survival function is far away from the centering parametric distribution family $S_{\btheta}$ and uncensored survival times are available. For example, for right censored data, the likelihood will involve $f_{\bx_{ij}}(t)=e^{\bx_{ij}'\bbeta+v_i} f_0(e^{\bx_{ij}'\bbeta+v_i}t)$, where $f_0(\cdot)$ is the density of a Polya tree. Note that $f_0(\cdot)$ consists of many big jumps when the precision parameter of the Polya trees is small, and hence a tiny change in $\bbeta$ imply a big jump in the likelihood value, leading to poor mixing. However, MCMC mixing issues are mitigated for interval censored data, since only the survival function $S_0(e^{\bx_{ij}'\bbeta+v_i} t)$ is involved in the likelihood and $S_0(t)$ is continuous. 

\begin{table}
	\caption{Simulation under MPT. Averaged bias (BIAS) and posterior standard deviation (PSD) of each point estimate, standard deviation (across 500 MC replicates) of the point estimate (SD-Est), coverage probability (CP) for the $95\%$ credible interval, and effective sample size (ESS) out of 5,000 with thinning=10 for each point estimate. }
	\label{simMPT:coeff}
	\centering
	{
		\begin{tabular}{ccrrccc}
			\hline
			\hline\noalign{\smallskip}
			Model & Parameter   & BIAS 		& PSD        & SD-Est      & CP	    & ESS \\
			\noalign{\smallskip}\hline\noalign{\smallskip}
			AFT	& $\beta_1=1$    & 0.002	& 0.094      & 0.071       & 0.986	& 1009 \\
			~	& $\beta_2=1$    & 0.002	& 0.050      & 0.038       & 0.988	& 1079 \\
			~ 	& $\tau^2=1$ 	 & 0.013	& 0.309 	 & 0.243	   & 0.976	& 3760\\
			\noalign{\smallskip}
			PH	& $\beta_1=1$    & -0.045	& 0.099      & 0.098       & 0.932	& 2887 \\
			~	& $\beta_2=1$    & -0.043	& 0.060      & 0.060       & 0.874	& 1794 \\
			~ 	& $\tau^2=1$ 	 & -0.084	& 0.318		 & 0.280	   & 0.954	& 3599 \\
			\noalign{\smallskip}
			PO	& $\beta_1=1$    & -0.014	& 0.149	     & 0.142       & 0.962  & 3579 \\
			~	& $\beta_2=1$    & -0.029	& 0.082      & 0.078       & 0.938	& 2561\\
			~ 	& $\tau^2=1$ 	 & -0.038	& 0.407		 & 0.346	   & 0.966	& 2903 \\
			\noalign{\smallskip}\hline
			
		\end{tabular}
	}
\end{table}

\subsection{Model Selection via LPML and DIC}\label{sec:sim:model:selection}
We next demonstrate via simulations that the LPML and DIC are reasonable criteria for model selection among AFT, PH and PO models. Arbitrarily censored survival data of size $n=740$ and $n=1850$ were generated from each of the three models with ICAR frailties using the same settings as \textbf{Simulation I}. For each model, 200 MC replicates were generated. We fitted each dataset using all three models with the default priors and the same MCMC settings as \textbf{Simulation I}. Table~\ref{sim:model:selection} (under log-logistic $S_{\btheta}(\cdot)$) presents the proportion (out of 200 MC replicates) of times each model is picked. The model picked is the one with largest LPML or smallest DIC. DIC and LPML yield very similar proportions for $n=740$ and identical results when $n=1850$, indicating that the two criteria are consistent for model comparison. When the true model is PH, DIC has a $3\%$ chance of picking PO under $n=740$, but is reduced to zero for the larger sample size $n=1850$.

\begin{table}[tbh] 
	\caption{Simulation for model selection via LPML and DIC. Proportion of times DIC or LPML selects each model when truth is known out of 200 replicated datasets.}
	\label{sim:model:selection}
	\centering
	\begin{tabular}{ccccccccccccc} 
		\hline
		&  & \multicolumn{3}{c}{log-logistic} &~& \multicolumn{3}{c}{Weibull} &~& \multicolumn{3}{c}{log-normal}\\
		&  & \multicolumn{3}{c}{Model picked} &~& \multicolumn{3}{c}{Model picked}&~& \multicolumn{3}{c}{Model picked}\\
		\cline{3-5} \cline{7-9} \cline{11-13}
		True model & Criteria & AFT & PH & PO &~& AFT & PH & PO &~& AFT & PH & PO \\ 
		\hline
		& &  \multicolumn{11}{c}{$n=740$} \\
		AFT	& DIC   & 1.000 & 0.000 & 0.000 &~& 1.000 & 0.000 & 0.000 &~& 1.000 & 0.000 & 0.000\\
		~	  & LPML & 1.000 & 0.000 & 0.000 &~& 1.000 & 0.000 & 0.000 &~& 1.000 & 0.000 & 0.000\\
		PH   & DIC   & 0.000 & 0.985 & 0.015 &~& 0.000 & 1.000 & 0.000 &~& 0.000 & 1.000 & 0.000\\
		~	  & LPML & 0.000 & 0.970 & 0.030 &~& 0.000 & 1.000 & 0.000 &~& 0.000 & 0.995 & 0.005\\
		PO   & DIC   & 0.000 & 0.000 & 1.000 &~& 0.000 & 0.075 & 0.925 &~& 0.000 & 0.000 & 1.000\\
		~	  & LPML & 0.000 & 0.000 & 1.000 &~& 0.000 & 0.025 & 0.975 &~& 0.000 & 0.000 & 1.000\\
		\hline
		& &  \multicolumn{11}{c}{$n=1850$} \\
		AFT	& DIC   & 1.000 & 0.000 & 0.000 &~& 1.000 & 0.000 & 0.000 &~& 1.000 & 0.000 & 0.000\\
		~	  & LPML & 1.000 & 0.000 & 0.000 &~& 1.000 & 0.000 & 0.000 &~& 1.000 & 0.000 & 0.000\\
		PH   & DIC   & 0.000 & 1.000 & 0.000 &~& 0.000 & 1.000 & 0.000 &~& 0.000 & 1.000 & 0.000\\
		~	  & LPML & 0.000 & 1.000 & 0.000 &~& 0.000 & 1.000 & 0.000 &~& 0.000 & 1.000 & 0.000\\
		PO   & DIC   & 0.000 & 0.000 & 1.000 &~& 0.000 & 0.010 & 0.990 &~& 0.000 & 0.000 & 1.000\\
		~	  & LPML & 0.000 & 0.000 & 1.000 &~& 0.000 & 0.010 & 0.990 &~& 0.000 & 0.000 & 1.000\\
		\hline
	\end{tabular}
\end{table}

\subsection{Sensitivity Analysis of The TBP's Centering Distribution}
The TBP prior is centered at a parametric family of distributions. The log-logistic  $S_{\btheta}(t)=\{1+(e^{\theta_1} t)^{\exp(\theta_2)}\}^{-1}$, the log-normal $S_{\btheta}(t)=1-\Phi\{(\log t+\theta_1)\exp(\theta_2)\}$, and the Weibull $S_{\btheta}(t)=1-\exp\left\{-(e^{\theta_1} t)^{\exp(\theta_2)} \right\}$ families are implemented in the software. We next demonstrate via simulations that posterior inference and model selection is not very sensitive to the choice of centering parametric family. Table~\ref{sim:model:selection} presents the proportion (out of 200 MC replicates) of times each model is picked under all settings when data are generated as in the previous simulation J.5. When the true model is PH with $n=740$, the Weibull centering distribution has a improved chance to pick the correct model than log-logistic, indicating that the Weibull slightly favors PH for this bimodal baseline $S_0$. As sample sizes increase, all three centering distributions give the same model selection results. 

\begin{figure}
	\centering
	\subfigure[]{ \includegraphics[width=0.3\textwidth]{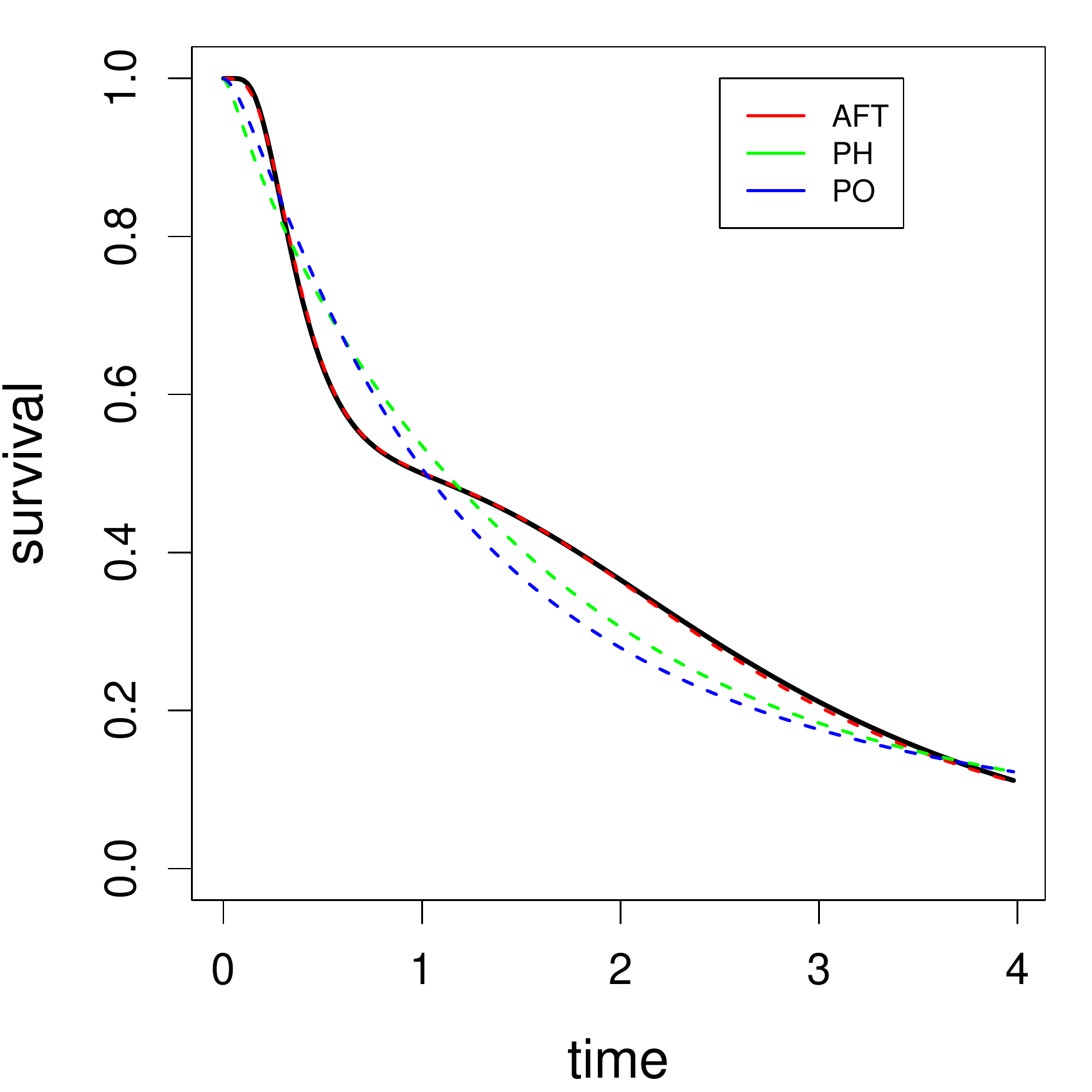} }
	\subfigure[]{ \includegraphics[width=0.3\textwidth]{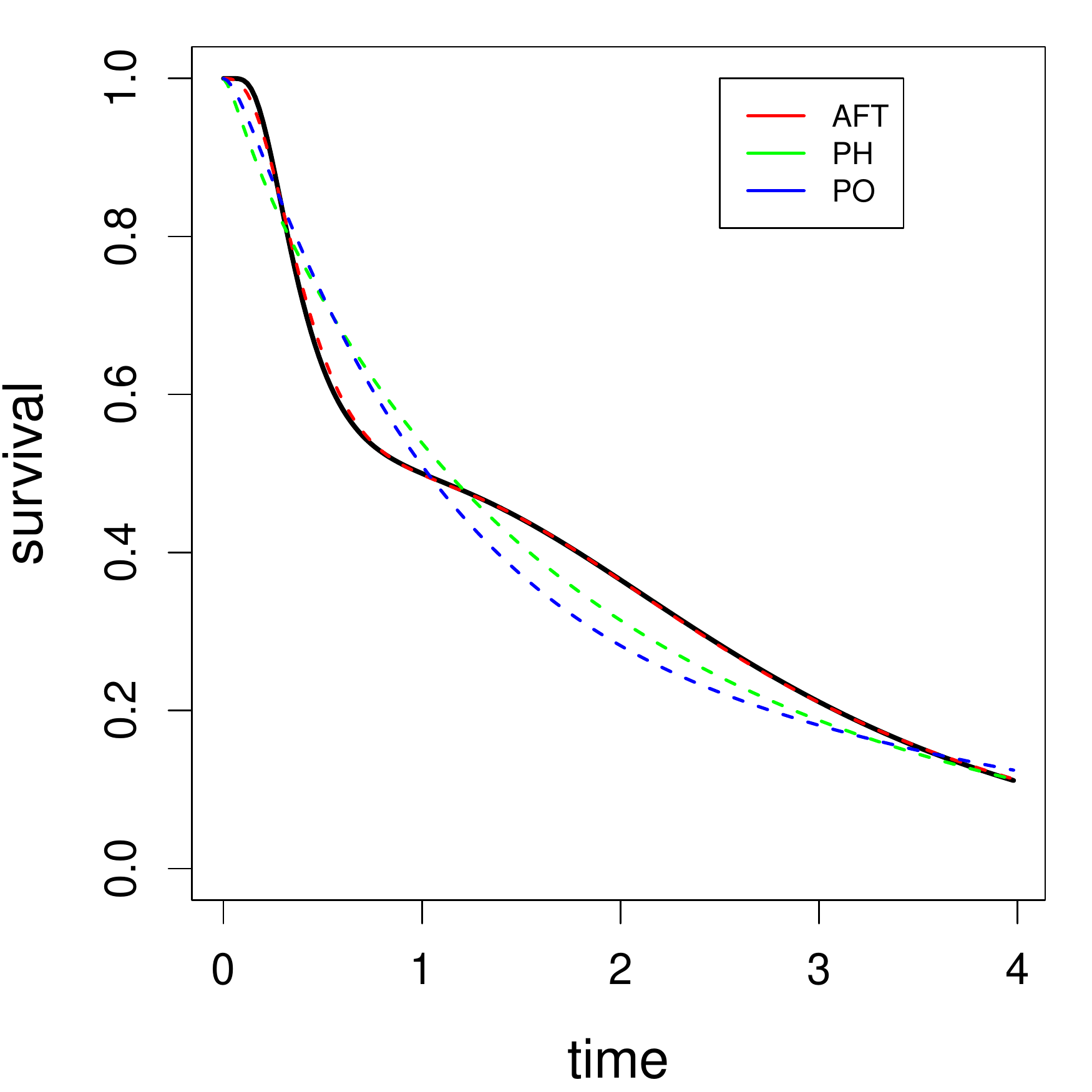} }
	\subfigure[]{ \includegraphics[width=0.3\textwidth]{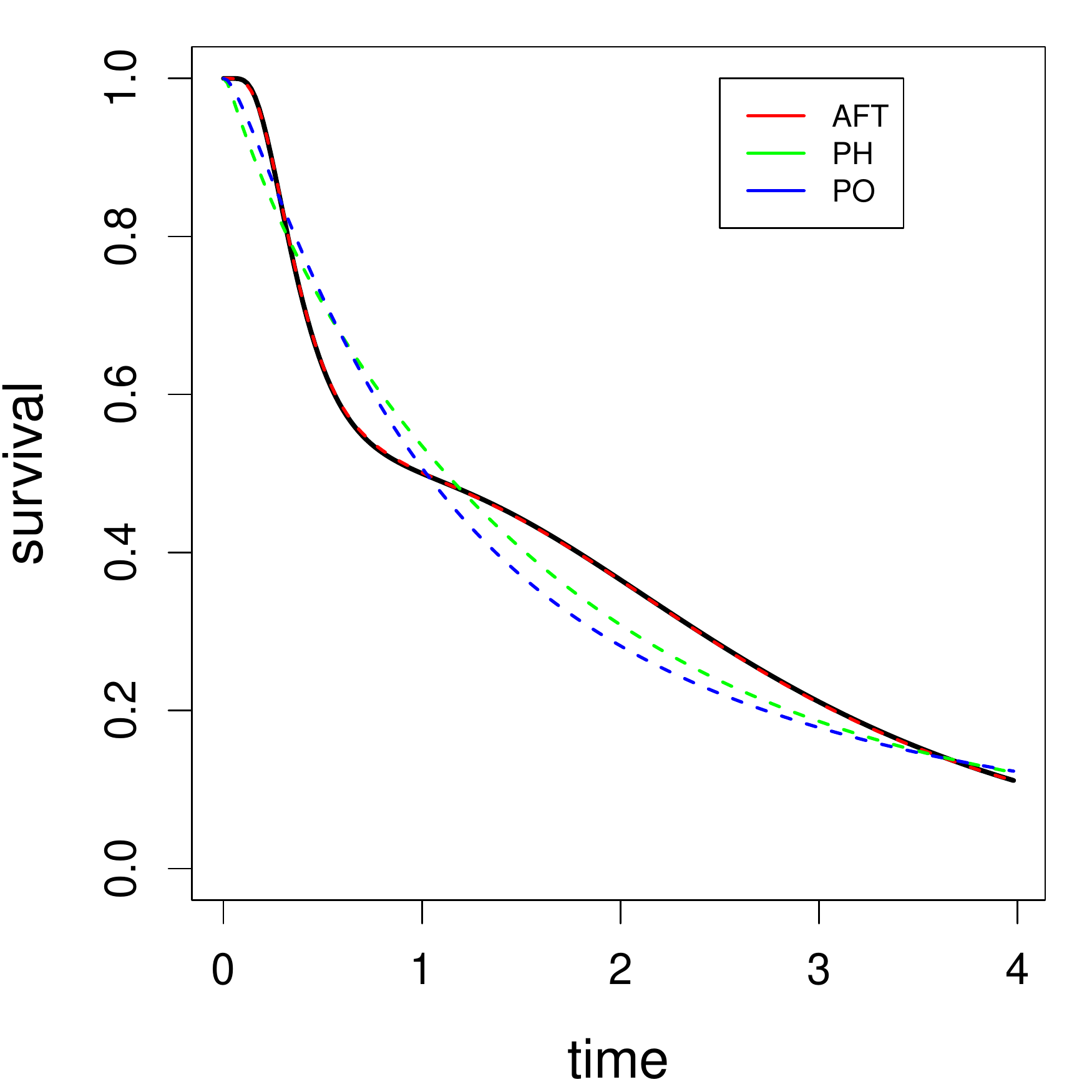} }
	\caption{Simulation for sensitivity analysis of the TBP's centering distribution when AFT is the true model. Mean, across the 200 MC replicates, of the posterior mean of the baseline survival functions under log-logistic (panel a), Weibull (panel b) and log-normal (panel c). The true curves are represented by continuous lines and the fitted curves are represented by dashed lines (red is for AFT, green is for PH and blue is for PO).}
	\label{sim:curv:comparision:aft}
\end{figure}

\begin{figure}
	\centering
	\subfigure[]{ \includegraphics[width=0.3\textwidth]{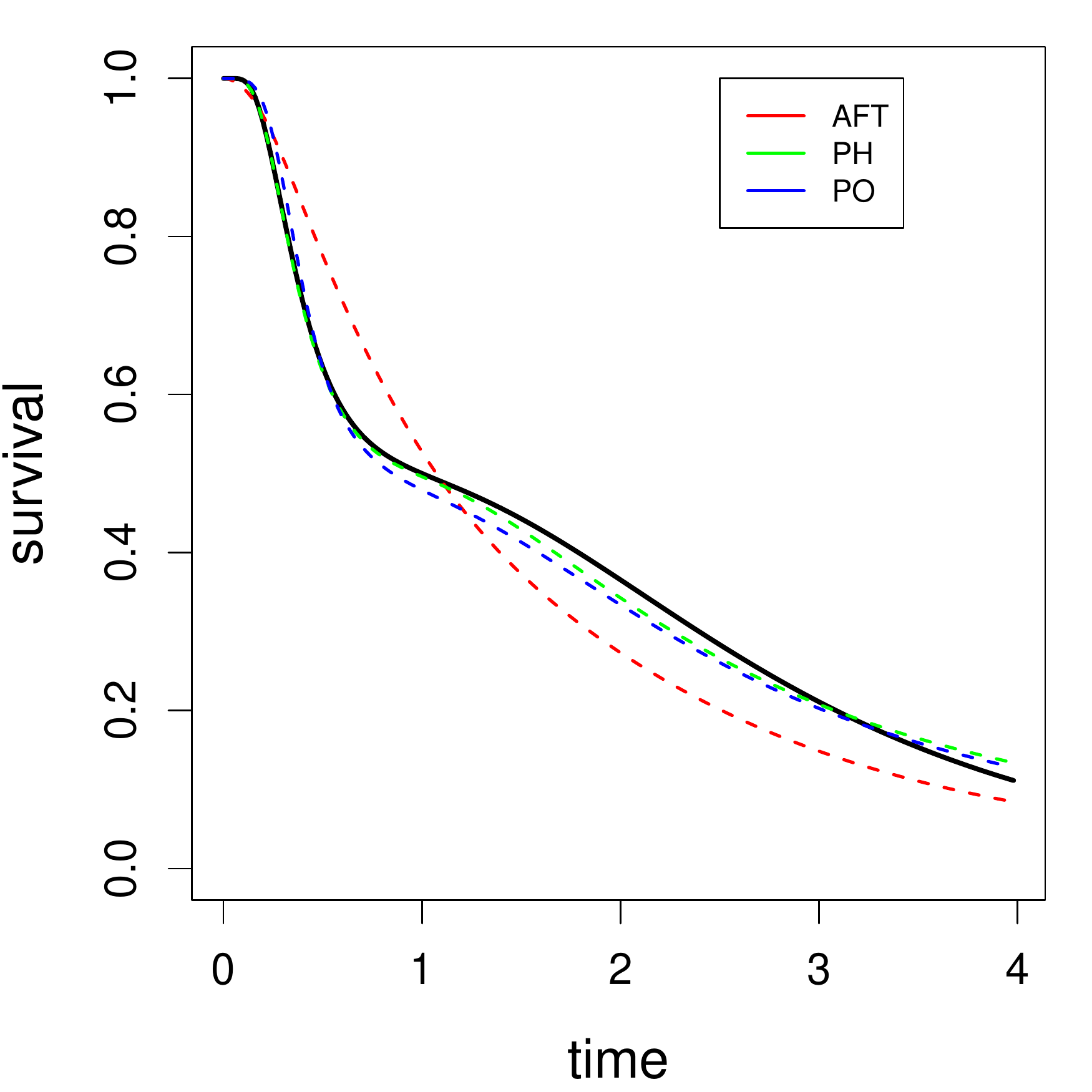} }
	\subfigure[]{ \includegraphics[width=0.3\textwidth]{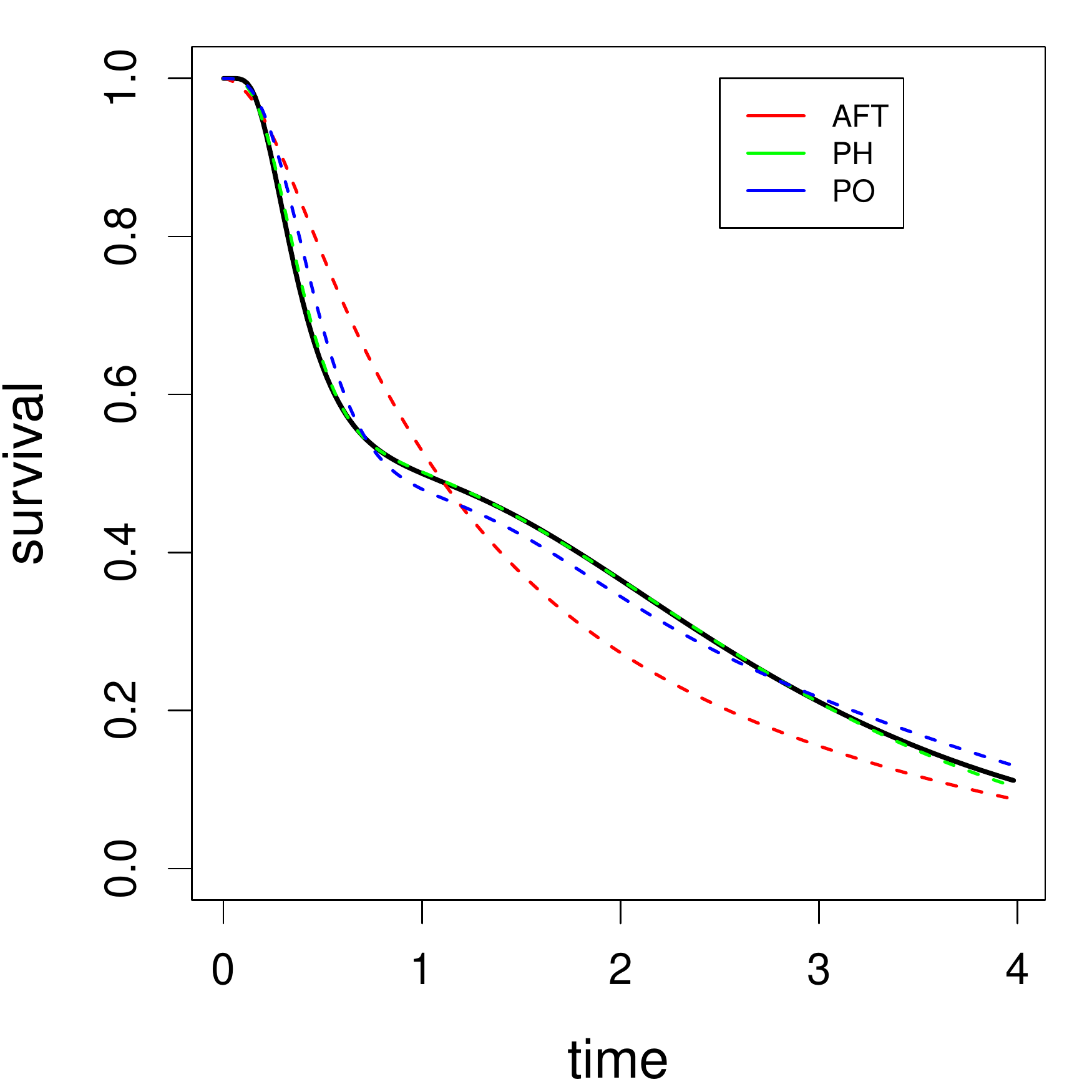} }
	\subfigure[]{ \includegraphics[width=0.3\textwidth]{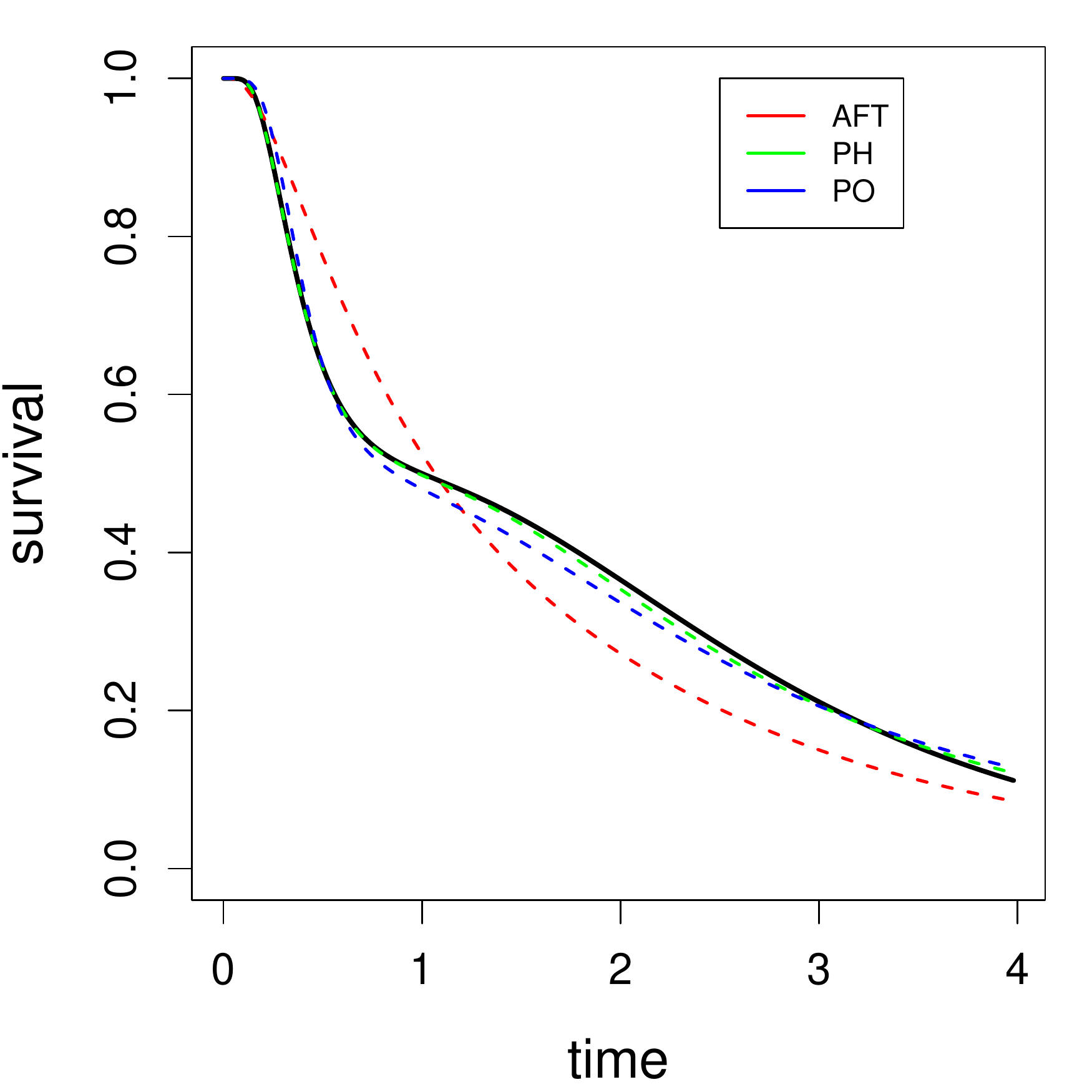} }
	\caption{Simulation for sensitivity analysis of the TBP's centering distribution when PH is the true model. Mean, across the 200 MC replicates, of the posterior mean of the baseline survival functions under log-logistic (panel a), Weibull (panel b) and log-normal (panel c). The true curves are represented by continuous lines and the fitted curves are represented by dashed lines (red is for AFT, green is for PH and blue is for PO).}
	\label{sim:curv:comparision:ph}
\end{figure}

\begin{figure}
	\centering
	\subfigure[]{ \includegraphics[width=0.3\textwidth]{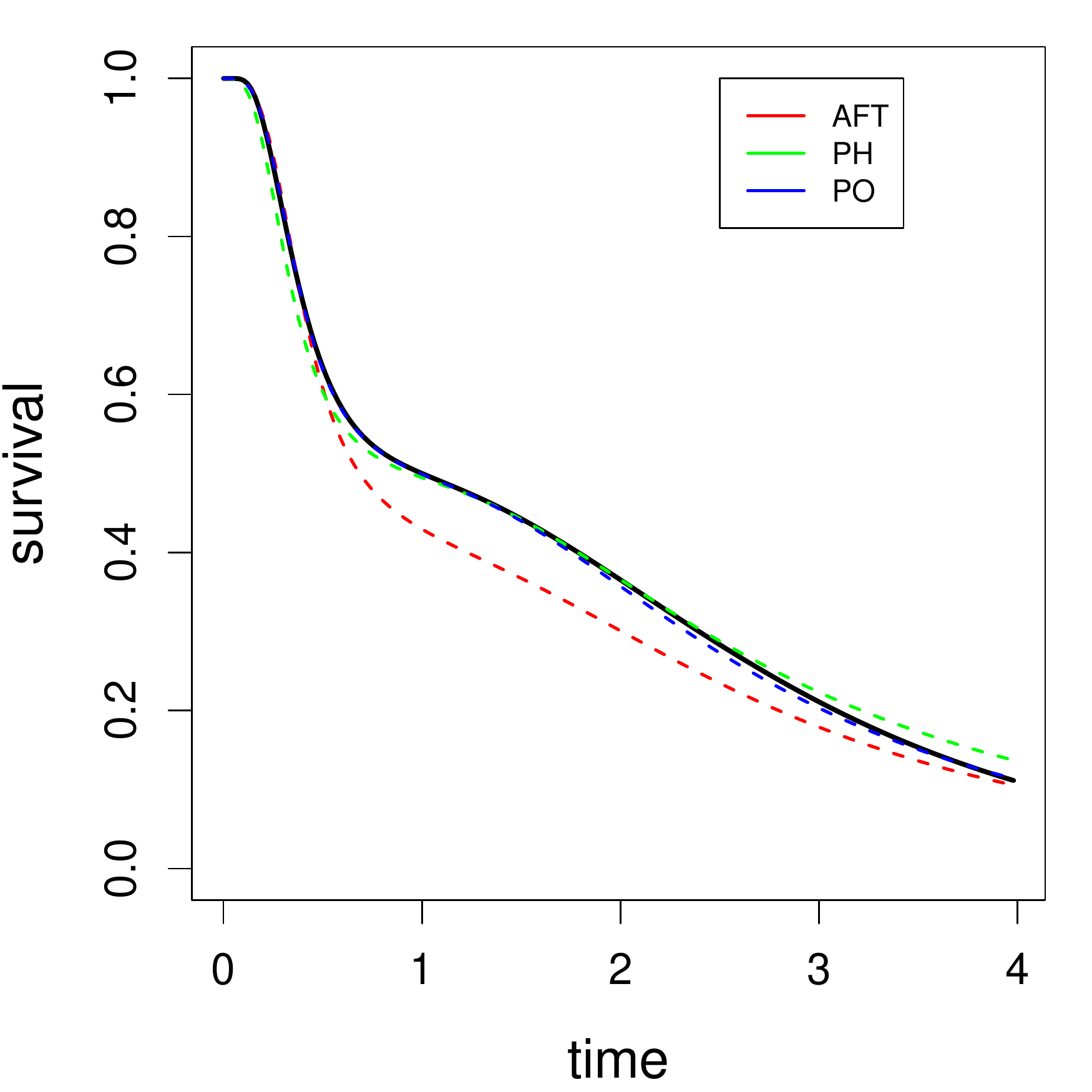} }
	\subfigure[]{ \includegraphics[width=0.3\textwidth]{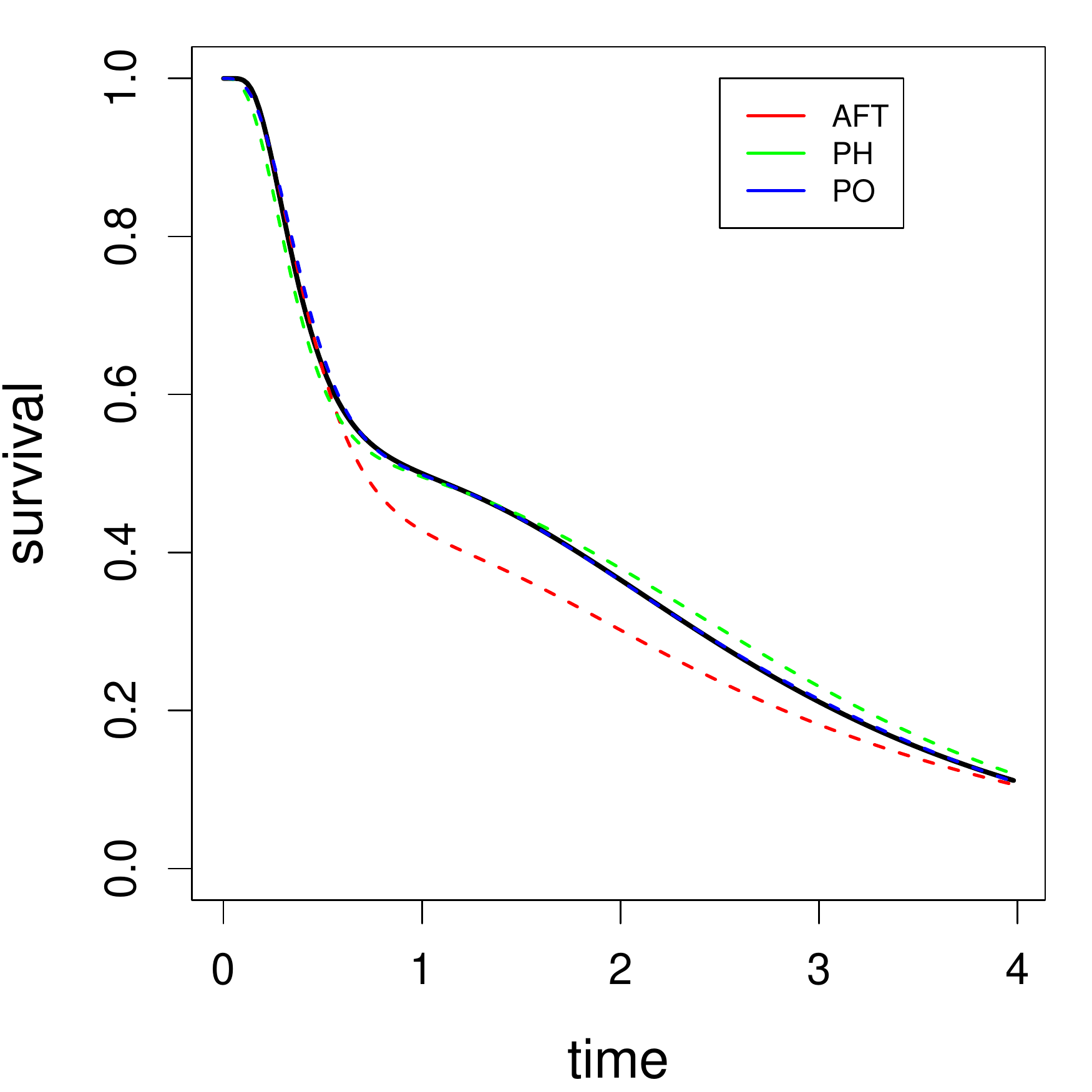} }
	\subfigure[]{ \includegraphics[width=0.3\textwidth]{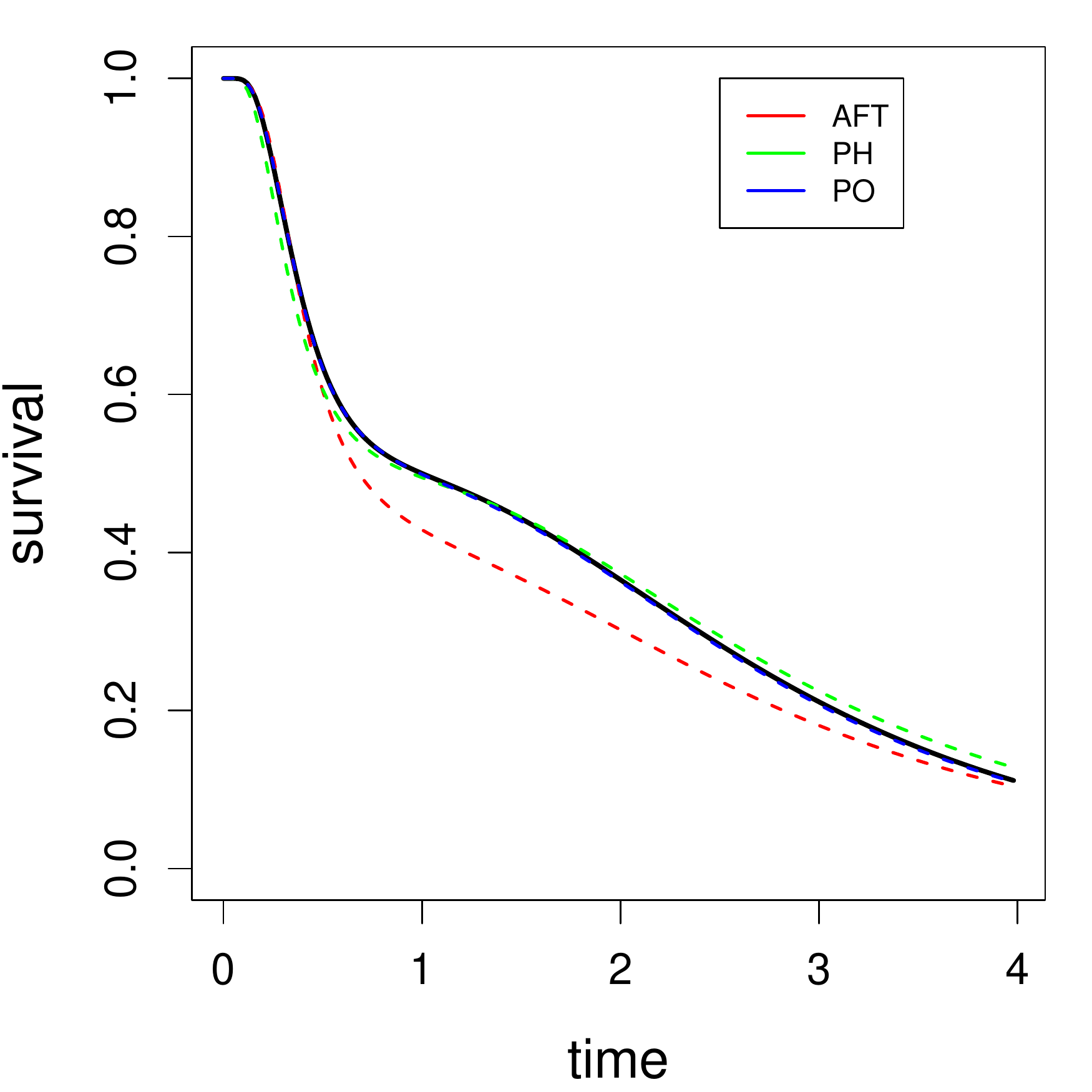} }
	\caption{Simulation for sensitivity analysis of the TBP's centering distribution when PO is the true model. Mean, across the 200 MC replicates, of the posterior mean of the baseline survival functions under log-logistic (panel a), Weibull (panel b) and log-normal (panel c). The true curves are represented by continuous lines and the fitted curves are represented by dashed lines (red is for AFT, green is for PH and blue is for PO).}
	\label{sim:curv:comparision:po}
\end{figure}

Figures~\ref{sim:curv:comparision:aft}, \ref{sim:curv:comparision:ph}, and \ref{sim:curv:comparision:po} present the averaged (across the 200 MC replicates) fitted baseline survival functions under three centering distribution families when the true model is AFT, PH and PO, respectively. Overall, the three families yield almost the same estimates regardless of what the true model is, although we do see that Weibull provides a slightly better estimate than the other two (Figure~\ref{sim:curv:comparision:ph}) when the true model is PH with the bimodal baseline $S_0$ and PH is used to fit the model. We also compared the inference results on the coefficient estimates (not shown), which resulted in very similar biases, coverage probabilities and effective sample sizes.

\newpage
\bibliographystyle{apalike}
\bibliography{bibliography}\label{lastpage}